\documentclass[journal,onecolumn,12pt]{IEEEtran}
\usepackage{graphicx}
  \usepackage{here}
  \usepackage{array}
  \usepackage{color,amsmath,here}
  \usepackage{amssymb}
  \usepackage{enumerate}
  \usepackage{ascmac}
  \usepackage{tascmac}
  \usepackage{fancybox}
  \usepackage{float}
  \usepackage{amsmath}
  \usepackage{mathrsfs}
  \usepackage{ascmac}

\definecolor{mygreen}{rgb}{0, 0.7, 0}
\definecolor{myyellow}{rgb}{0.7, 0.7, 0}
\definecolor{mypurple}{rgb}{0.42, 0, 0.84}

\newtheorem{assumption}{Assumption}
\newtheorem{proposition}{Proposition}

\newtheorem{remark}            {Remark}

\newcommand{\req}[1]{(\ref{#1})}

\usepackage{url}

\author{Tomonori Sadamoto$^{1}$\footnote{$^{1}$ Department of Systems and Control Engineering, Graduate School of
Engineering, Tokyo Institute of Technology; 2-12-1, Meguro, Tokyo, Japan. \texttt{\{sadamoto, ishizaki, imura\}@sc.e.titech.ac.jp}}, Aranya Chakrabortty$^{2}$\footnote{$^{2}$ Electrical \& Computer Engineering, North Carolina State University;
Raleigh, NC 27695. \texttt{achakra2@ncsu.edu}}, Takayuki Ishizaki$^{1}$, Jun-ichi Imura$^{1}$
}

\newif\ifPDF \ifx\pdfoutput\undefined\PDFfalse \else\ifnum\pdfoutput > 0\PDFtrue \else\PDFfalse \fi \fi
\ifPDF 
\usepackage[pdftex, plainpages = false, colorlinks=true, linkcolor=black, citecolor = green!50!blue, urlcolor = blue, filecolor=black, pagebackref=false, hypertexnames=false,  pdfpagelabels ]{hyperref}
\fi

\begin{document}
\title{Dynamic Modeling, Stability, and Control of Power Systems with Distributed Energy Resources}

\maketitle

%
\IEEEpeerreviewmaketitle


\begin{abstract}
This article presents a suite of new control designs for next-generation electric smart grids. The future grid will consist of thousands of non-conventional renewable generation sources such as wind, solar, and energy storage. These new components are collectively referred to as distributed energy resources (DER). The article presents a comprehensive list of dynamic models for DERs, and shows their coupling with the conventional generators and loads. It then presents several innovative control designs that can be used for facilitating large-scale DER integration. Ideas from decentralized retrofit control and distributed sparsity-promoting optimal control are used for developing these designs, followed by illustrations on an IEEE power system test model. 
\end{abstract}


\section{Introdution}
Significant infrastructural changes are currently being implemented on power system
networks around the world by maximizing the penetration of renewable energy, by installing new transmission lines, by adding flexible loads, by promoting independence in power production by disintegrating the grid into micro-grids, and so on \cite{efficiencystrengthening}. The shift of energy supply from large central generating stations to smaller producers such as wind farms, solar photovoltaic (PV) farms, roof-top PVs, and energy storage systems, collectively known as Distributed Energy Resources (DERs) or Inverter-Based Resources (IBRs), is accelerating at a very rapid pace. Hundreds of power electronic devices are being added, creating hundreds of new control points in the grid. This addition is complimented by an equal progress in sensing technology, whereby high-precision, high sampling-rate, GPS-synchronized dynamic measurements of voltages and currents are now available from sensors such as Phasor Measurement Units (PMUs) \cite{fardanesh1998multifunctional}. 
With all these transformational changes happening in the grid, operators are inclining to explore new control methods that go far beyond how the grid is controlled today. A list of these changes is summarized in Table~\ref{table_transition}.

In the current state-of-art, power system controllers, especially the ones that are responsible for transient stability and power oscillation damping, are all operated in a decentralized and uncoordinated fashion using local output feedback only. A survey of these controllers is provided in Appendix \ref{sidebar-surveyCofSG}. With rapid modernization of the grid, these local controllers, however, will not be tenable over the long-term. Instead, system-wide coordinated controllers will become essential. Such controllers where signals measured at one part of the grid are communicated to other remote parts for feedback are called wide-area controllers \cite{chakrabortty2013introduction}.

Wide-area control (WAC) alone, however, will not be enough either. It may improve the stability and dynamics performance of the legacy system, but will not be able to keep up with the unpredictable rate at which DERs are being added to the grid. Every time a new DER is added it will be almost impossible for an operator to retune all the wide-area control gains to accommodate the change in dynamics. DERs have high variability and intermittency, and need to be operated in a plug-and-play fashion. Accordingly their controllers need to be local, decentralized, and modular in both design and implementation. In other words, neither should the design of one DER controller depend on that of another nor should either of these two controllers need to be updated when a third DER is added in the future. The overall control architecture for the future grid needs to be a combination of these {\it decentralized} plug-and-play DER controllers and {\it distributed} wide-area controllers.

The objective of this article is twofold. The first objective is to present a suite of control methods, for developing this combined control architecture. The presentation will be based on recent results reported by the authors in \cite{ishizaki2016distributed} and \cite{jain2017online}. For brevity, the discussion will be limited to only one particular application - namely, adding damping to the oscillations in power flows after both small and large disturbances, also called {\it power oscillation damping} (POD) \cite{kundur1994power}. POD is one of the most critical real-time control problems in today's power grid, and its importance is only going to increase with DER integration. The applicability of the design, however, go far beyond just POD to many other grid control problems such as frequency control, voltage control, and congestion relief. Efficacy of the methods for enhancing transient stability as a bonus application will be illustrated via simulations. While many papers on decentralized DER control and distributed wide-area control exist in the literature (surveys on these two control methods are given in Appendix \ref{sidebar-surveyLC} and \ref{sidebar-surveyWAC}, respectively), very few have studied the simultaneous use of both. Moreover, most DER controllers reported so far lack the modularity and plug-and-play characteristic explained above. The control methods presented in this article address all of these challenges.
The second objective is to present a comprehensive list of mathematical models of the various components of a power grid ranging from synchronous generators, their internal controllers, loads, wind and solar farms, batteries to the power electronic device interfaces and associated control mechanisms for each of these components. While these models are individually well-cited in the literature, very few references so far have collected all of them together to understand the holistic dynamic behavior of an entire grid. The three main contributions of the article are summarized as follows: 
\begin{itemize}
 \item Present an end-to-end differential-algebraic model of a power system in its entirety - including, synchronous generators, wind farms, solar farms, energy storages, power electronic converters, and controllers for each device.
 \item Show how DERs and power electronic devices affect small-signal stability and dynamic performance of the grid.
 \item Present a two-layer control architecture for tomorrow's grid, where one layer consists of decentralized plug-and-play controllers for  power electronic converter control of DERs, and another layer consists of system-wide distributed controllers actuated through the generators. 
\end{itemize}

The rest of the article is organized as follows. The next section describes the dynamic model of a power system, integrated with different types of DERs. A general framework for modeling is provided first, followed by details of each individual component model. The section \ref{sec-impact} demonstrates the impacts of DERs on power system dynamics through numerical simulations. Motivated by these simulation results, the section \ref{sec-newapp} shows newly-developed decentralized DER control laws using the idea of {\it retrofit control} \cite{ishizaki2016distributed}, as well as distributed wide-area controllers for damping of low frequency oscillations using sparse optimal control \cite{jain2017online}. The effectiveness of this combined control strategy is demonstrated on the IEEE 68-bus power system with wind and solar farms. The article concludes with a list of open research problems. 

 \begin{table}
  \begin{center}
  \caption{Transitions in power grid infrastructure}
    \begin{tabular}{|p{28mm}|p{50mm}|p{70mm}|} \hline
      & \multicolumn{1}{c|}{Today} & \multicolumn{1}{c|}{Tomorrow}  \\ \hline \hline
    Generation units & Synchronous generators & + Renewable generation with power electronic converters, storage devices  \\ \hline
    Load Mechanisms & Conventional consumers                 & Prosumers (Producers + Consumers), PHEVs, smart buildings, load-side decision-making. \\ \hline
    Transmission & Moderately dense, one-directional power flow &  More dense, bi-directional power flow \\ \hline    
    Control      & Local control (AVR, PSS, FACTS, PI) & System-wide control, plug-and-play control \\ \hline
    Controller Design           & Model-driven, seldom retuned           & Measurement-driven, adaptive, reconfigurable  \\ \hline
     Monitoring System & Slow sensing (SCADA)                & Fast sensing (WAMS based on PMUs)                        \\ \hline
    Communication               & Low-bandwidth legacy communication                 & High-speed communication, more importance on cyber-security and privacy \\ \hline
    Data Processing                & Offline processing, low volumes of data                 & Real-time processing using cloud computing, massive volumes of data               \\ \hline 
   \end{tabular}
 \label{table_transition}
  \end{center}
 \end{table}

\begin{table}
\begin{center}
  \caption{Acronyms used in this article and their complete expressions.}
\begin{tabular}{|p{23mm}|l|p{105mm}|}\hline 
  Acronyms & Complete expression  \\\hline 
 PV &  Photovoltaic\\ 
 DER &  Distributed Energy Resources \\
 IBR & Inverter-Based Resources \\
 FACTS &  Flexible Alternating Current Transmission System \\
 PHEV &  Plug-in Hybrid Electric Vehicle \\
 SCADA & Supervisory Control And Data Acquisition \\
 WAMS & Wide Area Monitoring Systems \\
 GPS & Global Positioning System \\
 PMU & Phasor Measurement Unit \\
 WAC & Wide Area Control \\
 POD & Power Oscillation Damping  \\
 AVR & Automatic Voltage Regulator \\
 PSS & Power System Stabilizer \\
 DFIG & Doubly-Fed Induction Generator \\
 B2B & Back-to-Back \\
 RSC & Rotor-Side Converter \\
 GSC & Grid-Side Converter \\
 MPP & Maximum Power Point \\
 LFC & Load Frequency Control \\ \hline
\end{tabular}
 \label{nomen_acronym}
\end{center}
\end{table}

\section{Power System Models}\label{sec:model}
\begin{table}
  \caption{Nomenclature for power system models. The tie-line parameters for constructing ${\bf Y}$ of the IEEE 68-bus power model, which is a benchmark model used in the simulation, are available in \cite{pal2006robust}. All variables are considered to be in per unit unless otherwise stated. }
\begin{tabular}{|p{23mm}|l|p{105mm}|}\hline 
  Symbol &  Numerical value & Description \\\hline 
  $\bar{\omega}$ & $120\pi$ & base angular speed for a 60 Hz power system. Unit of $\bar{\omega}$ is rad/sec. \\
  $N$ & & number of buses \\
  ${\bf V}_k \in \mathbb C$ && $k$-th bus voltage \\
  $P_k$, $Q_k$ && active and reactive power injected from the $k$-th component\\
  $x_k$ & & state of the $k$-th component \\
  $u_k$ & & control input of the $k$-th component \\
  $\alpha_k$ && model parameter depending on operating point\\
  ${\bf Y} \in \mathbb C^{N \times N}$ && admittance matrix \\
  $\mathbb N_{\rm G}$, 
  $\mathbb N_{\rm L}$,
  $\mathbb N_{\rm W}$
  $\mathbb N_{\rm S}$,
  $\mathbb N_{\rm E}$,
  $\mathbb N_{\rm N}$ && index set of the buses connecting to generators, loads, wind
 farms, solar farms, energy storages, and that of non-unit buses. These sets are disjoint, and 
$\mathbb N_{\rm G} \cup \mathbb N_{\rm L} \cup  \mathbb N_{\rm W} \cup  \mathbb
N_{\rm S} \cup \mathbb N_{\rm E} \cup  \mathbb N_{\rm N} = \{1,\ldots,N\}$ \\ \hline
\end{tabular}
 \label{nomen_network}
\end{table}

\begin{figure}[t]
  \begin{center}
    \includegraphics[clip,width=165mm]{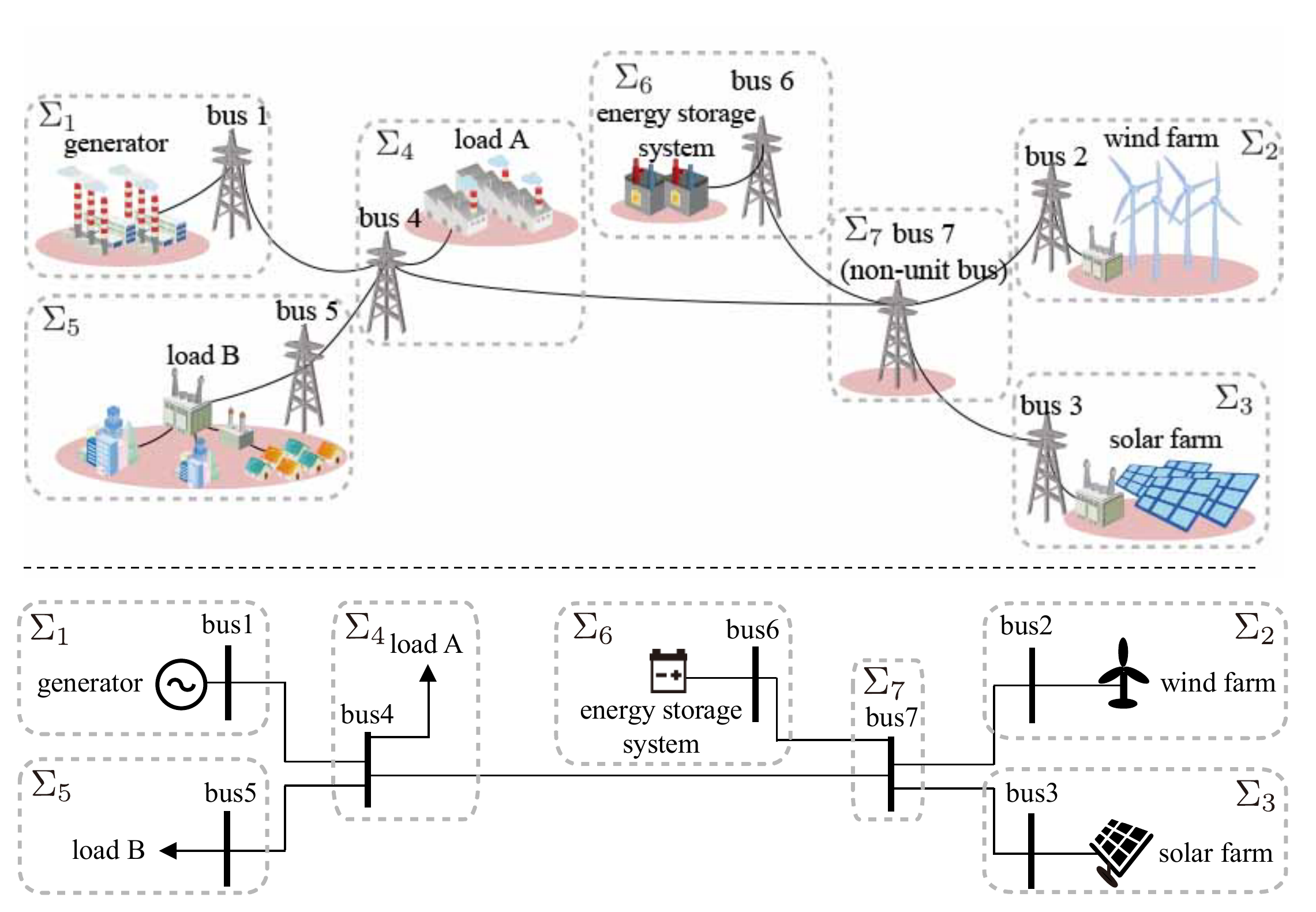}
    \caption{(Top) Illustrative example of a power system. (Bottom) Schematic diagram of the model.}
    \label{fig_example}
  \end{center}
\end{figure}

\begin{figure}[t]
  \begin{center}
    \includegraphics[clip,width=140mm]{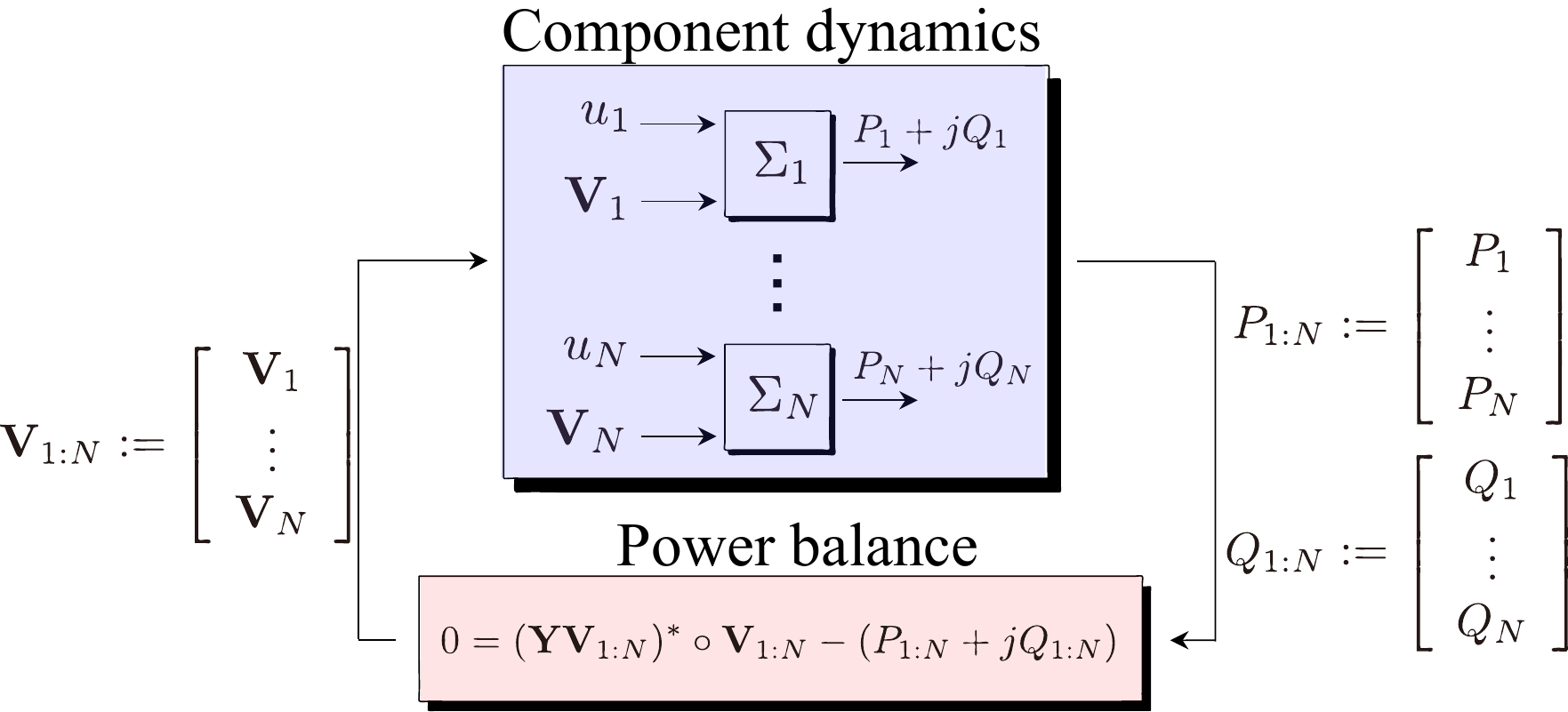}
    \caption{Signal-flow diagram of power system model \req{model_comp} and \req{inter}}
    \label{fig_signal}
  \end{center}
\end{figure}

First, the dynamic models of the four core components of a power system are developed - namely, generation, transmission, load, and energy storage. The generating units are classified into conventional power plants and DERs such as wind generators and PV generators. Each model follows from first-principles of physics.
Note that in reality a generation facility, whether that be conventional generation or wind/solar generation, and energy storage facilities contain many generating units and storage devices inside them.
In the following, the terms {\it generator}, {\it wind farm}, {\it solar farm}, and {\it energy storage system} are used to refer to an aggregate of those individual units representing the overall facility. Similarly, the term {\it load} is used to refer to an aggregate of all consumers inside the associated demand area.
Each aggregated unit comes with its own individual bus, such as a generator bus or a load bus. The buses are connected to each other through a network of power transmission lines. The power system may also contain buses where no generator, wind/solar farm, load, or energy storage system is connected. These buses are called non-unit buses. 
The term {\it component} is used to refer to either a unit with its bus or the non-unit bus. 
An example of these connected components is shown in Figure~\ref{fig_example}. 

As will be shown in the following, a general form for the dynamic model of the $k$-th component of a power system, whether that component be a generator, load, storage, wind farm, or solar farm, can be written as
\begin{equation}\label{model_comp}
 \Sigma_k: \left\{
 \begin{array}{ccl}
  \dot{x}_k &\hspace{-2mm}=&\hspace{-2mm} f_k(x_k, {\bf V}_k, u_k; \alpha_k), \\
  P_k+jQ_k&\hspace{-2mm}=&\hspace{-2mm} g_k(x_k, {\bf V}_k; \alpha_k), \\
 \end{array}
\right. 
\end{equation}
for $k \in \{1,\ldots, N\}$. The nomenclature of this model is summarized in Table~\ref{nomen_network}.
Details of the two functions $f_k(\cdot, \cdot, \cdot; \cdot)$ and $g_k(\cdot, \cdot; \cdot)$ for each component will be described shortly.
Throughout the article, complex variables will be written in bold fonts (for example ${\bf V}$). All symbols with superscript ${\star}$ will denote setpoints. 

The $N$ components are interconnected by a transmission network. 
Let ${\bf Y} \in \mathbb C^{N \times N}$ denote the admittance matrix of the network (for details of the construction of this matrix, please see Appendix \ref{sidebar-admittance}).
The power balance across the transmission lines follows from Kirchhoff's laws as
\begin{equation}\label{inter}
 0 = ({\bf Y}{\bf V}_{1:N})^* \circ {\bf V}_{1:N} - (P_{1:N} + jQ_{1:N}), \quad
\end{equation}
where $\circ$ is the element-wise multiplication, $*$ is the element-wise complex conjugate operator, ${\bf V}_{1:N}$, $P_{1:N}$, and $Q_{1:N}$ are the stacked representations of ${\bf V}_k$, $P_k$ and $Q_k$ for $k \in \{1,\ldots, N\}$. From \req{inter}, ${\bf V}_{1:N}$ is determined for a given $P_{1:N}$ and $Q_{1:N}$. The overall dynamics of a power system can be described by the combination of \req{model_comp} and \req{inter}. A signal-flow diagram of this model is shown in Figure~\ref{fig_signal}.

The power system model \req{model_comp}-\req{inter} is operated around its equilibrium. This is determined as follows. 
The steady-state value of $x_k$, ${\bf V}_k$, $P_k$, and $Q_k$, and parameter $\alpha_k$ in  \req{model_comp} must satisfy
\begin{equation}\label{model_comp_star}
 \left\{
 \begin{array}{ccl}
  0 &\hspace{-2mm}=&\hspace{-2mm} f_k(x_k^{\star}, {\bf V}_k^{\star}, 0; \alpha_k), \\
  P_k^{\star}+jQ_k^{\star} &\hspace{-2mm}=&\hspace{-2mm} g_k(x_k^{\star}, {\bf V}_k^{\star}; \alpha_k), \\
 \end{array}
\right. 
\end{equation}
and \req{inter}. The steady-state value of $u_k$ is assumed to be zero without loss of generality. 
A standard procedure for finding the steady-state values consists of two steps: {\it power flow calculation} and {\it initialization}, which are summarized as follows. \vspace{2mm}
\begin{description} 
 \item[{\bf Power Flow Calculation:}]\mbox{}\vspace{1mm}\\ 
	    Find ${\bf V}^{\star}_{1:N}$, $P_{1:N}^{\star}$, and $Q_{1:N}^{\star}$ satisfying \req{inter} and other constraints for individual components (see Appendix \ref{sidebar_pflow} for the details of these constraints).\vspace{2mm}
 \item[{\bf Initialization:}]\mbox{} \vspace{1mm}\\
	    For $k \in \{1,\ldots, N\}$, given ${\bf V}^{\star}_{k}$, $P^{\star}_{k}$, and $Q^{\star}_{k}$, find $x_k^{\star}$ and $\alpha_k$ satisfying \req{model_comp_star}. These solutions then serve as the initial conditions for the dynamic model \req{model_comp}-\req{inter} for any incoming contingency.
\end{description} \vspace{2mm}
Note that there exist an infinite number of solutions satisfying \req{inter}. However, ${\bf V}_k$, $P_k$ and $Q_k$ of some of the components are either known a priori or specified by  economic dispatch \cite{kundur1994power}. 
The details of this are described in Appendix \ref{sidebar_pflow}.
Once the steady-state values of ${\bf V}_{1:N}$, $P_{1:N}$ and $Q_{1:N}$ are obtained, in the second step the setpoints $x_k^{\star}$ and $\alpha_k$ of the $k$-th component  can be computed independently. The details of this initialization step will be described later in each subsection describing the detailed dynamics of the components.
The uniqueness of the solution ($x_k^{\star}$, $\alpha_k$) satisfying \req{model_comp_star} under a given triple $({\bf V}_k^{\star}, P_k^{\star}, Q^{\star}_k)$ depends on the component itself. In  fact, the equilibria for the generators, loads, and non-unit buses are unique, but those for wind farms, solar farms, and energy storage systems are not. The details of this uniqueness property will also be described in the following subsections.

Next, the state-space models of generators, loads, energy storage systems, wind farms, solar farms, and non-unit buses, conforming to the structure in \req{model_comp}, are derived.
For easier understanding, each subsection starts with a qualitative description of the respective component model followed by its state-space representation.
Some parts may refer to equations that appear later in the text. 
To simplify the notation, the subscript $k$ is omitted unless otherwise stated.

\subsection{Generators}\label{subsec:model_syncgens}
\begin{figure}[h!]
  \begin{center}
    \includegraphics[clip,width=165mm]{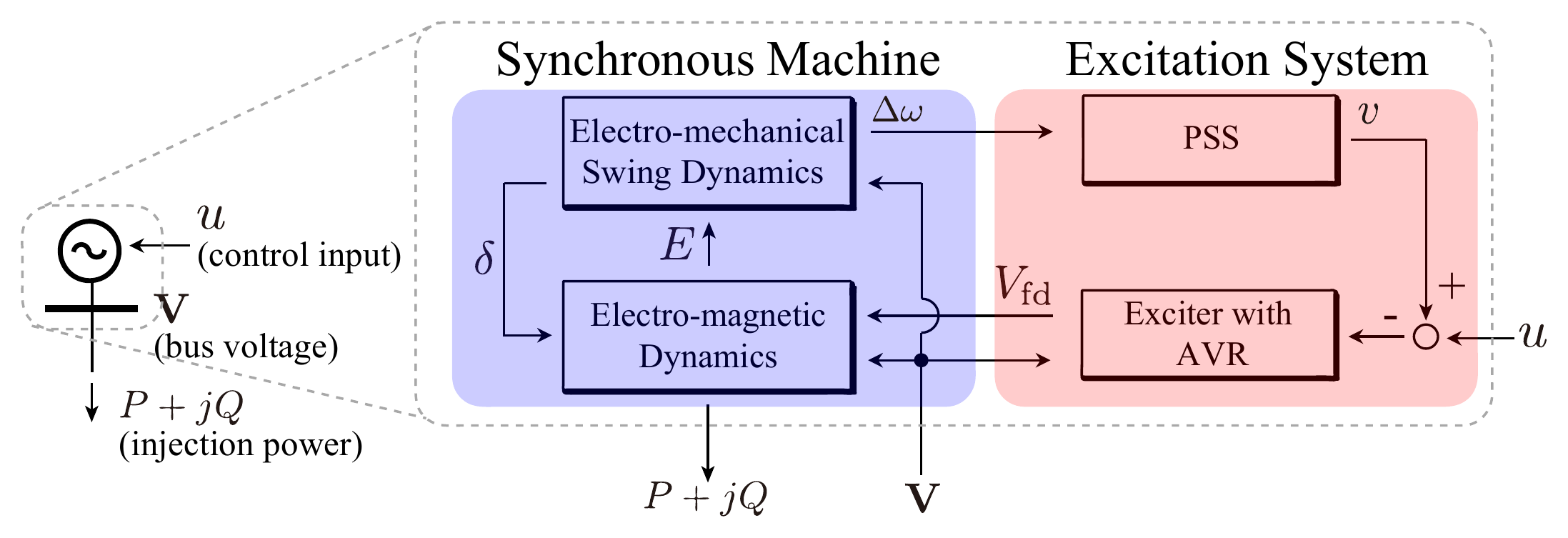}
    \caption{Signal-flow diagram of the model of a generator with its terminal bus, where the constant signals $P_{\rm m}^{\star}$, $V_{\rm fd}^{\star}$, and $|{\bf V}|^{\star}$ are omitted. }
    \label{fig_signal_gen}
  \end{center}
\end{figure}

A generator consists of a synchronous machine, an energy supply system (or a prime-mover), and an excitation system \cite{kundur1994power}. The excitation system induces currents in the excitation winding, and thereby magnetizes the rotor. The prime-mover generates mechanical power to rotate the rotor in this magnetic field. The synchronous machine converts the mechanical power to electrical power, which is transmitted to the rest of the grid. The dynamics of the prime-mover are usually ignored because of their slow time constant \cite{dib2009globally,tsolas1985structure,machowski2008power}. 

\vspace{2mm}
\subsubsection{Synchronous Machine}
While various types of synchronous machine models are available in the literature (for example see \cite{kundur1994power}), in this article a well-known model called the {\it one-axis model} or {\it flux-decay model}  is used. This model consists of the electro-mechanical swing dynamics \req{model_sync01} and the electro-magnetic voltage dynamics \req{model_sync02}. 
For simplicity, the mechanical power $P_{\rm m}^{\star}$ in  \req{model_sync01} is assumed to be constant. 


\vspace{2mm}
\subsubsection{Excitation System}
Typically, the excitation system consists of an exciter, an Automatic Voltage Regulator (AVR) that regulates the generator voltage magnitude to its setpoint value, and a Power System Stabilizer (PSS) that ensures the power system stability. The exciter with AVR is modeled as \req{AVR_PSS1}, where $u$ is a control input representing an additional voltage reference signal to the AVR. The PSS is taken as a typical speed-feedback type controller which consists of two stage lead-lag compensators and one highpass washout filter \cite{chow2004power}. 

The state-space representation of the overall generator model can be written as follows (definitions of $\bar{\omega}$, ${\bf V}$, $P+jQ$, $u$ are given in Table \ref{nomen_network} while those of the other symbols are provided in Table \ref{table_genmodel}):

\begin{table}
  \caption{Nomenclature for generator model. The values of the synchronous machine parameters below are typical, and rated at generator capacity. The parameters for the IEEE 68-bus test system, which is a benchmark power system model used later in this article, are available in \cite{pal2006robust}.} 
  \begin{tabular}{|l|l|p{114mm}|}\hline 
  Symbol &  Numerical value & Description \\\hline 
  $\delta$ &  & rotor angle relative to the frame rotating at $\bar{\omega}$. Unit of $\delta$ is (rad) \\
  $\Delta \omega$ & & frequency deviation, that is, rotor angular velocity relative to $\bar{\omega}$ \\
  $E$& & q-axis voltage behind transient reactance $X_{\rm d}'$ \\
  $V_{{\rm fd}}$& & field voltage\\
  $\zeta \in \mathbb R^{3}$& & PSS state\\
   $V_{\rm ef}$&  & exciter field voltage\\  
   $v$ & & PSS output\\
  $u$ & & additional voltage reference to AVR\\
  $M$ & 30 & inertia constant (sec)\\
  $d$& 0.1 & damping coefficient\\  
  $\tau_{{\rm do}}$& 0.1 & d-axis transient open-circuit time constant (sec)\\
  $X_{{\rm d}}$, $X_{{\rm q}}$& 1.8 & d- and q-axis synchronous reactance\\ 
  $X'_{{\rm d}}$& 0.3 & d-axis transient reactance\\
  $P_{{\rm m}}^{\star}$ & 8.0 & steady-state mechanical power\\ 
  $\tau_{{\rm e}}$ & 0.05 & time constant of exciter (sec)\\
  $K_{\rm a}$& 20 & AVR gain\\ 
  $K_{{\rm pss}}$& 150 & PSS gain\\
  $V_{{\rm fd}}^{\star}$& 1.0 & setpoints for the field voltage\\
  $|{\bf V}|^{\star}$& 1.0 & setpoints for the bus voltage magnitude\\
  $\tau_{{\rm pss}}$ & 10 & washout filter time constant (sec)\\
  $\tau_{{\rm L1}}$, $\tau'_{{\rm L1}}$ & 0.02, 0.07& lead-lag time constants of the first stage of PSS (sec)\\
  $\tau_{{\rm L2}}$, $\tau'_{{\rm L2}}$ & 0.02, 0.07& lead-lag time constants of the second stage of PSS (sec)\\\hline 
  \end{tabular}
  \label{table_genmodel}
\end{table}

\vspace{2mm}
\noindent
\underline{Synchronous machine: }\\ \noindent
$\bullet$ Electro-mechanical swing dynamics: 
\begin{flalign}\label{model_sync01}
 &
\left\{\hspace{-0.5mm}
  \begin{array}{rcl}
   \dot{\delta} &\hspace{-2mm}=&\hspace{-2mm} \bar{\omega} \Delta \omega,\\
   \Delta \dot{\omega} &\hspace{-2mm}=&\hspace{-2mm} \frac{1}{M}\left(P_{{\rm m}}^{\star} - d
    \Delta \omega - \frac{|{\bf V}|E}{X'_{{{\rm d}}}} \sin (\delta - \angle {\bf V})  + \frac{|{\bf V}|^2}{2}
   \left(\frac{1}{X'_{{\rm d}}} - \frac{1}{X_{{\rm q}}}\right)\sin (2\delta - 2\angle {\bf V})\right). 
  \end{array}
 \right.
 &
\end{flalign}
$\bullet$ Electro-magnetic dynamics: 
\begin{flalign}\label{model_sync02}
 &
 \left\{
 \begin{array}{ccl}
  \dot{E} &\hspace{-2mm}=&\hspace{-2mm} \frac{1}{\tau_{{\rm do}}}\left(-\frac{X_{{\rm
  d}}}{X'_{{{\rm d}}}} E + (\frac{X_{{\rm d}}}{X'_{{{\rm d}}}} -1) |{\bf V}| \cos (\delta - \angle {\bf V}) + V_{{\rm fd}} \right), \\
  P+jQ  &\hspace{-2mm}=&\hspace{-2mm}   \frac{E|{\bf V}|}{X'_{{{\rm d}}}} \sin (\delta - \angle {\bf V}) -
 \frac{|{\bf V}|^2}{2}
 \left(\frac{1}{X'_{{\rm d}}} - \frac{1}{X_{{\rm q}}}\right)\sin (2\delta - 2\angle {\bf V})  \\
  && + j\left(
 \frac{E|{\bf V}|}{X'_{{{\rm d}}}} \cos (\delta - \angle {\bf V}) - |{\bf V}|^2
   \left(
   \frac{\sin^2 (\delta - \angle {\bf V})}{X_{{\rm q}}} +
   \frac{\cos^2 (\delta - \angle {\bf V})}{X'_{{\rm d}}}
 \right)
	\right). 
 \end{array}
 \right.
 &
\end{flalign}
\\
\noindent
\underline{Excitation System: } \\ \noindent 
$\bullet$ Exciter with AVR:
\begin{flalign}\label{AVR_PSS1}
 &
  \dot{V}_{{\rm fd}} = \frac{1}{\tau_{{\rm e}}}
  \left(-V_{{\rm fd}}+V_{{\rm fd}}^{\star} + V_{\rm ef}\right), \quad V_{\rm ef} = K_{\rm a}(|{\bf V}| - |{\bf V}|^{\star}- v +u), 
 &
\end{flalign}
where $|{\bf V}|^{\star}$ represents the setpoint of $|{\bf V}|$. 

\noindent
$\bullet$ {PSS: }
\begin{flalign}\label{AVR_PSS2}
 & 
 \hspace{0mm} \dot{\zeta} = A_{\rm pss}\zeta + B_{\rm pss}\Delta \omega, \quad
 v = C_{\rm pss}\zeta + D_{\rm pss}\Delta \omega, 
 &
\end{flalign}
where
\begin{flalign}\label{ABCDpss}
 &\begin{array}{ll}
 A_{{\rm pss}} = \left[
\begin{array}{ccc}
 -\frac{1}{\tau_{{\rm pss}}}& 0& 0\\
 -\frac{K_{{\rm pss}}}{\tau_{{\rm pss}}\tau_{{\rm L1}}}(1-\frac{\tau'_{{\rm L1}}}{\tau_{{\rm L1}}}) & -\frac{1}{\tau_{{\rm L1}}}&0 \\
 -\frac{K_{{\rm pss}}\tau'_{{\rm L1}}}{\tau_{{\rm pss}}\tau_{{\rm L1}}\tau_{{\rm L2}}}(1-\frac{\tau'_{{\rm L2}}}{\tau_{{\rm L2}}}) & \frac{1}{\tau_{{\rm L2}}}(1-\frac{\tau'_{{\rm L2}}}{\tau_{{\rm L2}}})&-\frac{1}{\tau_{{\rm L2}}} \\
\end{array}
\right],& 
 B_{{\rm pss}} = \left[
\begin{array}{c}
 \frac{1}{\tau_{{\rm pss}}} \\
 \frac{K_{{\rm pss}}}{\tau_{{\rm pss}}\tau_{{\rm L1}}}(1-\frac{\tau'_{{\rm L1}}}{\tau_{{\rm L1}}}) \\
 \frac{K_{{\rm pss}}\tau'_{{\rm L1}}}{\tau_{{\rm pss}}\tau_{{\rm L1}}\tau_{{\rm L2}}}(1-\frac{\tau'_{{\rm L2}}}{\tau_{{\rm L2}}})\\
\end{array}
\right], \vspace{1mm}\\
 C_{{\rm pss}} =
  \left[
-\frac{K_{{\rm pss}}\tau'_{{\rm L1}}\tau'_{{\rm L2}}}{\tau_{{\rm pss}}\tau_{{\rm L1}}\tau_{{\rm L2}}}~~
 \frac{\tau'_{{\rm L2}}}{\tau_{{\rm L2}}}~~
1
\right], &
 D_{{\rm pss}} = \frac{K_{{\rm pss}}\tau'_{{\rm L1}}\tau'_{{\rm L2}}}{\tau_{{\rm pss}}\tau_{{\rm L1}}\tau_{{\rm L2}}}.
\end{array}
 &
\end{flalign}

Therefore, for $k \in \mathbb N_{\rm G}$, the model of the generator at the $k$-th bus can be written in the form of  \req{model_comp} with
\begin{equation}\label{defGENX}
 x_k := [\delta_k, \Delta \omega_k, E_k, V_{{\rm fd,k}}, \zeta_k^{\sf T}]^{\sf
 T} \in \mathbb R^7, \quad
 \alpha_k := [P_{{\rm m},k}^{\star}, V_{{\rm fd},k}^{\star}, |{\bf V}_k|^{\star}]^{\sf T}\in \mathbb R^3,
\end{equation}
and $f_k(\cdot, \cdot, \cdot; \cdot)$ and $g_k(\cdot, \cdot; \cdot)$ in \req{model_comp} follow from \req{model_sync01}-\req{ABCDpss}. 
The signal-flow diagram of this model is shown in Figure~\ref{fig_signal_gen}. For a given triple $({\bf V}_k^{\star}, P^{\star}_k, Q^{\star}_k)$, the pair $(x_k^{\star}, \alpha_k)$ satisfying \req{model_comp_star} is uniquely determined as  $x_k^{\star} = [\delta_k^{\star}, 0, E_k^{\star}, V_{{\rm fd},k}^{\star}, 0,0,0]^{\sf T}$
and $\alpha_k$ in \req{defGENX} with $P_{{\rm m},k}^{\star} = P_k^{\star}$ where
\begin{equation}
\begin{array}{l}
  \delta_k^{\star} = \angle {\bf V}_k^{\star} + 
   \arctan
   \left(\frac{P_k^{\star}}{Q_k^{\star} + \frac{ |{\bf V}_k|^{\star 2}}{X_{{\rm q},k}}}\right),\quad
    V_{{\rm fd},k}^{\star} = \frac{X_{{\rm d},k}}{X'_{{\rm d},k}}E_k^{\star} - \left(\frac{X_{{\rm d},k}}{X'_{{\rm d},k}}-1\right) |{\bf V}_k|^{\star}\cos(\delta_k^{\star} - \angle {\bf V}_k^{\star}),\\
 E_k^{\star} = \frac{ |{\bf V}_k|^{\star 4} + Q_k^{\star 2}X'_{{\rm d},k}X_{{\rm
  q},k} + Q_k^\star  |{\bf V}_k|^{\star 2}X'_{{\rm d},k} + Q_k^\star |{\bf V}_k|^{\star 2}X_{{\rm q},k} + P_k^{\star 2}X'_{{\rm d},k}X_{{\rm q},k}}{ |{\bf V}_k|^{\star}
  \sqrt{P_k^{\star 2}X_{{\rm q},k}^2 + Q_k^{\star 2}X_{{\rm q},k}^2 +
  2Q_k^\star  |{\bf V}_k|^{\star 2}X_{{\rm q},k} +  |{\bf V}_k|^{\star 4}}}. \\
\end{array}
\end{equation}
\begin{remark}
 Relationships between the one-axis model \req{model_sync01}-\req{model_sync02} and some other standard models of synchronous generators are shown in Appendix \ref{sidebar-relation}.
\end{remark}

\subsection{Non-unit Buses}\label{subsec:model_ncbus}
Non-unit buses are simply modeled by the Kirchhoff's power balance law, namely for $k \in \mathbb N_{\rm N}$,
\begin{equation}\label{non_model}
 P_k + jQ_k = 0. 
\end{equation}
In reference to \req{model_comp} this means $x_k$,  $u_k$, and  $\alpha_k$ are empty vectors.

\subsection{Loads}\label{subsec:model_loads}
Loads are commonly modeled by algebraic power balance equations, although extensive literature also exists for dynamic loads (for example, see \cite{hiskens1995load,hill1993nonlinear}). The well-known static load models are:
\begin{align}
& \hspace{-45mm} \mbox{\underline{constant~impedance~model:}} \hspace{23mm} P+jQ = (\bar{\bf z}^{-1} {\bf V})^{*}{\bf V}, \label{load_model}\\
& \hspace{-45mm} \mbox{\underline{constant~current~model:}}  \hspace{30.4mm}  P+jQ = \bar{\bf i}^{*}{\bf V},  \label{load_model2}\\
 & \hspace{-45mm} \mbox{\underline{constant~power~model:}}  \hspace{30.9mm}  P+jQ = \bar{P} + j\bar{Q},  \label{load_model3}
\end{align}
where $*$ is the complex conjugate operator, 
$\bar{{\bf z}} \in \mathbb C$, 
$\bar{{\bf i}} \in \mathbb C$, and 
$\bar{P} + j\bar{Q} \in \mathbb C$ are constant. 
Therefore, for $k \in \mathbb N_{\rm L}$, in reference to \req{model_comp} a load at the $k$-th bus can be represented by $x_k$ and $u_k$ being empty vectors, $\alpha_k$ being either $\bar{\bf z}_k$, $\bar{\bf i}_k$, or $\bar{P}_k + j\bar{Q}_k$, and the output equation being either \req{load_model}, \req{load_model2} or \req{load_model3}. For the simulations later in this article, constant impedance loads will be used. For a given triple $({\bf V}_k^{\star}, P^{\star}_k, Q^{\star}_k)$, the load impedance $\bar{\bf z}_{k}$ will be uniquely calculated such that $P_k^{\star}+jQ_k^{\star} = (\bar{\bf z}_k^{-1} {\bf V}^{\star}_k)^{*}{\bf V}^{\star}_k$. 

\subsection{Wind Farms}\label{subsec:model_wind}
\begin{figure}[t]
\vspace{-4mm}
   \begin{center}
    \includegraphics[clip,width=165mm]{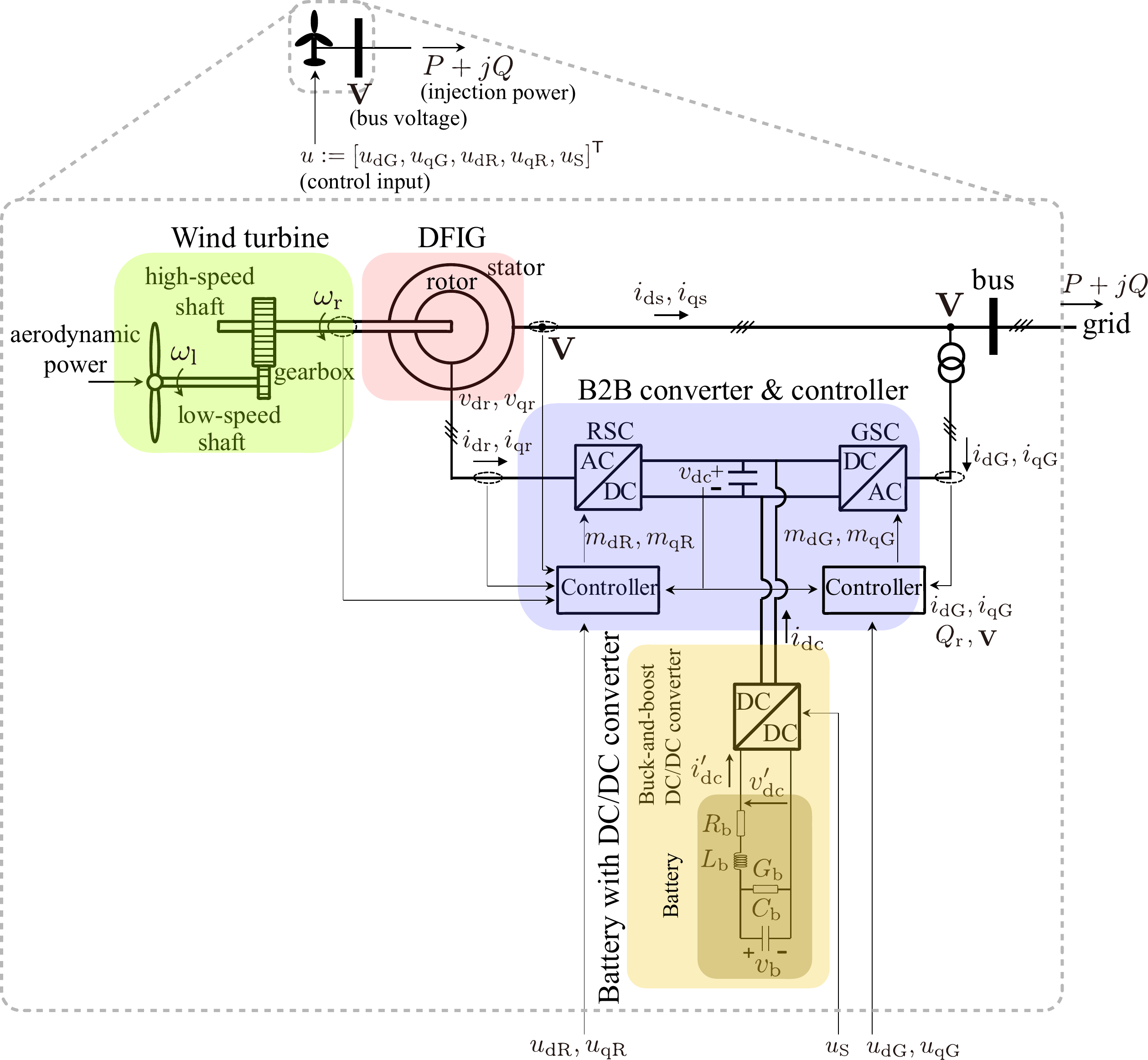}
    \caption{Physical structure of the model of a wind farm with its terminal bus}
    \label{fig:windmodel}
   \end{center}
 \vspace{-4mm}
\end{figure}
\begin{figure}[t]
  \begin{center}
    \includegraphics[clip,width=165mm]{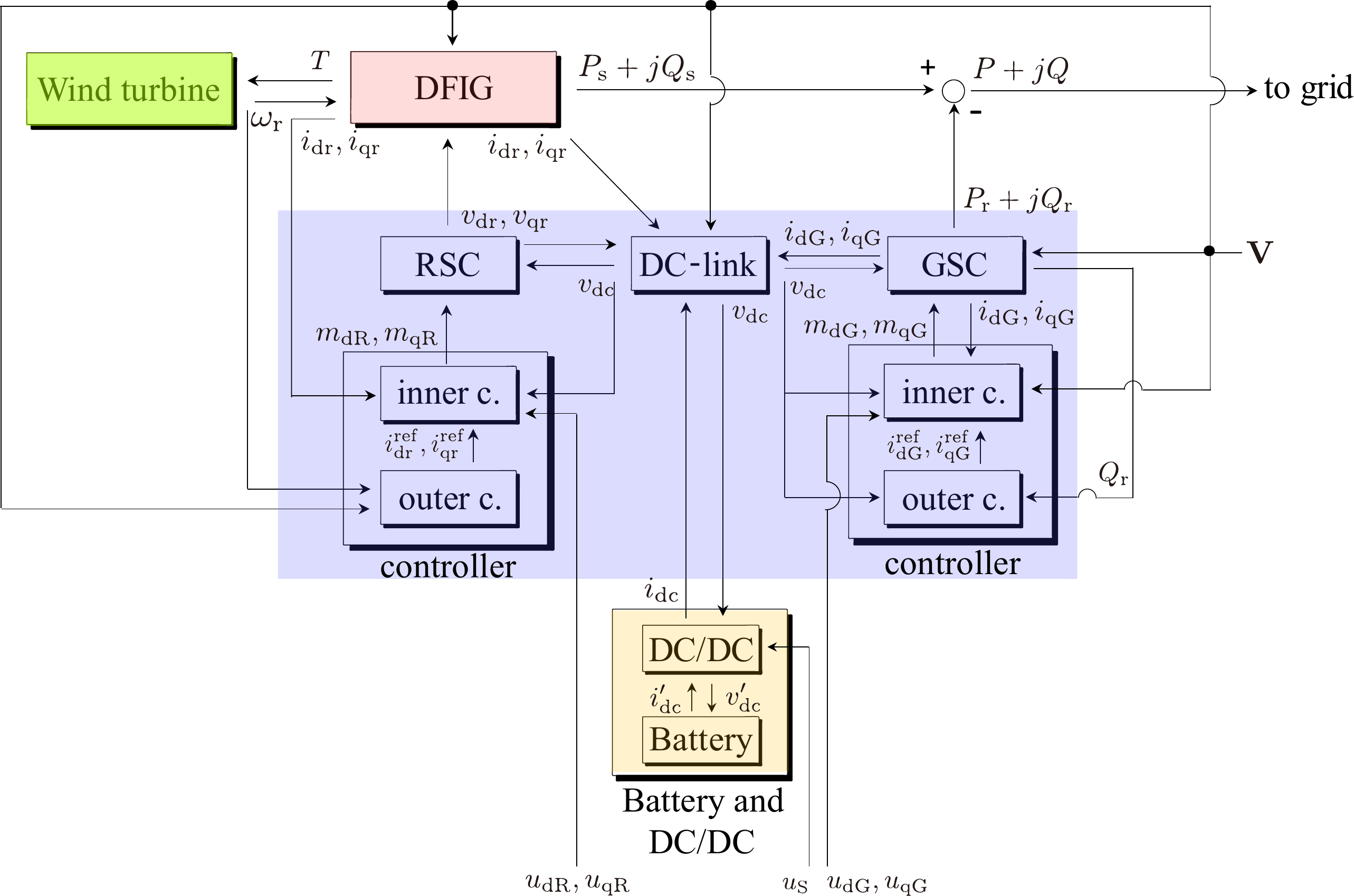}
    \caption{Signal-flow diagram of the model of a wind farm with its terminal bus, where constant signals $P_{\rm a}$, $v_{\rm dc}^{\star}$, $Q_{\rm r}^{\star}$, $|{\bf V}|^{\star}$, $\omega_{\rm r}^{\star}$ are omitted. }
    \label{fig:sigflow_wind}
  \end{center}
  \vspace{-4mm}
\end{figure}
\begin{table}[h!]
  \caption{Nomenclature for the wind farm model. The signs of the currents are positive when flowing in the direction of the corresponding arrows in Figure~\ref{fig:windmodel}. The values of the models parameters of a 2MW 690V wind turbine, DFIG are shown in \cite{sloth2010active,ugalde2013state}. In the following list, values rated at 100MW, which is the system capacity used in the simulations, are shown.}
  \begin{tabular}{|l|l|p{102mm}|}\hline 
   Symbol & Numerical value & Description \\\hline \hline
   \multicolumn{3}{|c|}{Wind turbine}\\ \hline
   $\omega_{\rm l}$, $\omega_{\rm r}$&& angular velocity of low-speed shaft and high-speed shaft\\
   $\theta_{\rm T}$&& torsion angle (rad)\\
   $P_{\rm a}$& $2.45 \times 10^{-5}$ & aerodynamic power input depending on wind speed\\
   $J_{\rm l}$, $J_{\rm r}$ & $1.95 \times 10^4, 0.138$ & inertia coefficients of the low-speed and high-speed shafts (sec)\\
   $B_{\rm l}$, $B_{\rm r}$& $9.87, 0.001$ & friction coefficients of the low-speed and high-speed shafts \\
   $K_{\rm c}$&$508.9$ & torsional stiffness (1/rad)\\
   $d_{\rm c}$&$337.76$& damping coefficient of turbine\\
   $N_{\rm g}$& 90 &  gear ratio \\
   $\bar{\omega}_{\rm m}$ &$60\pi$& mechanical synchronous frequency (rad/sec) \\\hline
   \multicolumn{3}{|c|}{DFIG}\\ \hline
   $i_{\rm dr}$, $i_{\rm qr}$&& d- and q-axis rotor currents\\
   $i_{\rm ds}$, $i_{\rm qs}$&&  d- and q-axis stator currents\\ 
   $v_{\rm dr}$, $v_{\rm qr}$&&  d- and q-axis rotor voltages\\
   $T$&& electromechanical torque converted by DFIG \\
   $P_{\rm s}+jQ_{\rm s}$&& power flowing from DFIG to bus\\
   $\gamma_{\rm W}$&& number of wind generators inside the farm \\ 
   $X_{\rm m}$& $197.64$ & magnetizing reactance\\ 
   $X_{\rm ls}$, $X_{\rm lr}$& $4.620, 4.976$ & stator and rotor leakage reactance\\
   $R_{\rm s}$, $R_{\rm r}$& $0.244, 0.274$ & stator and rotor resistance\\ \hline
   \multicolumn{3}{|c|}{GSC and its controller}\\ \hline
   $i_{\rm dG}$, $i_{\rm qG}$&&  d- and q-axis currents flowing from AC side to DC side\\
   $m_{\rm dG}$, $m_{\rm qG}$&&  d- and q-axis duty cycles\\
   $P_{\rm r}+jQ_{\rm r}$&& power flowing from bus to GSC\\
   $Q_{\rm r}^{\star}$& $0.001$ & steady-state value of $Q_{\rm r}$\\
     $v_{\rm dc}^{\star}$& $2.03$ & steady-state DC link voltage\\
     $\chi_{\rm dG}$, $\chi_{\rm qG}$&& inner-loop controller state\\
   $\zeta_{\rm dG}$, $\zeta_{\rm qG}$&& outer controller state\\
      $i_{\rm dG}^{\rm ref}$, $i_{\rm qG}^{\rm ref}$&& reference signal of $i_{\rm dG}$ and $i_{\rm qG}$\\
   $u_{\rm dG}$, $u_{\rm dG}$&& additional control input signals\\
   $L_{\rm G}$, $R_{\rm G}$& $633.46, 0.05$ & inductance and resistance of GSC\\
      $K_{\rm P, dG}$, $K_{\rm P, qG}$ & $0.1, 0.01$ & P gains of outer controller \\
   $K_{\rm I, dG}$, $K_{\rm I, qG}$ & $1\times10^{-4}, 0.001$ & I gains of outer controller \\
   $\tau_{\rm G}$& $0.1$ & GSC current dynamics' time constant to be designed\\ 
  \end{tabular}
  \label{table_wind}
\end{table}
\begin{table}[h!]
    \begin{tabular}{|l|p{28mm}|p{104mm}|}
 \multicolumn{3}{|c|}{RSC and its controller}\\ \hline
$m_{\rm dR}$, $m_{\rm qR}$&& d- and q-axis duty cycles \\
$\chi_{\rm dR}$, $\chi_{\rm qR}$&& inner-loop controller state \\
   $i_{\rm dr}^{\rm ref}$, $i_{\rm qr}^{\rm ref}$&& reference signal of $i_{\rm dr}$ and $i_{\rm qr}$ \\
   $u_{\rm dR}$, $u_{\rm dR}$ && additional control input signals \\
   $K_{\rm P, dR}$, $K_{\rm P, qR}$ & $ 0.01, 0.1$ & P gains of outer controller \\
   $\kappa_{\rm P, dR}$, $\kappa_{\rm P, qR}$& $5, 5$ & P gains of the inner-loop controller\\
   $\kappa_{\rm I, dR}$, $\kappa_{\rm I, qR}$& $1, 1$& I gains of the inner-loop controller\\  \hline
 \multicolumn{3}{|c|}{DC link}\\ \hline
 $v_{\rm dc}$&& DC-side voltage\\
 $C_{\rm dc}$& $448.7$ &  DC link capacitance\\
 $G_{\rm sw}$& $1.2\times 10^{-5}$&  conductance representing switching loss of the B2B converter \\ \hline
     \multicolumn{3}{|c|}{Buck-and-Boost DC/DC converter}\\ \hline
     $i_{\rm dc}$&& current injecting from DC/DC converter into DC link\\
     $v_{\rm dc}'$&& voltage at battery side \\
     $S$& $2$ & step down/up gain\\ \hline 
     \multicolumn{3}{|c|}{Battery}\\ \hline
     $v_{\rm b}$&&  battery voltage\\ 
     $i_{\rm dc}'$&& current injected from the battery \\
     $C_{\rm b}$ & $8.97 \times 10^3$ & battery capacity \\
     $G_{\rm b}$ & $1.2 \times 10^{-5}$ & battery conductance\\
     $R_{\rm b}$, $L_{\rm b}$& $0.21, 79.1$ &  resistance and inductance of battery circuit\\ \hline
    \end{tabular}
\end{table}
A wind farm model typically consists of a wind turbine, a doubly-fed induction generator (DFIG), and a back-to-back (B2B) converter with associated controllers. A battery with DC/DC converter can be added to the B2B converter if needed. 
Figure~\ref{fig:windmodel} shows the physical architecture of a wind farm with its bus while Figure~\ref{fig:sigflow_wind} shows a signal-flow diagram of the model. When the battery is not connected, the current $i_{\rm dc}$ in Figure~\ref{fig:windmodel} is regarded as zero. The symbols for the wind farm model  are listed in Table~\ref{table_wind}. 


\vspace{2mm}
\subsubsection{Wind Turbine}
The wind turbine, as shown in Figure~\ref{fig:windmodel}, converts aerodynamic power coming from the wind to mechanical power that is transmitted to the DFIG. The turbine is typically modeled as a one-inertia or two-inertia model (the latter is followed in this article) consisting of a low-speed shaft, a high-speed shaft, and a gearbox \cite{sloth2010active}.
For simplicity, the aerodynamic power $P_{\rm a}$ in the two-inertia dynamics \req{turbine_model} is assumed to be constant as wind speeds usually change  slowly. 

\vspace{2mm}
\subsubsection{Doubly-Fed Induction Generator (DFIG)}
The DFIG shown in Figure~\ref{fig:windmodel} converts the mechanical power from the turbine into electrical power. The DFIG consists of a three-phase rotor and a three-phase stator. 
The stator is connected to the wind bus to transmit the electrical power into the grid while the rotor is connected to the B2B converter with associated controllers that control the rotor winding voltage. 
The stator and rotor are coupled electro-magnetically, which is reflected in the dynamics of the stator and rotor currents expressed in a rotating d-q reference frame \cite{ugalde2013state}.

\vspace{2mm}
\subsubsection{Back-to-Back (B2B) Converter with its Controllers}
The B2B converter  is used for regulating the DFIG rotor voltages $v_{\rm dr}$, $v_{\rm qr}$ as well as the reactive power flowing from the stator to the converter. 
The B2B converter consists of two three-phase voltage source converters, namely, the rotor-side converter (RSC) and the grid-side converter (GSC), linked via a common DC line \cite{ortega2015generalized}.
Each of the converters is equipped with a controller.
The explanation of the models of GSC, RSC, and their controllers  is as follows.

$\bullet$ Following \cite{ortega2015generalized}, the GSC dynamics is expressed as the variation of the AC-side current in d-q reference frame. 

$\bullet$ The GSC controller consists of an inner-loop controller and an outer-loop controller \cite{ortega2015generalized}. The objective of the outer-loop controller is to generate a reference signal of the GSC currents $i_{\rm dG}, i_{\rm qG}$ for regulating both of the DC link voltage $v_{\rm dc}$ and the reactive power $Q_{\rm r}$ flowing into the GSC to their respective setpoints. The outer-loop controller is designed as a PI controller as  \req{outer}. 
The inner-loop controller aims at regulating $i_{\rm dG}, i_{\rm qG}$ to the generated reference signals $i_{\rm dG}^{\rm ref}$, $i_{\rm qG}^{\rm ref}$ by the control of the duty cycles $m_{\rm dG}$, $m_{\rm qG}$.
Following \cite{ortega2015generalized}, in this article the controller is designed such that the transfer function from  $i_{\rm dG}^{\rm ref}$ (or $i_{\rm qG}^{\rm ref}$) to $i_{\rm dG}$ (or $i_{\rm qG}$) is a desired first-order system $1/(\tau_{\rm G} s + 1)$ when the duty cycles are not saturated. The controller is implemented as \req{con_duty}.

\vspace{0mm}
 $\bullet$ 
 The RSC model is described as
 \begin{equation}
   v_{\rm dr} \hspace{-0.2mm}=\hspace{-0.2mm} \frac{L_{\rm R}}{\bar{\omega}}\dot{i}_{\rm dr} + R_{\rm R}i_{\rm dr} - L_{\rm R}i_{\rm qr} + \frac{m_{\rm dR}}{2}v_{\rm dc}, \quad
   v_{\rm qr}\hspace{-0.2mm}=\hspace{-0.2mm} \frac{L_{\rm R}}{\bar{\omega}}\dot{i}_{\rm qr} +R_{\rm R}i_{\rm qr} + L_{\rm R}i_{\rm dr} + \frac{m_{\rm qR}}{2}v_{\rm dc}, 
\label{rsc}
 \end{equation}
 where $i_{\rm dr}$ and $i_{\rm qr}$ are the DFIG rotor currents,
 $m_{\rm dR}$ and $m_{\rm qR}$ are the duty cycles of the RSC, and
 $v_{\rm dc}$ is the DC link voltage. 
 In this article, the RSC resistance and inductance are considered to be negligible, that is, $R_{\rm R} = L_{\rm R}= 0$.  This assumption is always satisfied by incorporating the two into the DFIG rotor circuit. Thus, the RSC model used in this article is described as \req{simple_rsc}. 
 

\vspace{0mm}
 $\bullet$ The RSC is equipped with an inner-loop controller and an outer-loop controller.
 The outer-loop controller generates reference signals for the DFIG rotor currents $i_{\rm dr}$ and $i_{\rm qr}$ for regulating the stator voltage magnitude and the high-speed shaft speed $\omega_{\rm r}$ to their setpoints while the inner-loop controller aims at regulating the RSC currents \cite{anaya2014offshore}. This control action is actuated through the control of the duty cycles of the B2B converter. 

 \vspace{0mm}
 $\bullet$ The RSC and GSC are connected by a DC link equipped with a capacitor whose dynamics is derived from the power balance through the B2B converter \cite{ortega2015generalized}. 

 \vspace{2mm}
 \subsubsection{Battery and DC/DC converter}
A battery is used for charging or discharging of electricity whenever needed. The battery comes with a DC/DC converter that steps up/down the battery terminal voltage. Both devices are also sometimes used for suppressing the fluctuations in the output power $P + jQ$ by controlling the DC/DC converter. The model for each is described as follows.

\vspace{0mm}
  $\bullet$ The DC/DC converter is modeled by buck (step-down) and boost (step-up) models. These models are widely available in the literature \cite{sira2006control}. When the converter dynamics are sufficiently fast, simpler models where the output voltage and current are explicit functions of the duty ratio  can be derived. In this article this simple model is used.

 \vspace{0mm}
$\bullet$ The battery circuit is shown as the dark yellow part in Figure~\ref{fig:windmodel} \cite{ortega2015generalized}. 
Its dynamics can be represented as the variation of the battery voltage $v_{\rm b}$ and the output current $i_{\rm dc}'$. 

 \vspace{2mm}
\subsubsection{Interconnection to Grid}\label{subsub_connection}
The net active and reactive power injected by the wind farm to the grid are determined as the sum of the power leaving from the stator and that consumed by the B2B converter. 

 \vspace{2mm}
\noindent
\underline{Wind turbine: } 
\begin{flalign}\label{turbine_model}
 &\scalebox{1.0}{$\displaystyle
\left\{\hspace{-1.5mm}
\begin{array}{rcl}
 J_{{\rm l}} \dot{\omega}_{{\rm l}} &\hspace{-2.5mm}=&\hspace{-2.5mm}
  -(d_{{\rm c}}+B_{{\rm l}})\omega_{{\rm l}} + \frac{d_{{\rm c}}}{N_{{\rm g}}}\omega_{{\rm r}}
  -K_{{\rm c}}\theta_{{\rm T}} + \frac{P_{{\rm a}}}{\omega_{{\rm l}}},\\
 J_{{\rm r}} \dot{\omega}_{{\rm r}} &\hspace{-2.5mm}=&\hspace{-2.5mm} \frac{d_{{\rm c}}}{N_{{\rm g}}}\omega_{{\rm l}} -
  \left(\frac{d_{{\rm c}}}{N_{{\rm g}}^2}+B_{{\rm r}}\right)\omega_{{\rm r}} +
  \frac{K_{{\rm c}}}{N_{{\rm g}}}\theta_{{\rm T}} - T_{},\\
 \dot{\theta}_{{\rm T}} &\hspace{-2.5mm}=&\hspace{-2.5mm} \bar{\omega}_{\rm m}\left(\omega_{{\rm l}} - \frac{1}{N_{{\rm g}}}\omega_{{\rm r}}\right),
\end{array}
\right. $}
 &
\end{flalign}
where $T$ is defined in \req{wind_current}. \\

 \vspace{2mm}
\noindent
\underline{DFIG: }
 \begin{flalign}\label{wind_current}
 &  \scalebox{1.0}{$\displaystyle \left\{
\begin{array}{rcl}
 \dot{i} &\hspace{-2.5mm}=&\hspace{-2.5mm} A_{{\rm i}}(\omega_{{\rm r}})i + G_{{\rm i}}[{\rm Re}({\bf V}), {\rm Im}({\bf V})]^{\sf T} + B_{{\rm i}}[v_{\rm dr}, v_{\rm qr}]^{\sf T}, \vspace{2mm}\\
 T &\hspace{-2.5mm}=&\hspace{-2.5mm} \displaystyle X_{{\rm m}} \left(i_{{\rm ds}}i_{{\rm qr}} - i_{{\rm qs}}i_{{\rm dr}}\right),\\
 P_{\rm s} + jQ_{\rm s} &\hspace{-2.5mm}=&\hspace{-2.5mm} \gamma_{\rm W} ({\rm Re}({\bf V})i_{\rm ds} + {\rm Im}({\bf V})i_{{\rm qs}}) + j \gamma_{\rm W}({\rm Im}({\bf V})i_{\rm ds} - {\rm Re}({\bf V})i_{{\rm qs}}),\\
\end{array}
\right.\hspace{-1mm} ~
 i := 
\left[
\begin{array}{c}
 i_{{\rm dr}}\\ i_{{\rm qr}}\\ i_{{\rm ds}}\\ i_{{\rm qs}}\\
\end{array}
  \right],$}
 &
\end{flalign}
where $\omega_{\rm r}$ is defined in \req{turbine_model},  $v_{\rm dr}$ and $v_{\rm qr}$ in \req{simple_rsc}, and 
\begin{flalign}\label{ABCDelectrical}
 &  \hspace{-1.5mm} \scalebox{1.0}{$\displaystyle
 \begin{array}{rcl}
 A_{{\rm i}}(\omega_{{\rm r}}) &\hspace{-2.5mm} =&\hspace{-2.5mm}  \displaystyle\frac{1}{\beta}\left[\hspace{-1mm}
\begin{array}{cccc}
 -R_{{\rm r}}X_{{\rm s}}& \beta - \omega_{{\rm r}}X_{{\rm s}}X_{{\rm r}} & R_{{\rm s}}X_{{\rm m}} & -\omega_{{\rm r}} X_{{\rm s}}X_{{\rm m}}\\
 -\beta + \omega_{{\rm r}}X_{{\rm s}}X_{{\rm r}} & -R_{{\rm r}}X_{{\rm s}} & \omega_{{\rm r}}X_{{\rm s}}X_{{\rm m}} & R_{{\rm s}}X_{{\rm m}} \\
 R_{{\rm r}}X_{{\rm m}} & \omega_{{\rm r}}X_{{\rm r}}X_{{\rm m}} & -R_{{\rm s}}X_{{\rm r}} & \beta+\omega_{{\rm r}}X_{{\rm m}}^2\\
 -\omega_{{\rm r}}X_{{\rm r}}X_{{\rm m}} & R_{{\rm r}}X_{{\rm m}} & -\beta-\omega_{{\rm r}}X_{{\rm m}}^2 &  -R_{{\rm s}}X_{{\rm r}}
\end{array}\hspace{-1mm}
\right]\hspace{-0.5mm},~ B_{{\rm i}} \hspace{-0.5mm}=\hspace{-0.5mm}
 \displaystyle \frac{1}{\beta}\left[\hspace{-1mm}
\begin{array}{cc}
 -X_{{\rm s}} & 0\\
 0 & -X_{{\rm s}} \\
 X_{{\rm m}} & 0 \\
 0 &X_{{\rm m}}\\
\end{array}\hspace{-1mm}
\right],\\
  G_{{\rm i}} &\hspace{-2mm} =&\hspace{-2mm}
   \displaystyle \frac{1}{\beta}\left[\hspace{-1mm}
\begin{array}{cccc}
X_{\rm  m} & 0 & -X_{\rm r} & 0\\
0 &X_{{\rm m}}& 0 &-X_{{\rm r}}\\
 \end{array}\hspace{-1mm}
\right]^{\sf T}\hspace{-2.0mm}, ~X_{\rm s} := X_{\rm m} + X_{\rm ls},
   ~ X_{\rm r} := X_{\rm m} + X_{\rm lr}, 
   ~ \beta := X_{\rm s}X_{\rm r} - X_{\rm m}^2. 
 \end{array}
   $}
 &
\end{flalign}

 \vspace{2mm}
\noindent
\underline{GSC: }
\begin{flalign}\label{fsc}
 &
 \left\{
  \begin{array}{rcl}
   \frac{L_{\rm G}}{\bar{\omega}}\dot{i}_{\rm dG}&\hspace{-2mm}=&\hspace{-2mm} -R_{\rm G}i_{\rm dG} + L_{\rm G}i_{\rm qG} + {\rm Re}({\bf V}) - \frac{m_{\rm dG}}{2}v_{\rm dc},\\
   \frac{L_{\rm G}}{\bar{\omega}}\dot{i}_{\rm qG}&\hspace{-2mm}=&\hspace{-2mm} -R_{\rm G}i_{\rm qG} - L_{\rm G}i_{\rm dG} + {\rm Im}({\bf V}) - \frac{m_{\rm qG}}{2}v_{\rm dc},\vspace{1mm}\\
   P_{\rm r} + jQ_{\rm r} &\hspace{-2mm}=&\hspace{-2mm} \gamma_{\rm W} ({\rm Re}({\bf V})i_{\rm dG} + {\rm Im}({\bf V})i_{{\rm qG}}) + j \gamma_{\rm W}({\rm Im}({\bf V})i_{\rm dG} - {\rm Re}({\bf V})i_{{\rm qG}}),\\
  \end{array}
 \right.
 &
\end{flalign}
where $m_{\rm dG}$ and $m_{\rm qG}$ are defined in \req{con_duty}, and $v_{\rm dc}$ in \req{dclink}. 

 \vspace{2mm}
\noindent
\underline{Outer-Loop controller of GSC: }
  \begin{flalign}\label{outer}
   & \hspace{-2mm}
 \left\{
\begin{array}{rcl}
 \dot{\zeta}_{\rm dG}&\hspace{-2mm}=&\hspace{-2mm} K_{\rm I, dG}(v_{\rm dc} - v_{\rm dc}^{\star}),\\
 i_{\rm dG}^{\rm ref}&\hspace{-2mm}=&\hspace{-2mm} K_{\rm P, dG}(v_{\rm dc} - v_{\rm dc}^{\star}) + \zeta_{\rm dG},\\
\end{array}
 \right. ~
 \left\{
\begin{array}{rcl}
 \dot{\zeta}_{\rm qG}&\hspace{-2mm}=&\hspace{-2mm} K_{\rm I, qG}(Q_{\rm r} - Q_{\rm r}^{\star}),\\ 
 i_{\rm qG}^{\rm ref}&\hspace{-2mm}=&\hspace{-2mm} K_{\rm P, qG}(Q_{\rm r} - Q_{\rm r}^{\star}) + \zeta_{\rm qG},\\
\end{array}
 \right. 
   &
  \end{flalign}
where $Q_{\rm r}$ and $v_{\rm dc}$ are defined in \req{fsc} and \req{dclink}.
  
  \vspace{2mm}
  \noindent
\underline{Inner-Loop controller of GSC: }
  \begin{flalign}\label{con_duty}
   &
\begin{array}{l}
 \left\{
  \begin{array}{rcl}
   \tau_{\rm G} \dot{\chi}_{\rm dG} &\hspace{-2mm}=&\hspace{-2mm} i_{\rm dG}^{\rm ref} - i_{\rm dG}, \\
   m_{\rm dG} &\hspace{-2mm}=&\hspace{-2mm} {\rm sat}\left(\frac{2}{v_{\rm dc}}  \left({\rm Re}({\bf V}) + L_{\rm G}i_{\rm qG} - R_{\rm G}\chi_{\rm dG} - \frac{L_{\rm G}}{\bar{\omega}\tau_{\rm G}}(i_{\rm dG}^{\rm ref} - i_{\rm dG}) + u_{\rm dG}\right)\right), \\ 
  \end{array}
  \right. \\
 \left\{
  \begin{array}{rcl}
   \tau_{\rm G} \dot{\chi}_{\rm qG} &\hspace{-2mm}=&\hspace{-2mm} i_{\rm qG}^{\rm ref} - i_{\rm qG}, \\
   m_{\rm qG} &\hspace{-2mm}=&\hspace{-2mm} {\rm sat}\left(\frac{2}{v_{\rm dc}}  \left({\rm Im}({\bf V}) - L_{\rm G}i_{\rm dG} - R_{\rm G}\chi_{\rm qG} - \frac{L_{\rm G}}{\bar{\omega}\tau_{\rm G}}(i_{\rm qG}^{\rm ref} - i_{\rm qG}) + u_{\rm qG}\right) \right),
  \end{array}
 \right.
\end{array}  &
  \end{flalign}
where
$i_{\rm dG}$ and $i_{\rm qG}$ are defined in \req{fsc}, 
$i_{\rm dG}^{\rm ref}$ and $i_{\rm qG}^{\rm ref}$ in \req{outer}, 
$v_{\rm dc}$ in \req{dclink}, and ${\rm sat}(\cdot)$ is a saturation function whose output is restricted within the range of $[-1, 1]$. 
  
 \vspace{2mm}
  \noindent
\underline{RSC: } 
 \begin{flalign}\label{simple_rsc}
  &
   v_{\rm dr} = \frac{m_{\rm dR}}{2}v_{\rm dc}, \quad  v_{\rm qr} = \frac{m_{\rm qR}}{2}v_{\rm dc},
  &
 \end{flalign}
 where $m_{\rm dR}$ and $m_{\rm qR}$ are defined in \req{outer2}, and $v_{\rm dc}$ in \req{dclink}. \\

 \vspace{2mm}
 \noindent
\underline{Outer-Loop controller of RSC: }
  \begin{flalign}\label{outerRSC}
   &
 i_{\rm dr}^{\rm ref} = K_{\rm P, dR}(|{\bf V}| - |{\bf V}|^{\star}),\quad
   i_{\rm qr}^{\rm ref} = K_{\rm P, qR}(\omega_{\rm r} - \omega_{\rm r}^{\star}), 
   &
 \end{flalign}
 where $\omega_{\rm r}$ is defined in  \req{turbine_model}. \\

  \vspace{2mm}
 \noindent
\underline{Inner-Loop controller of RSC: }
 \begin{flalign}\label{outer2}
  &
   \scalebox{1.00}{$\displaystyle
\begin{array}{l}
 \left\{
\begin{array}{rcl}
 \dot{\chi}_{\rm dR} &\hspace{-2mm}=&\hspace{-2mm} \kappa_{\rm I, dR}(i_{\rm dr} - i_{\rm dr}^{\rm ref}),\\
 m_{\rm dR}&\hspace{-2mm}=&\hspace{-2mm} {\rm sat}
  \left(
   \frac{2}{v_{\rm dc}}\left(\kappa_{\rm P, dR}(i_{\rm dr} - i_{\rm dr}^{\rm ref}) + \chi_{\rm dR} + u_{\rm dR}\right)
   \right),\\
\end{array}
 \right. \\
 \left\{
  \begin{array}{rcl}
 \dot{\chi}_{\rm qR}&\hspace{-2mm}=&\hspace{-2mm} \kappa_{\rm I, qR}(i_{\rm qr} - i_{\rm qr}^{\rm ref}),\\
 m_{\rm qR}&\hspace{-2mm}=&\hspace{-2mm} {\rm sat}
  \left(
   \frac{2}{v_{\rm dc}}\left(\kappa_{\rm P, qR}(i_{\rm qr} - i_{\rm qr}^{\rm ref}) + \chi_{\rm qR} + u_{\rm qR}\right)
   \right),\\
\end{array}
 \right.
\end{array}
  $}
  &
\end{flalign}
where $i_{\rm dr}$ and $i_{\rm qr}$ are defined in \req{wind_current},
$i_{\rm dr}^{\rm ref}$ and $i_{\rm qr}^{\rm ref}$ in \req{outerRSC}, and
$v_{\rm dc}$ in \req{dclink}. 

 \vspace{2mm}
\noindent
\underline{DC link: }
\begin{flalign}\label{dclink}
 \left.
  \begin{array}{rcl}
   \frac{C_{\rm dc}}{\bar{\omega}}\dot{v}_{\rm dc} &\hspace{-2mm}=&\hspace{-2mm} \frac{1}{2v_{\rm dc}}
    \left(
     {\rm Re}({\bf V})i_{\rm dG} + {\rm Im}({\bf V})i_{\rm qG} +
     v_{\rm dr}i_{\rm dr} + v_{\rm qr}i_{\rm qr} -
     R_{\rm G}(i_{\rm dG}^2 + i_{\rm qG}^2)
    \right) - G_{\rm sw}v_{\rm dc} + \frac{1}{2}i_{\rm dc}, 
  \end{array}
 \right.
\end{flalign}
where $i_{\rm dG}$ and $i_{\rm qG}$ are defined in \req{fsc}, 
$v_{\rm dr}$ and $v_{\rm qr}$ in \req{simple_rsc}, 
$i_{\rm dr}$ and $i_{\rm qr}$ in \req{wind_current}, 
 and $i_{\rm dc}$ in \req{buckboost2}. When the battery and DC/DC are not connected, $i_{\rm dc} = 0$. 

 \vspace{2mm}
 \noindent
\underline{Buck-and-Boost DC/DC Converter: }
  \begin{flalign}\label{buckboost2}
   & v_{\rm dc}' = p(S + u_{\rm S})v_{\rm dc}, \quad i_{\rm dc} = p(S + u_{\rm S})i_{\rm dc}' , \quad p(x) =
   \left\{
   \begin{array}{ll}
    x & {\rm if}~x \geq 0,\\
    0 & {\rm otherwise},
   \end{array}
   \right.
  &
 \end{flalign}
 where $v_{\rm dc}$ and $i_{\rm dc}'$ are defined in \req{dclink} and \req{batter_ACDC_wind}.
 
 \vspace{2mm}
 \noindent
\underline{Battery: }
  \begin{flalign}\label{batter_ACDC_wind}
   &
   \left\{
  \begin{array}{rcl}
   \frac{C_{\rm b}}{\bar{\omega}}\dot{v}_{\rm b} &\hspace{-2mm}=&\hspace{-2mm} -i_{\rm dc}' - G_{\rm b}v_{\rm b},\\
   \frac{L_{\rm b}}{\bar{\omega}}\dot{i}_{\rm dc}' &\hspace{-2mm}=&\hspace{-2mm} v_{\rm b} - R_{\rm b}i_{\rm dc}' - v_{\rm dc}',\\
  \end{array}
 \right. 
   &
  \end{flalign}
where $v_{\rm dc}'$ is defined in \req{buckboost2}. 
 
 \vspace{2mm}
  \noindent
\underline{Interconnection to grid}
 \begin{flalign}\label{wind_power}
  &
  P+jQ = (P_{\rm s} -P_{\rm r})  + j (Q_{\rm s} - Q_{\rm r}),
&
 \end{flalign}
 where $P_{\rm s}$ and $Q_{\rm s}$ are defined in \req{wind_current}, and
 $P_{\rm r}$ and $Q_{\rm r}$ in \req{fsc}.


In reference to \req{model_comp}, the wind farm model with the battery and DC/DC converter can be summarized as:
\begin{equation}\label{total_wind}
\scalebox{0.96}{$\displaystyle
\begin{array}{rcl}
 x_k &\hspace{-2mm}:=&\hspace{-2mm} [\omega_{{\rm l},k}, \omega_{{\rm r},k}, \theta_{{\rm T},k}, i_{k}^{\sf T}, i_{{\rm G},k}^{\sf T}, \chi_{{\rm G},k}^{\sf T}, \zeta_{{\rm G},k}^{\sf T}, \chi_{{\rm R},k}^{\sf T}, v_{{\rm dc},k}, v_{{\rm b},k}, i_{{\rm dc},k}']^{\sf T} \in \mathbb R^{18}\vspace{1mm},\\
  \zeta_{\rm G,k}  &\hspace{-2mm}:=&\hspace{-2mm} [\zeta_{{\rm dG},k}, \zeta_{{\rm qG},k}]^{\sf T},\quad
   i_{\rm G,k} := [i_{{\rm dG},k}, i_{{\rm qG},k}]^{\sf T}, \quad 
   \chi_{\bullet,k} := [\chi_{{\rm d\bullet},k}, \chi_{{\rm q\bullet},k}]^{\sf T},   \quad \bullet \in \{\rm G, R\},\\
 u_k &\hspace{-2mm}:=&\hspace{-2mm} [u_{{\rm dG},k}, u_{{\rm qG},k}, u_{{\rm dR},k}, u_{{\rm qR},k}, u_{{\rm S},k}]^{\sf T} \in \mathbb R^5, \quad
  \alpha_k := [v_{{\rm dc},k}^{\star}, Q_{{\rm r},k}^{\star}, |{\bf V}_{k}|^{\star}, \omega_{{\rm r},k}^{\star}]^{\sf T} \in \mathbb R^4,
\end{array}
 $}
\end{equation}
and $f_k(\cdot, \cdot, \cdot ; \cdot)$ and $g_k(\cdot, \cdot ; \cdot)$ in \req{model_comp}  follow from \req{turbine_model}-\req{wind_power} for $k \in \mathbb N_{\rm W}$. The steady-state $x_k$ is determined as follows. Note that, given the total generated power $P_{k}^{\star} + jQ_k^{\star}$, there exists a degree of freedom for determining $P_{{\rm r},k}^{\star} + jQ_{{\rm r},k}^{\star}$ and $P_{{\rm s},k}^{\star} + jQ_{{\rm s},k}^{\star}$ satisfying \req{wind_power} in steady state. Thus, not only the triple $({\bf V}_k^{\star}, P_k^{\star}, Q_k^{\star})$ but also the pair $(P_{{\rm r},k}^{\star},Q_{{\rm r},k}^{\star})$ needs to be known. In this setting, the pair $(x_k^{\star}, \alpha_k)$ satisfying \req{model_comp_star} is uniquely determined. 

\subsection{Solar Farms}\label{subsec:model_PV}
\begin{figure}[t]
  \begin{center}
    \includegraphics[clip,width=165mm]{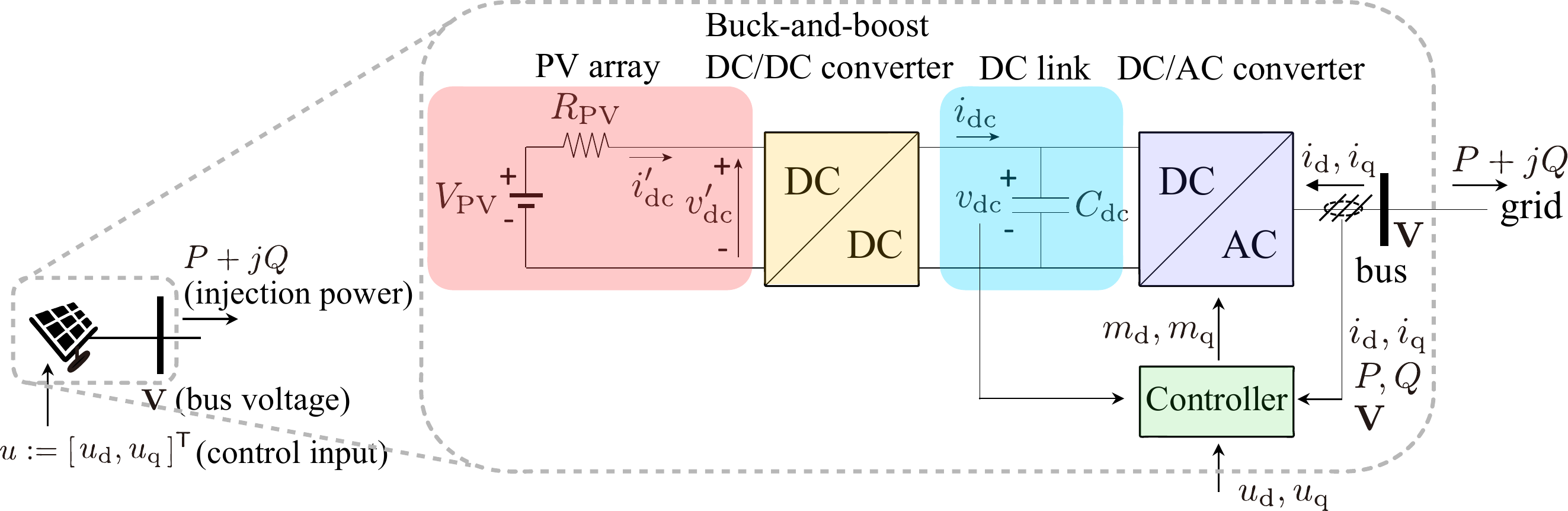}
    \caption{Equivalent circuit of the model of a solar farm and its terminal bus}
    \label{fig_solarMODEL}
  \end{center}
\end{figure}

\begin{figure}[t]
  \begin{center}
    \includegraphics[clip,width=165mm]{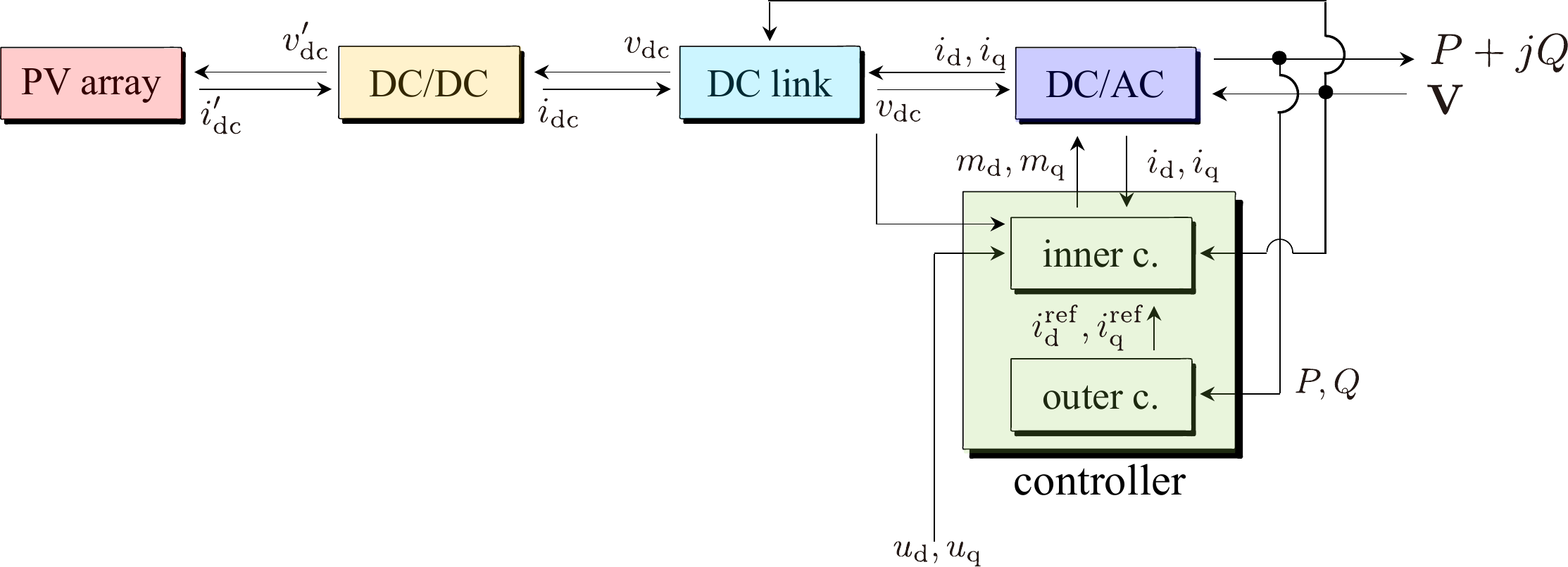}
    \caption{Signal-flow diagram of the model of a solar farm and its terminal bus, where the constant signal $P^{\star}$, $Q^{\star}$, and $S$ are omitted.}
    \label{fig:sigflow_PV}
  \end{center}
\end{figure}

\begin{figure}[h!]
  \begin{center}
    \includegraphics[clip,width=165mm]{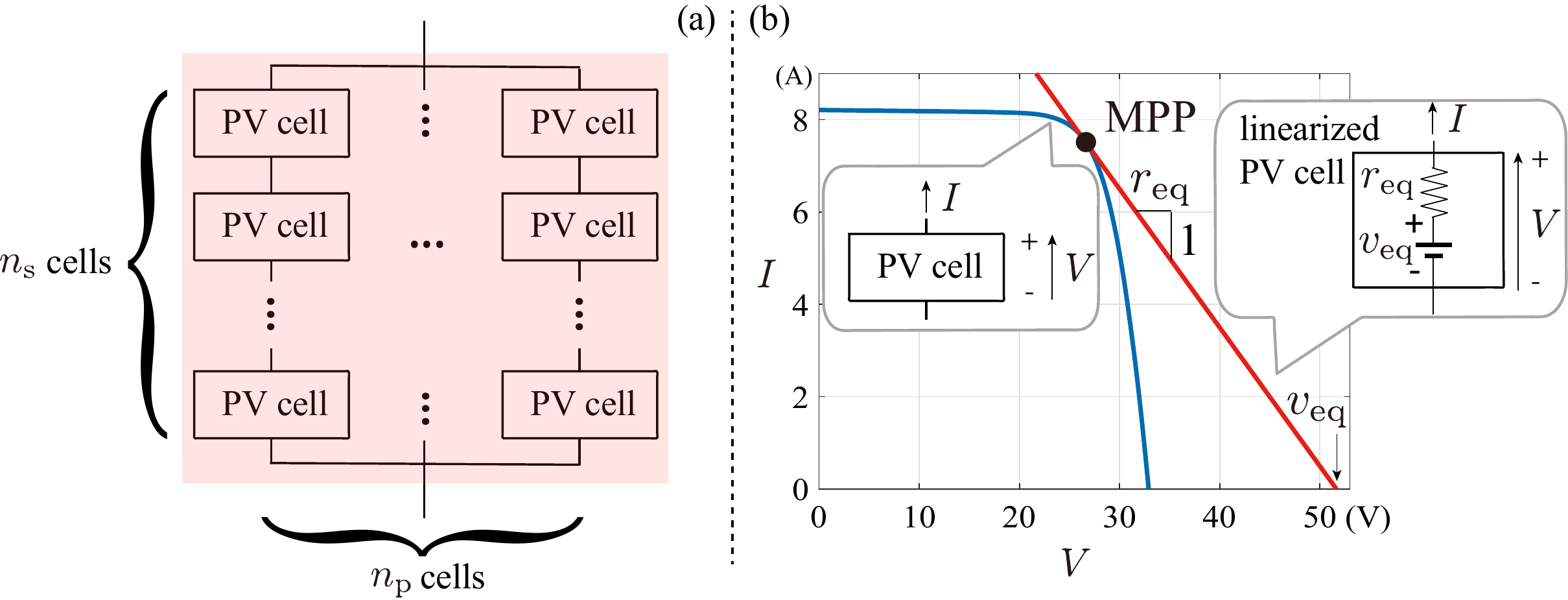}
    \caption{(a) PV array structure (b) I-V characteristics of PV cell KC200GT}
    \label{fig_PVarray}
  \end{center}
\end{figure}

\begin{table}[h!]
  \caption{Nomenclature for the solar farm model. The signs of the currents are positive when flowing in the direction of the corresponding arrows in Figure~\ref{fig_solarMODEL}.
  The values of the PV array parameters are the case where $n_{\rm s}=11$ and $n_{\rm p}=1000$ with the KC200GT PV cell \cite{villalva2009comprehensive}. The value of $S$ below is a typical one because that will change depending on power system operation conditions, as shown in \req{equivSOLAR}. The values of all parameters in per unit are rated at 100MW. }
  \begin{tabular}{|l|l|p{110mm}|}\hline 
   Symbol &  Numerical value & Description \\\hline \hline
   \multicolumn{3}{|c|}{DC/AC converter and its controller}\\ \hline
   $i_{\rm d}$, $i_{\rm q}$&&  d- and q-axis currents flowing from AC-side to DC-side \\ 
   $m_{\rm d}$, $m_{\rm q}$&&  d- and q-axis duty cycles \\
   $P+jQ$&& power injecting from solar bus \\
   $\chi_{\rm d}$, $\chi_{\rm q}$&& inner-loop controller state \\
   $\zeta_{\rm d}$, $\zeta_{\rm q}$&& outer-loop controller state \\
   $i_{\rm d}^{\rm ref}$, $i_{\rm q}^{\rm ref}$&& reference signal of $i_{\rm d}$ and $i_{\rm q}$ \\
   $u_{\rm d}$, $u_{\rm q}$&& additional control input signals on duty cycles\\
   $\gamma_{\rm PV}$&&  number of PV generators inside farm \\
   $L_{\rm ac}$, $R_{\rm ac}$& $39.59, 0.05$& inductance and resistance of DC/AC converter \\
   $P^{\star}+jQ^{\star}$& $0.02\gamma_{\rm PV}+j0$ & steady-state power injecting from solar bus \\
   $K_{\rm P, d}$, $K_{\rm I, d}$ &$-0.01, -0.1$& PI gains of d-axis outer-loop controller \\
   $K_{\rm P, q}$, $K_{\rm I, q}$ &$0.01, 0.1$& PI gains of q-axis outer-loop controller \\
   $\tau_{\rm ac}$& 0.7 & design parameter representing time constant of converter current dynamics \\ \hline
   \multicolumn{3}{|c|}{DC link}\\ \hline
   $v_{\rm dc}$&& DC link voltage \\
   $C_{\rm dc}$& $44.87$ & DC link capacitance\\ 
   $G_{\rm sw}$& $1.19 \times 10^{-4}$  & conductance representing switching loss of DC/AC converter\\  \hline
   \multicolumn{3}{|c|}{Buck-and-Boost DC/DC converter}\\ \hline
   $i_{\rm dc}$&& current flowing from DC/DC converter to DC link \\
   $v_{\rm dc}'$&& voltage at PV array side \\
   $S$& $0.144$ & step down/up gain so that solar farm is operated at MPP\\ \hline 
   \multicolumn{3}{|c|}{PV array}\\ \hline
   $i_{\rm dc}'$&& current flowing from the PV array to the DC link \\
   $R_{\rm PV}$ & $7.687$ & series resistance inside PV array model\\ 
   $V_{\rm PV}$ & $0.823$ & voltage of constant voltage source inside PV array\\ \hline
  \end{tabular}
\end{table}
 

A solar farm model consists of a PV array, a buck-and-boost DC/DC converter, a DC/AC converter with a controller, and a DC link \cite{villalva2009comprehensive}, as shown in Figure~\ref{fig_solarMODEL}.
The signal-flow diagram for the system is shown in Figure~\ref{fig:sigflow_PV}. The dynamics of the DC/AC converter, its controller, and DC link are similar to those in the wind farm model, given in \req{acdc_pv}-\req{dclink_pv}. The models of the PV array and DC/DC converter are described as follows. 

 \vspace{2mm}
\subsubsection{PV array}
 The PV array is a parallel interconnection of $n_{\rm p}$ circuits, each of which contains $n_{\rm s}$ series-connected PV cells, as shown in Figure~\ref{fig_PVarray} (a). Each PV cell is assumed to be identical. Typically, a PV cell has nonlinear I-V characteristics, as shown by the blue line in Figure~\ref{fig_PVarray} (b) \cite{villalva2009comprehensive}. Assuming that the PV cell is operated around the so-called {\it maximum power point (MPP)}  where the cell output power is maximized, the I-V curve around this point can be approximated by a linear function as shown by the red line.  In that case, the PV array can be modeled as a series connection of a constant voltage source with value $V_{\rm PV} := n_{\rm s}v_{\rm eq}$ and a resistance whose value is $R_{\rm PV} := (n_{\rm s}/n_{\rm p})r_{\rm eq}$. This PV array model is described as \req{pvarray}.

  \vspace{2mm}
 \subsubsection{DC/DC converter}
 A buck-and-boost DC/DC converter is used to ensure PV cell operation around MPP.
 This can be done by determining the converter step down/up gain $S$ such that  the steady-state PV array output voltage $v_{\rm dc}'$ and current $i_{\rm dc}'$ coincide with the maximum point on the I-V curve. While the gain can be dynamically regulated by controllers, for simplicity, the gain $S$ is supposed to be constant. The DC/DC converter model is described as \req{dcdc_PV}.


 
 \vspace{2mm}
 \noindent
\underline{PV array: }
\begin{flalign}\label{pvarray}
 &
 i_{\rm dc}' = \frac{V_{\rm PV} - v_{\rm dc}'}{R_{\rm PV}},
 &
\end{flalign}
where $v_{\rm dc}'$ is defined in \req{dcdc_PV}. \\

 \vspace{2mm}
 \noindent
 \underline{Buck-and-Boost DC/DC converter: }
\begin{flalign}\label{dcdc_PV}
 &
 v_{\rm dc}' = Sv_{\rm dc}, \quad  i_{\rm dc} = Si_{\rm dc}',
 &
\end{flalign}
where $v_{\rm dc}$ and $i_{\rm dc}'$ are defined in \req{dclink_pv} and \req{pvarray}. \\
\underline{DC/AC converter: }
\begin{flalign}\label{acdc_pv}
 &
 \left\{
  \begin{array}{rcl}
   \frac{L_{\rm ac}}{\bar{\omega}}\dot{i}_{\rm d}&\hspace{-2mm}=&\hspace{-2mm} -R_{\rm ac}i_{\rm d} + L_{\rm ac}i_{\rm q} + {\rm Re}({\bf V}) - \frac{m_{\rm d}}{2}v_{\rm dc},\\
   \frac{L_{\rm ac}}{\bar{\omega}}\dot{i}_{\rm q}&\hspace{-2mm}=&\hspace{-2mm} -R_{\rm ac}i_{\rm q} - L_{\rm ac}i_{\rm d} + {\rm Im}({\bf V}) - \frac{m_{\rm q}}{2}v_{\rm dc},\vspace{1mm}\\
   P + jQ &\hspace{-2mm}=&\hspace{-2mm} -\gamma_{\rm PV} ({\rm Re}({\bf V})i_{\rm d} + {\rm Im}({\bf V})i_{{\rm q}}) - j \gamma_{\rm PV}({\rm Im}({\bf V})i_{\rm d} - {\rm Re}({\bf V})i_{{\rm q}}),\\
  \end{array}
 \right.
 &
\end{flalign}
where $m_{\rm d}$ and $m_{\rm q}$ are defined in \req{con_duty_acdc}, and $v_{\rm dc}$ in \req{dclink_pv}. \\

 \vspace{2mm}
  \noindent
\underline{Outer-Loop controller of DC/AC converter: }
  \begin{flalign}\label{outer_acdc}
   & 
 \left\{
\begin{array}{rcl}
 \dot{\zeta}_{\rm d}&\hspace{-2mm}=&\hspace{-2mm} K_{\rm I, d}(P^{\star} - P),\\
 i_{\rm d}^{\rm ref}&\hspace{-2mm}=&\hspace{-2mm} K_{\rm P, d}(P^{\star} - P) + \zeta_{\rm d},\\
\end{array}
 \right. \quad
 \left\{
   \begin{array}{rcl}
 \dot{\zeta}_{\rm q}&\hspace{-2mm}=&\hspace{-2mm} K_{\rm I, q}(Q^{\star} - Q),\\ 
 i_{\rm q}^{\rm ref}&\hspace{-2mm}=&\hspace{-2mm} K_{\rm P, q}(Q^{\star} - Q_{\rm }) + \zeta_{\rm q},\\
   \end{array}
 \right.  &
  \end{flalign}
  where $P$ and $Q$ are defined in \req{acdc_pv}.\\

   \vspace{2mm}
  \noindent
\underline{Inner-Loop controller of DC/AC converter: }
  \begin{flalign}\label{con_duty_acdc}
   &
   \begin{array}{l}
 \left\{
  \begin{array}{rcl}
   \tau_{\rm ac} \dot{\chi}_{\rm d} &\hspace{-2mm}=&\hspace{-2mm} i_{\rm d}^{\rm ref} - i_{\rm d}, \\
   m_{\rm d} &\hspace{-2mm}=&\hspace{-2mm} {\rm sat}\left(\frac{2}{v_{\rm dc}}  \left({\rm Re}({\bf V}) + L_{\rm ac}i_{\rm q} - R_{\rm ac}\chi_{\rm d} - \frac{L_{\rm ac}}{\bar{\omega}\tau_{\rm ac}}(i_{\rm d}^{\rm ref} - i_{\rm d})\right) + u_{\rm d}\right),\\ 
  \end{array}
  \right. \\
 \left\{
  \begin{array}{rcl}
   \tau_{\rm ac} \dot{\chi}_{\rm q} &\hspace{-2mm}=&\hspace{-2mm} i_{\rm q}^{\rm ref} - i_{\rm q}, \\
   m_{\rm q} &\hspace{-2mm}=&\hspace{-2mm} {\rm sat}\left(\frac{2}{v_{\rm dc}}  \left({\rm Im}({\bf V}) - L_{\rm ac}i_{\rm d} - R_{\rm ac}\chi_{\rm q} - \frac{L_{\rm ac}}{\bar{\omega}\tau_{\rm ac}}(i_{\rm q}^{\rm ref} - i_{\rm q})\right) + u_{\rm q}\right),
  \end{array}
 \right.
   \end{array}&
  \end{flalign}
  where
  $i_{\rm d}$ and $i_{\rm q}$ are defined in \req{acdc_pv},
  $i_{\rm d}^{\rm ref}$ and $i_{\rm q}^{\rm ref}$ in \req{outer_acdc}, and
  $v_{\rm dc}$  in \req{dclink_pv}. \\

  \vspace{2mm}
  \noindent
\underline{DC link: }
 \begin{flalign}\label{dclink_pv}
  &
 \left.
  \begin{array}{rcl}
   \frac{C_{\rm dc}}{\bar{\omega}}\dot{v}_{\rm dc} &\hspace{-2mm}=&\hspace{-2mm} \frac{1}{2v_{\rm dc}}
    \left(
     {\rm Re}({\bf V})i_{\rm d} + {\rm Im}({\bf V})i_{\rm q} +
     v_{\rm dc}i_{\rm dc} -
     R_{\rm ac}(i_{\rm d}^2 + i_{\rm q}^2)
    \right) - G_{\rm sw}v_{\rm dc}, 
  \end{array}
  \right.
  &
\end{flalign}
where $i_{\rm d}$ and $i_{\rm q}$ are defined in \req{acdc_pv}, and $i_{\rm dc}$ in \req{dcdc_PV}. 

In reference to \req{model_comp} the solar farm model can be summarized as:
\begin{equation}\label{equibsolar}
\scalebox{1.0}{$\displaystyle
\begin{array}{rcl}
 x_k &\hspace{-2mm}:=&\hspace{-2mm} [i_{{\rm d},k}, i_{{\rm q},k}, \chi_{{\rm d},k}, \chi_{{\rm q},k}, \zeta_{{\rm d},k}, \zeta_{{\rm q},k}, v_{{\rm dc},k}]^{\sf T} \in \mathbb R^{7}, \vspace{1mm}\\
 u_k &\hspace{-2mm}:=&\hspace{-2mm} [u_{{\rm d},k}, u_{{\rm q},k}]^{\sf T} \in \mathbb R^2, \quad
 \alpha_k := [P_{k}^{\star},  Q_{k}^{\star}, S_{k}]^{\sf T} \in \mathbb R^3,
\end{array}
 $}
\end{equation}
and $f_k(\cdot, \cdot, \cdot ; \cdot)$ and $g_k(\cdot, \cdot ; \cdot)$ in \req{model_comp} follow from \req{pvarray}-\req{dclink_pv} for $k \in \mathbb N_{\rm S}$.  The steady-state value of $x_k$ and $S_k$ can be found as follows. 
Suppose that $(v_{{\rm dc},k}'^{\star}, i_{{\rm dc},k}'^{\star})$ is at the MPP.
Given ${\bf V}_k$, $P_k$, $Q_k$, $v_{{\rm dc},k}'^{\star}$ and $i_{{\rm dc},k}'$,
the pair $(x_k^{\star}, \alpha_k)$ satisfying \req{model_comp_star}  are then uniquely determined  as
 $ x_k^{\star} = [i_{{\rm d},k}^{\star}, i_{{\rm q},k}^{\star}, \chi_{{\rm d},k}^{\star}, \chi_{{\rm q},k}^{\star}, \zeta_{{\rm d},k}^{\star}, \zeta_{{\rm q},k}^{\star}, v_{{\rm dc},k}^{\star}]^{\sf T}$ 
 and $\alpha_k$ in \req{equibsolar} where
 \begin{equation}\label{equivSOLAR}
  \begin{array}{rcl}
   \left[
    \begin{array}{c}
     \zeta_{{\rm d},k}^{\star} \\
     \zeta_{{\rm q},k}^{\star} 
    \end{array}
   \right] &\hspace{-2mm}=&\hspace{-2mm}
     \left[
    \begin{array}{c}
     \chi_{{\rm d},k}^{\star} \\
     \chi_{{\rm q},k}^{\star} 
    \end{array}
   \right] = 
     \left[
    \begin{array}{c}
     i_{{\rm d},k}^{\star} \\
     i_{{\rm q},k}^{\star} 
    \end{array}
   \right] = \frac{1}{|{\bf V}_k^{\star}|^2}
     \left[
    \begin{array}{cc}
     -{\rm Re}({\bf V}_{k}^{\star}) & -{\rm Im}({\bf V}_{k}^{\star}) \\
     -{\rm Im}({\bf V}_{k}^{\star}) & {\rm Re}({\bf V}_{k}^{\star}) 
    \end{array}
   \right]    \left[
    \begin{array}{c}
     \frac{P_{k}^{\star}}{\gamma_{{\rm PV},k}} \\
     \frac{Q_{k}^{\star}}{\gamma_{{\rm PV},k}}
    \end{array}
   \right], \\
   v_{{\rm dc},k}^{\star} &\hspace{-2mm}=&\hspace{-2mm}
    \sqrt{\frac{v_{{\rm dc},k}'^{\star}i_{{\rm dc},k}'^{\star} - \left(\frac{P_k^{\star}}{\gamma_{{\rm PV},k}} +
   R_{{\rm ac},k}({i^{\star}_{{\rm d},k}}^{2} + {i^{\star}_{{\rm q},k}}^{2})\right)}{2G_{{\rm sw},k}}},\quad
    S_k = \frac{v_{{\rm dc},k}'^{\star}}{v_{{\rm dc},k}^{\star}}. 
  \end{array}
 \end{equation}

\subsection{Energy Storage Systems}\label{subsec:model_storage}
 
\begin{figure}[t]
  \begin{center}
    \includegraphics[clip,width=165mm]{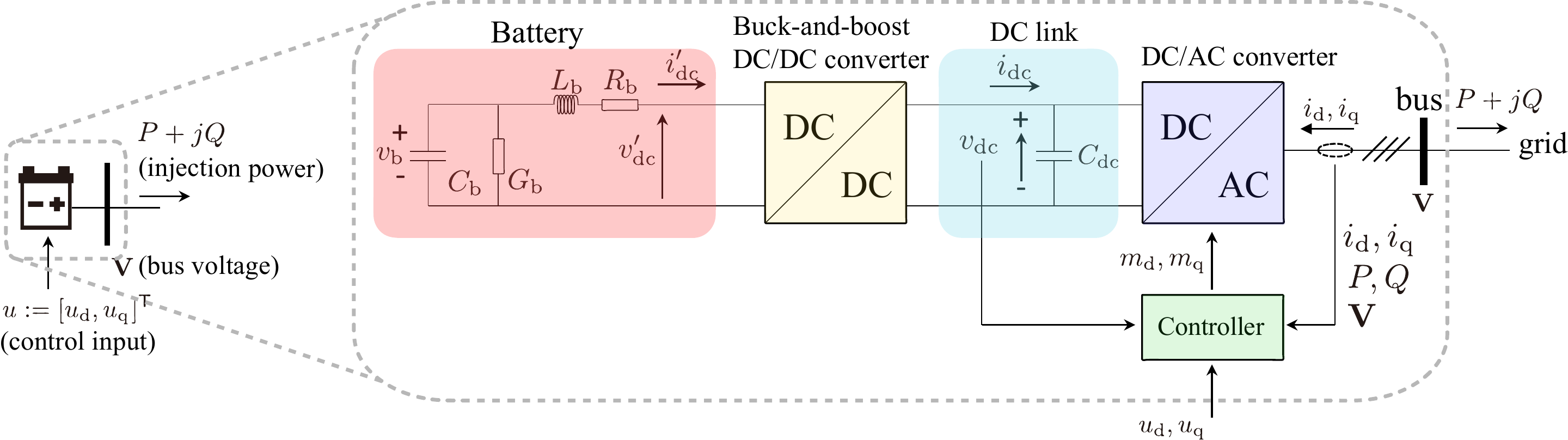}
    \caption{Physical structure of the model of an energy storage system with its terminal bus }
    \label{fig_storage}
  \end{center}
\end{figure}

 The energy storage system consists of a battery, a buck-and-boost DC/DC converter, a DC/AC converter, and a controller, as shown in Figure~\ref{fig_storage}. The basic functions of these four components are to charge/discharge electricity, to step down/up the battery terminal voltage, to rectify the three-phase current to a DC current, and to regulate the DC voltage in between the converters, respectively. When the energy storage system is connected to DC line, the DC/AC converter is not needed.  The dynamics of the DC/AC converter, its controller, DC/DC converter, and DC link are similar to those described in equations \req{acdc_pv}, \req{outer_acdc}-\req{con_duty_acdc}, \req{dcdc_PV}, and \req{dclink_pv}, respectively.


\section{Impact of Distributed Energy Resources on Power System Dynamics and Stability}
\label{sec-impact}
\begin{figure}[h!]
  \begin{center}
    \includegraphics[clip,width=160mm]{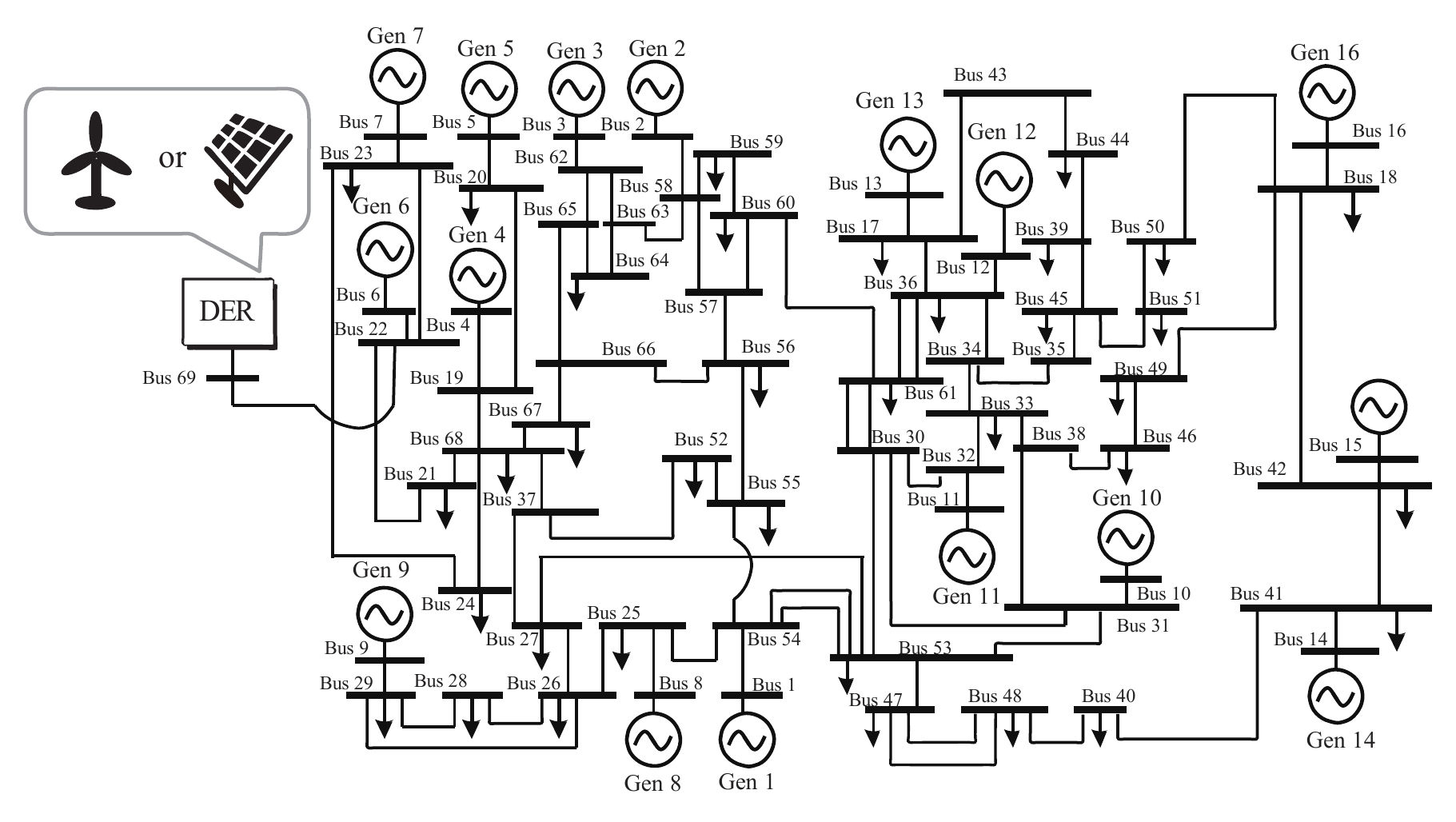}
    \caption{IEEE 68-bus, 16-machine power system model with one DER. The downward arrows represent load extractions.}
    \label{fig:example}
  \end{center}
\end{figure}

\begin{figure}[t]
  \begin{center}
    \includegraphics[clip,width=100mm]{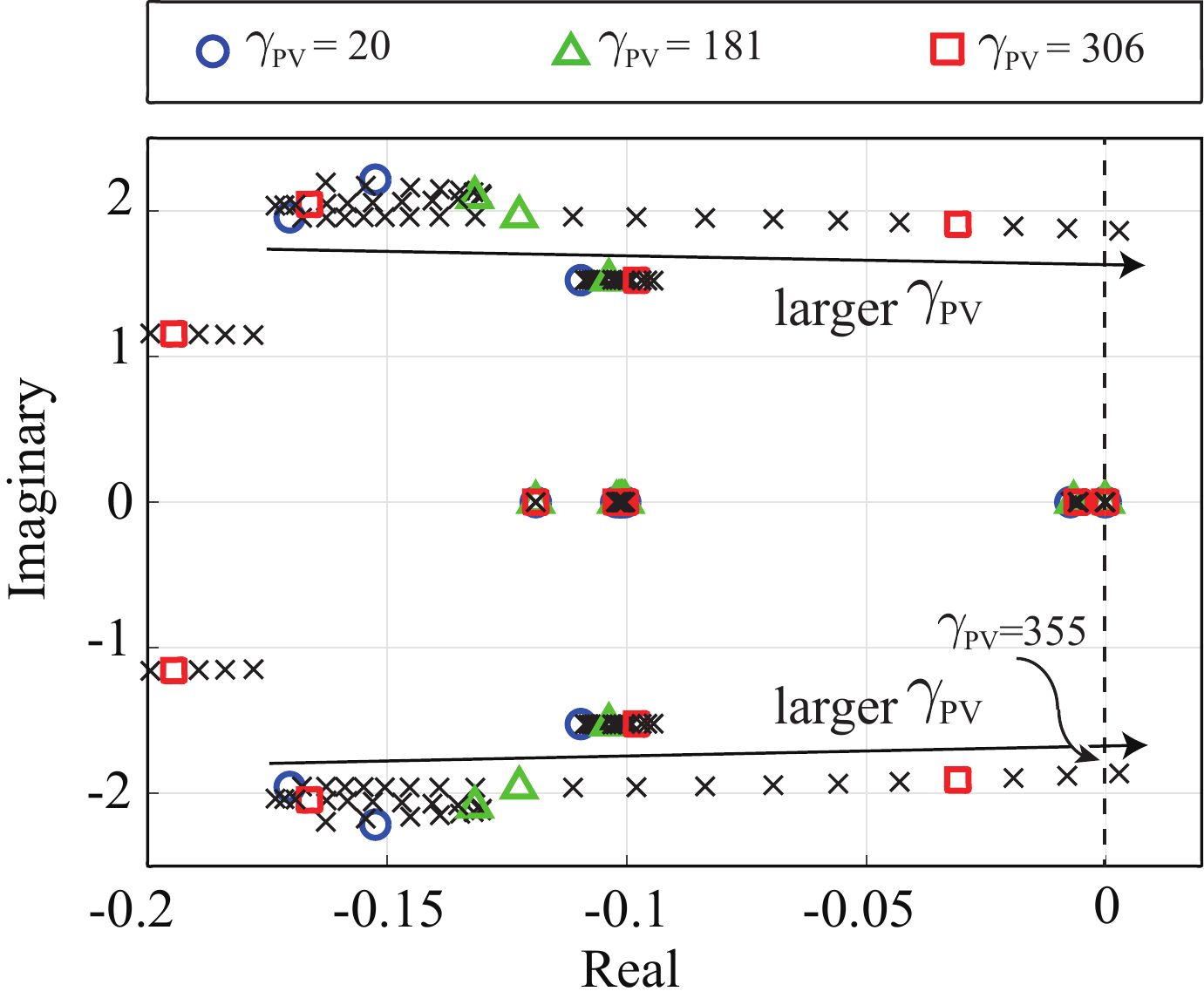}
    \caption{Variation of 13 dominant eigenvalues of PV-integrated power system. }
    \label{fig:PVeig}
  \end{center}
\end{figure}

\begin{figure}[t]
  \begin{center}
    \includegraphics[clip,width=150mm]{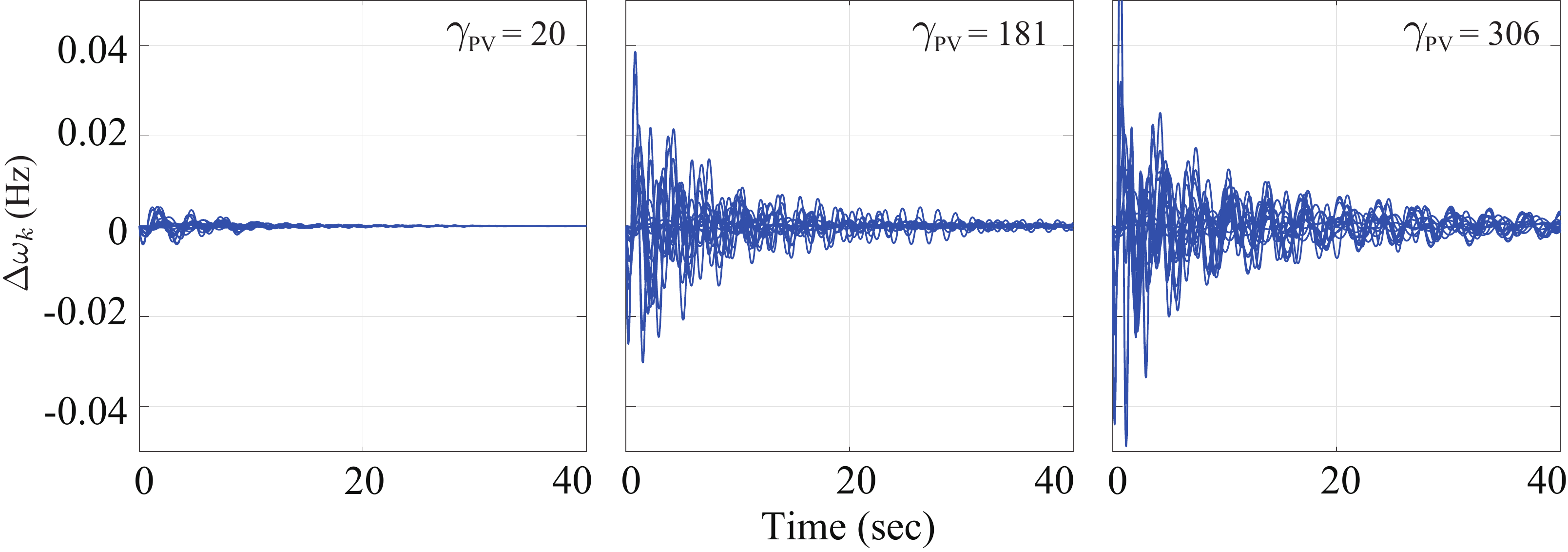}
    \caption{Trajectories of the frequency deviations of all synchronous generators. }
    \label{fig:PVtj}
  \end{center}
\end{figure}

\begin{figure}[t]
  \begin{center}
    \includegraphics[clip,width=100mm]{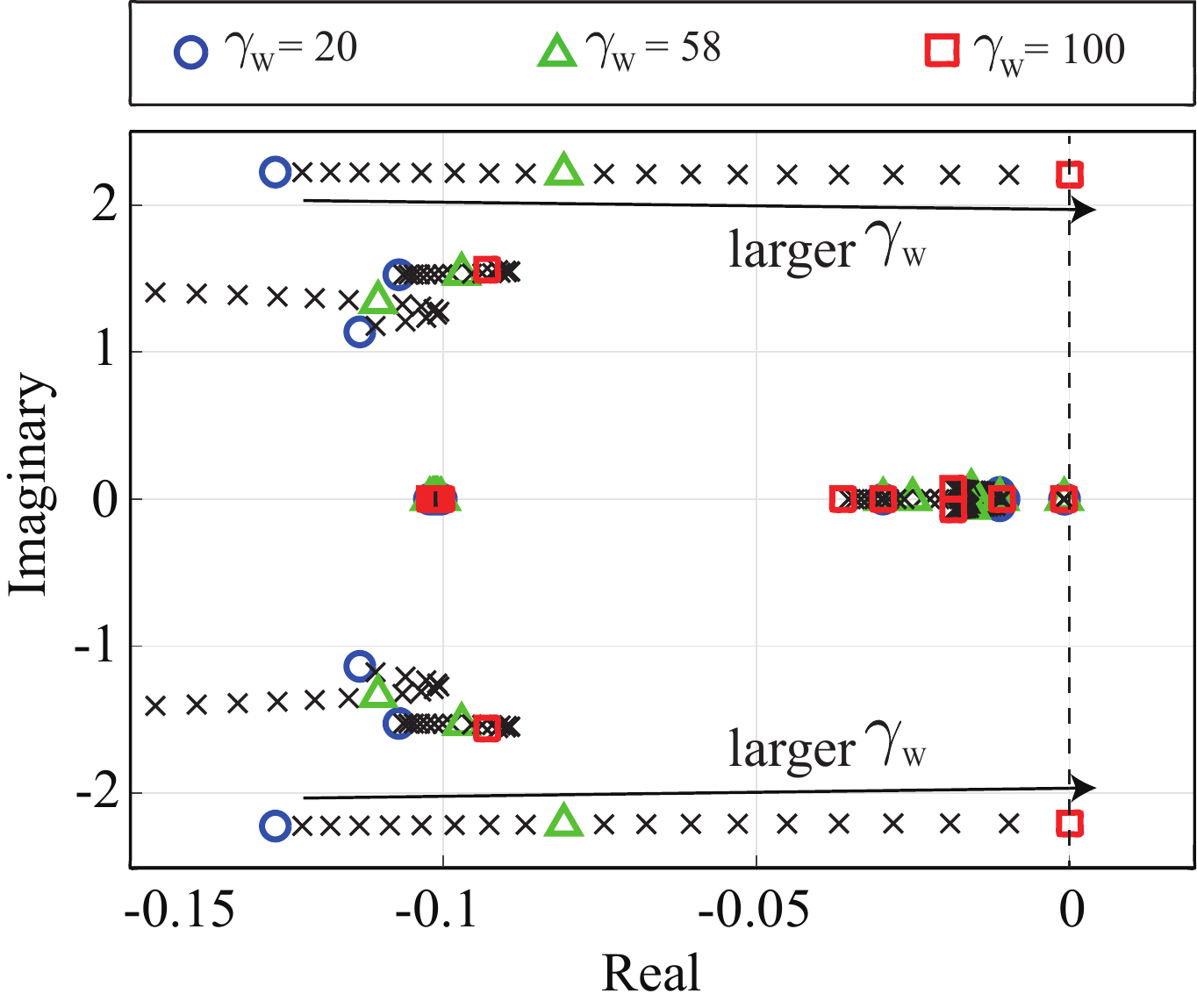}
    \caption{Variation of 14 dominant eigenvalues of wind-integrated power system.}
    \label{fig:windeig}
  \end{center}
\end{figure}

\begin{figure}[t]
  \begin{center}
   \includegraphics[clip,width=150mm]{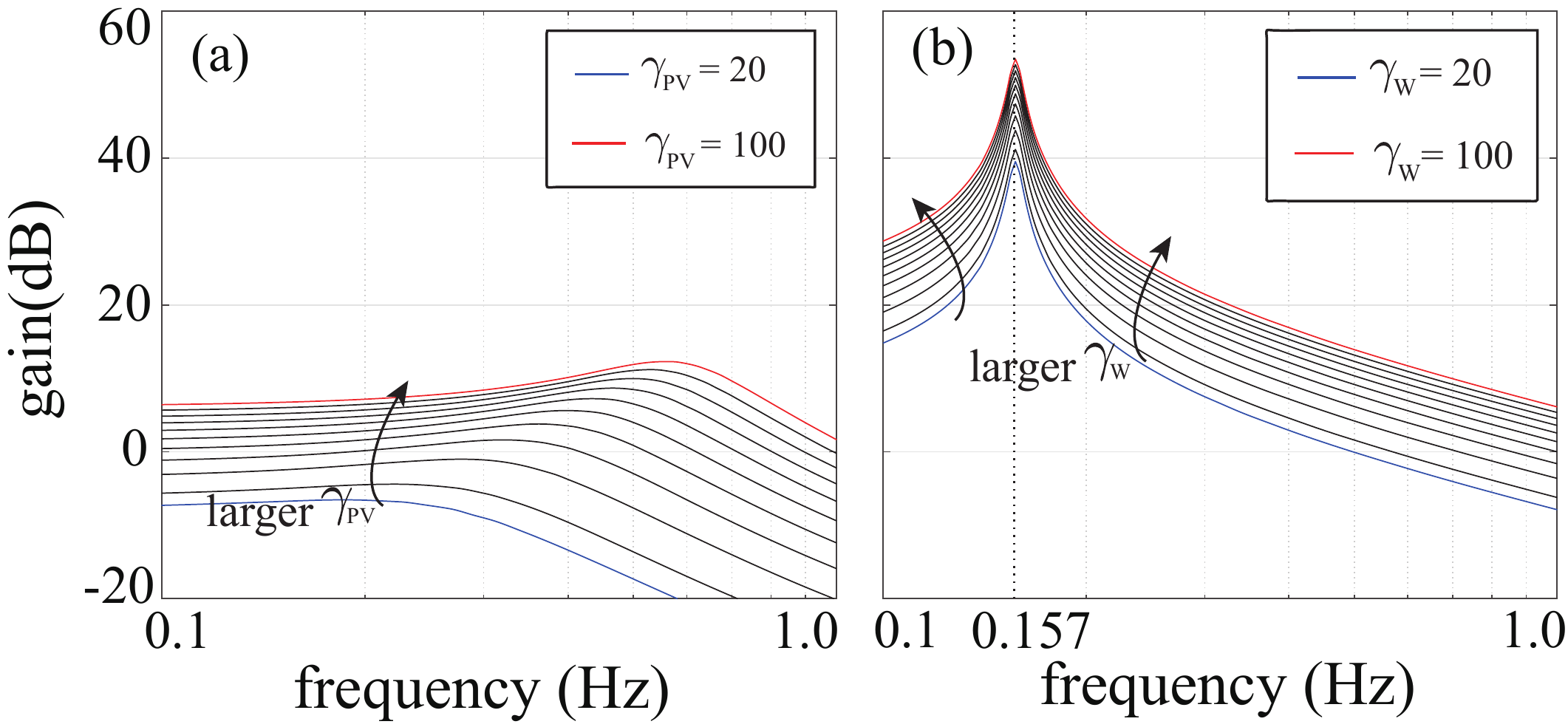}
   \caption{Singular value plot of the frequency response from $[{\rm Re}({\bf V}), {\rm Im}({\bf V})]^{\sf T}$ to $[P, Q]^{\sf T}$ for the (a) linearized solar farm model, and (b) linearized wind farm model}
    \label{fig:bode_wind}
  \end{center}
\end{figure}

\begin{figure}[t]
  \begin{center}
    \includegraphics[clip,width=90mm]{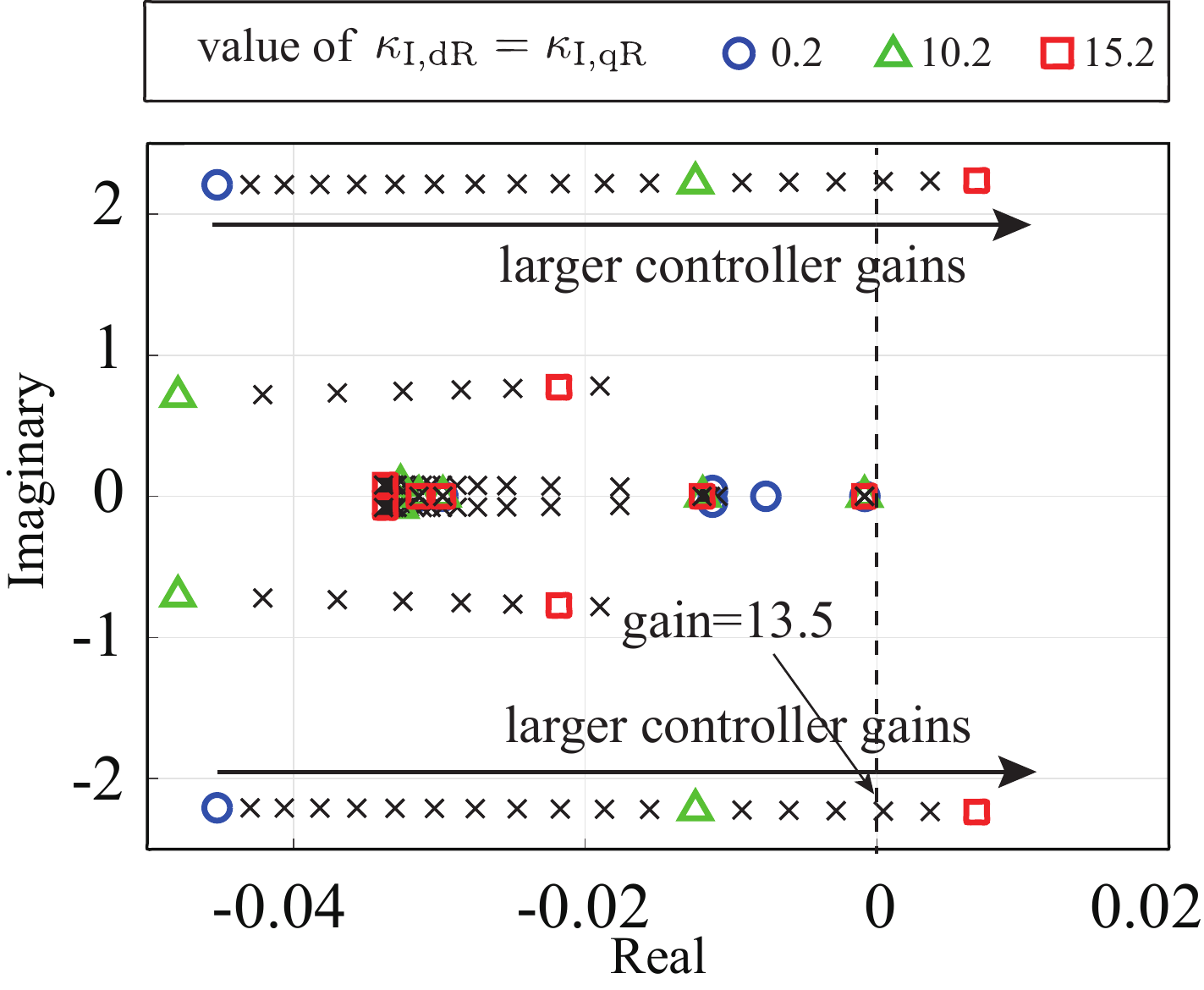}
    \caption{Variation of ten dominant eigenvalues of wind-integrated power system by changing the integral gains of the RSC controller of the B2B converter}
    \label{fig_PI}
  \end{center}
\end{figure}

Given a DER-integrated power system model \req{model_comp}-\req{inter}, the question is - how does the penetration of DERs and their controllers dictate the stability and dynamic performance of the grid? This section demonstrates these impacts using numerical simulations of the IEEE 68-bus power system model \cite{canizares2015benchmark}. The MATLAB codes for these simulations can be found in \cite{tomonori2018matlab}. The network diagram is shown in Figure \ref{fig:example}. Each individual component is modeled by the equations listed in the previous section. The DER bus, denoted as Bus 69, connects to Bus 22. 
The reactance between Bus 22 and 69 represents the transformer for stepping down the grid voltage to the DER voltage. Its value is taken as $j0.01$. 

First, consider the DER to be a solar farm as in \req{pvarray}-\req{dclink_pv} with $\mathbb N_{\rm S} = \{69\}$. The other bus indices $\mathbb N_{\rm G}$, $\mathbb N_{\rm L}$, $\mathbb N_{\rm N}$ are shown in Figure~\ref{fig:example}. 
The model of this PV-integrated power system is the combination of \req{model_comp}-\req{inter} where $\Sigma_k$ for $k \in \mathbb N_{\rm G}$, $\mathbb N_{\rm L}$, $\mathbb N_{\rm N}$ and $\mathbb N_{\rm S}$ are defined as \req{model_sync01}-\req{ABCDpss}, \req{load_model}, \req{non_model}, and \req{pvarray}-\req{dclink_pv}. Note that $\gamma_{\rm PV}$ is the number of PV generators inside the farm. A question here is - how does $\gamma_{\rm PV}$ affect small-signal stability of the grid? Small-signal stability is defined as the stability of the grid model linearized around its equilibrium \cite{kundur1994power}. The procedure to obtain the linearized  model is as follows: first, compute an equilibrium of the entire system \req{model_comp}-\req{inter}, as summarized in the previous section; second, linearize the individual component dynamics \req{model_comp} and the interconnection equation \req{inter} about this equilibrium. Note that generators, solar farms, and wind farms are dynamical systems while loads, non-unit buses, and the interconnection are static systems. Thus, the d- and q-axis voltages of all buses are redundant states. By eliminating these redundant states from the linearized differential algebraic equation (DAE) model, a linearized ordinary differential equation (ODE) model can be  obtained. This method of elimination is referred to as Kron reduction. Figure~\ref{fig:PVeig} shows the 13 dominant eigenvalues of this linearized power system model at a desired equilibrium. The eigenvalues around $-0.17\pm j2$ for $\gamma_{\rm PV} = 20$ start moving to the right as the value of $\gamma_{\rm PV}$ is increased,
and finally cross the imaginary axis when $\gamma_{\rm PV}> 355$, resulting in an unstable system.
Each PV generator is rated at 2 MW; therefore $\gamma_{\rm PV} = 355$ means that the net steady-state power output of the solar farm is $P_{69}^{\star} = 710$ MW, which is 3.85\% of the total generated power of the system. This may look like a small percentage, but in terms of the stability limit the amount of solar penetration is quite close to critical.
This pole shift happens due to the fact that the equilibrium changes with $\gamma_{\rm PV}$. 
When a fault (modeled as an impulse function causing the initial conditions of $i_{{\rm d},k}$ and $i_{{\rm q},k}$ in \req{acdc_pv} to move from their equilibrium values) is induced oscillations in the transient response of the states can easily be seen.
Figure~\ref{fig:PVtj} shows the frequency deviation of all 16 synchronous generators for the cases where $\gamma_{\rm PV} = 20, 181, 306$. The results indicate that as $\gamma_{\rm PV}$ increases the PV-integrated power system, without any DER control,  becomes oscillatory with poor damping.

Next, the solar farm at Bus 69 is replaced by a wind farm without  a battery or a DC/DC converter. 
Figure~\ref{fig:windeig} shows the first 14 dominant eigenvalues of the linearized wind-integrated power system at a desired equilibrium.
The eigenvalues around $-0.13\pm 2.1j$ for $\gamma_{\rm W} = 20$ start moving to the right as the value of $\gamma_{\rm W}$ is increased, and finally cross the imaginary axis when $\gamma_{\rm W} > 100$, resulting in an unstable system. 
Each wind generator is rated at 2 MW; therefore, $\gamma_{\rm W} = 100$ means that the net steady-state power output of the farm is $P_{69}^{\star} = 200$ MW. Thus, compared to the PV penetration, the wind penetration in this case poses a greater threat to small-signal stability.
To investigate the difference between the two, the singular value plot of the frequency response of each model from the d- and q-axis bus voltages $[{\rm Re}({\bf V}), {\rm Im}({\bf V})]^{\sf T}$ to the injected power $[P, Q]^{\sf T}$ is shown in Figure~\ref{fig:bode_wind} (a) and (b), respectively.
The figure shows that the wind farm model has a resonance peak at 0.157 Hz, and the amplitude of the peak increases as $\gamma_{\rm W}$ is increased. This is an interesting observation since 0.157 Hz lies in the range of frequencies for the low-frequency (0.1 Hz to 2 Hz) oscillations of the synchronous generators,  commonly called inter-area oscillations \cite{sauer2017power}.
The wind injection at Bus 69,  thus, stimulates an inter-area mode in this case. The resonance mode actually stems from the internal characteristics of the DFIG dynamic model. Details of this phenomenon can be found in \cite{sadamoto2017retrofit}. The PV model, on the other hand, does not show any such resonance peak. 

One potential way to combat the poorly damped oscillation would be to tune the PI gains of the converter controller \req{outer}-\req{con_duty} and \req{outerRSC}-\req{outer2}.
However, such tuning must be done extremely carefully with full knowledge of the entire grid model, since high values of these gains can jeopardize closed-loop stability.
This is shown in Figure~\ref{fig_PI}, where the first ten dominant eigenvalues of the linearized closed-loop model for $\gamma_{\rm w}=80$ are shown.
The integrator gains that are more than 13.5 end up destabilizing the power system.
This is because
high-gain controllers stimulate the negative coupling effect between the DER and the rest of the grid. 
These observations show how many of the power system damping controllers used in today's grid can easily become invalid tomorrow.  Much more systematic control mechanisms need to be built for the future grid to accommodate deep DER penetration while increasing flexibility and robustness.

\section{New Approaches for Controlling Power Systems with Distributed Energy Resources}
\label{sec-newapp}

\subsection{Local Control of Distributed Energy Resources}\label{subsec-localC}
To counteract the destabilizing effects that may be caused by deep penetration of DERs in a power grid, as shown in the previous section, a local control mechanism for each individual DER needs to be built.
A brief survey of local controllers used in today's grid, both with and without DERs, is summarized in Appendix \ref{sidebar-surveyLC}. 
One drawback of existing approaches, however, is that although the controller implementation is decentralized, their design is not necessarily so. This means that the controllers are designed jointly based on full knowledge of the entire power system  model. As DER penetration is growing at an unpredictably high rate, grid operators must make sure that if more DERs are installed in the future the existing  controllers would not need to be retuned or redesigned from scratch. In other words, the DER controllers to be designed must have {\it plug-and-play} capability.

A control method called {\it retrofit control}, recently proposed in \cite{ishizaki2016distributed,sadamoto2017retrofit} can fulfill this objective. Unlike conventional methods listed in Appendix \ref{sidebar-surveyLC}, this method has an inherent plug-and-play property, and therefore is ideal for local control of DERs. 
A brief summary of this approach is presented as follows.
The dynamical system $\dot{x} = f(x, u)$, where $u$ is input, is said to be stable if the autonomous system under $u=0$ is asymptotically stable.  
Consider a power system integrated with solar and wind farms. For $k \in \mathbb N_{\rm S}\cup\mathbb N_{\rm W}$, let the dynamic model of the DER connected to the $k$-th bus be rewritten as
\begin{equation}\label{sigma_wind}
 \dot{x}_k = A_k x_k + B_k u_k + \tilde{f}_k(x_k, {\bf V}_k, u_k),
\end{equation} 
where 
\[
 A_k := \frac{\partial f_k}{\partial x_k}, \quad
 B_k := \frac{\partial f_k}{\partial u_k}, \quad
 \tilde{f}_k(x_k, {\bf V}_k, u_k) := f_k(x_k, {\bf V}_k, u_k; \alpha_k) -  (A_k x_k + B_k u_k),
\]
and $f_k(\cdot, \cdot, \cdot ; \cdot)$ follows from \req{pvarray}-\req{dclink_pv} for solar farms (when $k \in \mathbb N_{\rm S}$), and from \req{turbine_model}-\req{wind_power} for wind farms (when $k \in \mathbb N_{\rm W}$). The following assumptions are imposed on the power system. 

 \vspace{2mm}
\begin{assumption}\label{assump1}\vspace{0mm}
 The power system model \req{model_comp}-\req{inter} is stable.
\end{assumption}
\begin{assumption}\label{assump2}\vspace{0mm}
 The DER state vector $x_k$ and its bus voltage ${\bf V}_k$ are measurable for each DER.
\end{assumption}
 \vspace{2mm}
 
Assumption 1 can be guaranteed by ensuring that the grid, without any additional controllers, is stable by properly tuning pre-existing controllers such as PSSs. Since PSS guarantees stable operation of all power systems in practice, one can see that this is a fair assumption for a retrofit controller to work in reality. 

Assumption 2 is made to simplify the design; the availability of $x_k$ can be relaxed to output feedback case.
The goal is to design a decentralized controller for each DER. The two main requirements from this controller are:

\vspace{0mm}
\begin{enumerate}
\item[i)] The controllers should preserve closed-loop system stability, and also
 improve the damping of the generator frequency deviations and line flows. \vspace{1mm}
\item[ii)] Each of the controller should depend only on local state feedback
	 from its DER, and not on any states from the
	 rest of the grid including other DERs. The controller should
	 also be designed independent of the model of the rest of the system. 
\end{enumerate}\vspace{2mm}
Property (ii) implies that the controllers should be modular by design, and decentralized by implementation.

\begin{figure}[t]
\begin{center}
\includegraphics[width=150mm]{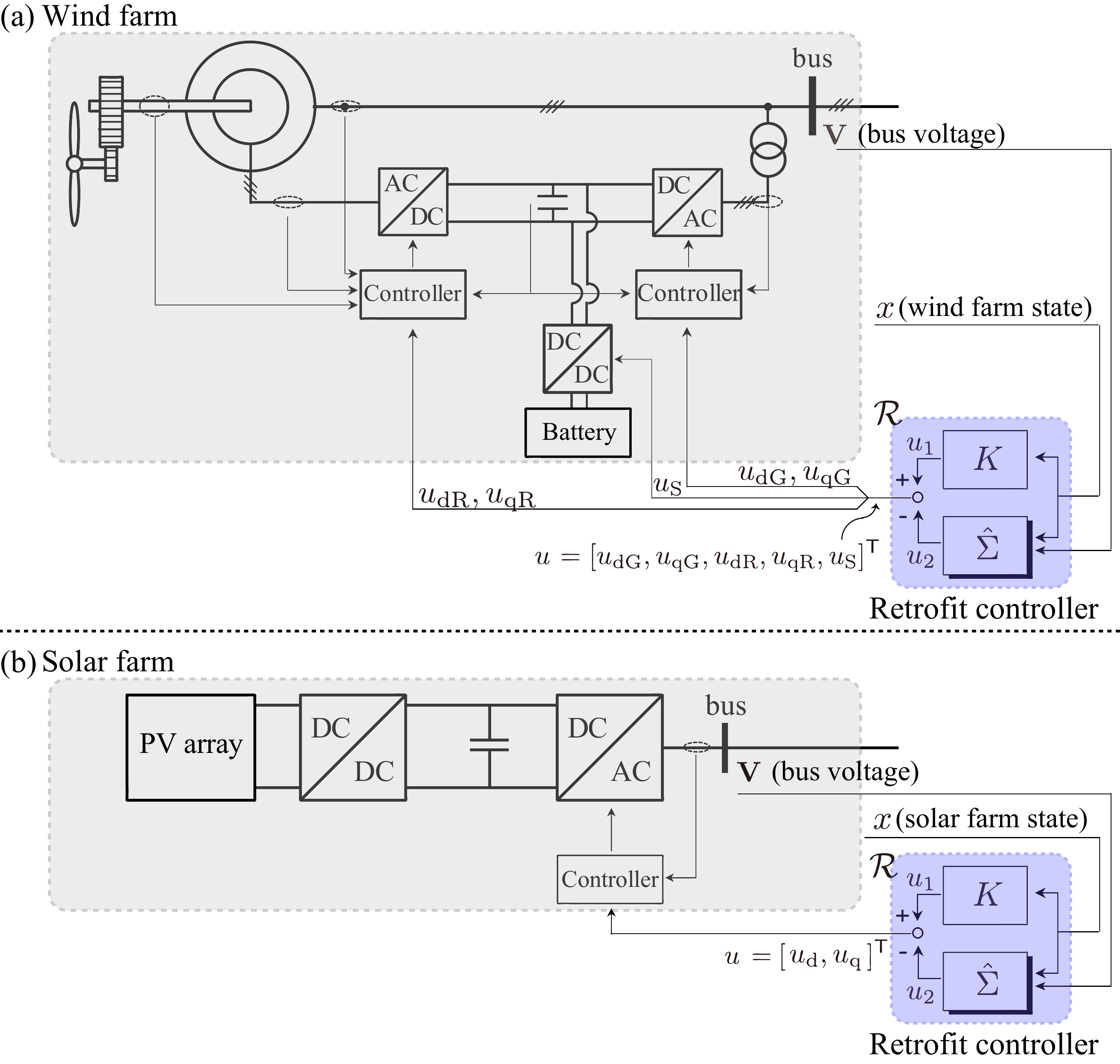}
\end{center}
\vspace{-0mm}
 \caption{Signal-flow diagram of DER with the retrofit controller $\mathcal R_k$ in \req{retrocon}. Figures (a) and (b) correspond to case where the DER is a wind farm and a solar farm, respectively. In each subfigure, the bus index $k$ is omitted. } 
\vspace{-0mm}
\label{fig:implementation}
\end{figure}

The input $u_k$ in \req{sigma_wind} is considered to be composed of two parts, namely
\begin{equation}\label{retroU}
 u_k = u_{1,k} + u_{2,k}.
\end{equation}
The component $u_{1,k}$ is designed under the assumption that $\tilde{f}_k(\cdot, \cdot, \cdot) = 0$ in \req{sigma_wind}. In other words, the design of $u_{1,k}$ ignores the nonlinearity in \req{sigma_wind} including the dependence on ${\bf V}_k$. 
Note that this assumption is only made to simplify the design of $u_k$. They do not influence the actual implementation of the control. 
Then, $u_{1,k}$ in \req{retroU} can be simply designed as
\begin{equation}\label{stateF}
 u_{1,k} = K_kx_k,
\end{equation}
where $K_k$ makes $A_k + B_kK_k$ Hurwitz. However, in reality this assumption will not be true. Thus, if $u_k=u_{1,k}$ is implemented,
then this control will pose serious threat to stability by neglecting the
dynamics following from $\tilde{f}_k(x_k, {\bf V}_k, u_k)$, and by neglecting the dynamics of the rest of the grid excluding the DER via ${\bf V}_k$, both of which will be stimulated by the control.  In order to prevent this stimulation, a compensation signal
$u_{2,k}$ is designed for \req{retroU} using a dynamic compensator
\begin{equation}\label{syscon}
\begin{array}{l}
\hat{\Sigma}_k:
\left\{
 \begin{array}{ccl}
     \dot{\hat{x}}_k &\hspace{-2mm}=&\hspace{-2mm} A_k \hat{x}_k + \tilde{f}_k(x_k, u_k, {\bf V}_k)\\
    u_{2,k} &\hspace{-2mm}=&\hspace{-2mm} - K_k\hat{x}_k, 
 \end{array}
\right. \quad \hat{x}_k(0) = x_k^{\star}. 
\end{array}
\end{equation}
The final controller is written as the combination of \req{retroU},
\req{stateF}, and \req{syscon} as 
\begin{equation}\label{retrocon}
 \mathcal R_k :
  \left\{
   \begin{array}{rcl}
   \dot{\hat{x}}_k &\hspace{-2mm}=&\hspace{-2mm} A_k \hat{x}_k + \tilde{f}_k(x_k, u_k, {\bf V}_k)\\
    u_k &\hspace{-2mm}=&\hspace{-2mm} K_kx_k - K_k\hat{x}_k,
   \end{array}
  \right. \quad \hat{x}_k(0) = x_k^{\star}. 
\end{equation}
Equation \req{retrocon} is called a {\it retrofit controller}. The
following proposition holds for this controller. 

 \vspace{2mm}
\begin{proposition}\label{lemfun}\vspace{0mm}
 Let Assumptions~\ref{assump1} and \ref{assump2} hold. 
 Then, the interconnection of the power system \req{model_comp}-\req{inter} and retrofit controllers $\mathcal R_k$ in \req{retrocon} for $k \in \mathbb N_{\rm S}\cup\mathbb N_{\rm W}$ is stable for any $K_k$ such that $A_k + B_kK_k$ is Hurwitz. 
\end{proposition}\vspace{2mm}

The proof of Proposition 1 is shown in \cite{ishizaki2016distributed,sadamoto2017retrofit}.
The signal-flow diagram of a DER equipped with the retrofit controller \req{retrocon} is shown in Figure~\ref{fig:implementation}.
Proposition 1 shows that the retrofit controllers satisfy the stability requirement in Property (i) listed earlier. Equation \req{retrocon} shows that the controllers satisfy Property (ii), that is,  $\mathcal R_k$ can be designed by using
information of only  $f_k(\cdot, \cdot, \cdot ;\cdot)$ and $x_k^{\star}$ of the corresponding DER, which makes it modular; $\mathcal R_k$ can be implemented by using feedback from
only $x_k$ and ${\bf V}_k$, which makes it decentralized. Neither the model nor the states of the ``rest of the grid'' are needed for designing or implementing $\mathcal R_k$. The controller gain $K_k$ can be anything as long as $A_k + B_kK_k$ is Hurwitz. For example, given the DER model \req{sigma_wind}, first design the state-feedback gain as $K_k = -R_k^{-1}B_k^{\sf T}X_k$ where $X_k$ is a positive-definite matrix satisfying 
\[
X_kA_k^{\sf T} +A_kX_k - X_kB_kR_k^{-1}B_kX_k + W_k = 0,
\]
with suitable weight matrices $R_k$ and $W_k$, and thereafter design $\mathcal R_k$ in \req{retrocon}. This LQR-based retrofit controller will be used in numerical simulations later.
Note that the initialization of this controller can also be done in a decentralized manner. 
The initial state $\hat{x}_k(0)$ is taken as the equilibrium of the DER to be controlled, that is, $\hat{x}_k(0) = x_k^{\star}$, where $x_k^{\star}$ can be computed in advance from the $k$-th component dynamics under a solution of the power flow calculation. In other words, the initialization of the controller can be done without taking into account any of the other components. 

The section is closed by stating two important properties of the retrofit controller. 

1. Not just stability, the plug-and-play action of retrofit control can also be used to improve the closed-loop dynamic performance of the grid by proper selection of $K_k$ in \req{retrocon}, which can help in attenuating transient oscillations in the power flows. This will be shown in numerical simulations later. For theoretical details on this point please see \cite{ishizaki2016distributed,sadamoto2017retrofit}.

2. Retrofit controllers are most sensitive to faults that occur either at or near the DER bus. This is because closer a fault is to the DER bus, more significant will be the change in the DER initial state from its equilibrium. If, on the other had, a fault occurs far away from the DER bus so as to cause practically no change in its initial state, then the retrofit controller will show no effect. This property implies that retrofit controllers can be added to or removed from the grid in a plug-and-play fashion without creating any sensitivity to other retrofit
controllers. This modularity property of retrofitting will be illustrated shortly by numerical simulations of the IEEE 68-bus system. For more theoretical details on this point, please see \cite{sadamoto2017retrofit}.



\subsection{Wide-Area Control}

Retrofit control is ideal for handling local disturbances. A separate layer of control is needed for handling disturbances that cause system-wide impacts on the entire grid. These controllers are called wide-area controllers. They are commonly actuated through additional control loops in the PSSs of synchronous generators. Ideally speaking, one can consider using retrofit control for designing wide-area controllers as well. However, the assets in the legacy grid excluding DERs do not fluctuate much over longer terms, and therefore do not necessarily need a plug-and-play type modular control in excess of what is already provided by the conventional PSS. Thus, in practice, retrofitting may be an overkill for WAC. A typical WAC problem is formulated as follows. 
Consider a power system consisting of generators, loads, non-unit buses, wind/solar farms. The entire system dynamics is described as \req{model_comp} and \req{inter}. 
The variables $P_{k}+jQ_{k}$ and ${\bf V}_{k}$ for $k \in \{1,\ldots, N\}$ are {\it auxiliary} variables which can be eliminated to convert the differential-algebraic model into an ordinary differential equation model by a process called Kron reduction \cite{kundur1994power}. By linearizing the Kron-reduced model about a desired equilibrium, the model of the power system is written in a compact form as
\begin{equation}\label{lqr2}
 \dot{\tilde{x}}_{\rm G} = A_{\rm G}\tilde{x}_{\rm G} + B_{\rm G}u_{\rm G} + R_{{\rm G}}\tilde{x}_{\rm D},\quad
 \dot{\tilde{x}}_{\rm D} = A_{\rm D}\tilde{x}_{\rm D} + B_{\rm D}u_{\rm D} + R_{{\rm D}}\tilde{x}_{\rm G},
\end{equation}
where $\tilde{x}_{\rm G} \in \mathbb R^{7|\mathbb N_{\rm G}|}$ and $\tilde{x}_{\rm D} \in \mathbb R^{7|\mathbb N_{\rm S}|+18|\mathbb N_{\rm W}|}$ are the stacked representations of the generator state error and the DER state error relative to the equilibrium $x_{k}^{\star}$ for $k \in \mathbb N_{\rm G}$ and $k \in \mathbb N_{\rm S}\cup\mathbb N_{\rm W}$, respectively, and
$u_{\rm G} \in \mathbb R^{|\mathbb N_{\rm G}|}$ and $u_{\rm D} \in \mathbb R^{|\mathbb N_{\rm S}| + |\mathbb N_{\rm W}|}$ are the stacked representations of $u_k$ for $k \in \mathbb N_{\rm G}$ and $k \in \mathbb N_{\rm S}\cup\mathbb N_{\rm W}$, respectively.
The $k$-th element of $u_{\rm G}$ represents a signal actuated through the PSS, as explained in \req{AVR_PSS1}.
Ideally speaking, both $u_{\rm G}$ and $u_{\rm D}$ should be used for WAC.
However, since in the current state-of-art the penetration of DERs in most power systems is still quite small most wide-area controller design are based on the generator models only, ignoring the dynamics of the DERs. In other words, consider a model of the form 
\begin{equation}\label{lqr4}
\dot{\xi}_{\rm G} = A_{\rm G}\xi_{\rm G} + B_{\rm G}u_{\rm G}, \quad \xi(0) = \tilde{x}_{\rm G}(0).
\end{equation}
A WAC problem for POD can then be defined as finding a gain matrix $K_{\rm G}$ such that 
\begin{equation}\label{pss0}
 u_{\rm G} = K_{\rm G}\xi_{\rm G}, \quad K_{\rm G} \in \mathcal S,
\end{equation}
minimizes
\begin{equation}\label{lqr1}
J := \int_0^\infty \big(\xi_{\rm G}^{\sf T}(t)Q\xi_{\rm G}(t)+u_{\rm G}^{\sf T}(t)Ru_{\rm G}(t)\big)dt,
\end{equation}
for a given positive-semidefinite matrix $Q$ and a positive-definite matrix $R$,  subject to \req{lqr4}.
In \req{pss0}, $\mathcal S \subseteq \mathbb R^{|\mathbb N_{\rm G}| \times 7|\mathbb N_{\rm G}|}$ represents admissible controllers encapsulating the distributed nature of the controller.
Alternative formulations have also been proposed, an overview of which is summarized in Appendix \ref{sidebar-surveyWAC}.
Once $K_{\rm G}$ is designed, the wide-area control is implemented as $u_{\rm G} = K_{\rm G}\tilde{x}_{\rm G}$ in \req{lqr2} by setting up a sparse communication network between the designated set of generators.
For simplicity, the communication is assumed to be ideal, that is, it does not have any delays or packet losses. Works on WAC under communication delays or packet losses can be found in papers such as \cite{wang2012wide,soudbakhsh2017delay}. 
The generator state vector $\tilde{x}_{\rm G}$ is assumed to be available (for state estimation of $\tilde{x}_{\rm G}$ using PMU measurements and decentralized Kalman filters, please see \cite{fardanesh1998multifunctional,huang2012state,ghahremani2011dynamic,farantatos2009pmu,do2009forecasting}).
Owing to the wide availability of PMU data, utilities these days have reasonably  good models for their operating regions. The independent system operators also have fairly accurate power system models. For a robust implementation of WAC these  models should be updated every few hours using fresh PMU data as the operating point changes, followed by an update in $K_{\rm G}$.


The goal of the constraint in \req{pss0} is to promote sparsity in $K_{\rm G}$ for minimizing the density of the underlying communication network without sacrificing closed-loop performance much.
The usual philosophy for constructing $\mathcal S$ is as follows.
For a natural number $L \leq |\mathbb N_{\rm G}|$, consider a set of groups $\{\mathcal G_{l}\}_{l \in \mathbb \{1,\ldots,L\}}$ such that $\bigcup_{l \in \{1,\ldots,L\}} \mathcal G_{l} = \{1,\ldots, |\mathbb N_{\rm G}|\}$. Note that the groups are not necessarily disjoint, namely, there may exist pair $(l, l')$ such that $\mathcal G_{l} \cap \mathcal G_{l'} \not= \emptyset$.
Let $K_{{\rm G},ij} \in \mathbb R^{1\times 7}$ denote the $(i,j)$-block matrix of $K_{\rm G}$. 
Define $\mathcal S$ as
\begin{equation}\label{defSS}
 \mathcal S := \left\{
K_{\rm G} : K_{{\rm G}, ij} = 0, ~ \mbox{there does not exists~} l\in\{1,\ldots,L\},~\mbox{~s.t.~} i \in \mathcal G_l  \land j \in \mathcal G_l
\right\}.
\end{equation}
The problem then is to find $K_{\rm G}$ in \req{pss0} minimizing \req{lqr1} under the constraint $\mathcal S$ as in \req{defSS}. 
The (sub)optimal set of groups $\{\mathcal G_{l}\}_{l \in \mathbb \{1,\ldots,L\}}$ and the structured feedback gain $K_{\rm G}$ can be constructed in different ways depending on the objective of the controller.
For POD problems operators are often interested in damping only the inter-area oscillation modes. In that case, $\mathcal G_l$ can be chosen in the following way, as recently shown in \cite{jain2017online}. Modeling the fault as an impulse input, let the small-signal impulse response of any generator frequency deviation be written as 
\begin{equation}
 \Delta \tilde{\omega}_{i}(t) = \underbrace{\sum_{j=1}^{\kappa}({\boldsymbol \alpha}_{ij}\exp({{\boldsymbol \lambda}_{j}t}) + {\boldsymbol \alpha}_{ij}^*\exp({{\boldsymbol \lambda}_{j}^*t}))}_{\mbox{inter-area modes}} +\underbrace{\sum_{i=\kappa+1}^n(
  {\boldsymbol \beta}_{ij}\exp({\boldsymbol \rho}_{j}t) +   {\boldsymbol \beta}_{ij}^*\exp({\boldsymbol \rho}_{j}^*t))}_{\mbox{other modes}}, \label{trd}
\end{equation}
where $\kappa$ is the number of local modes (explained later), $n$ is the dimension of the entire system, $\boldsymbol \alpha_{ij}, \boldsymbol \lambda_i, {\boldsymbol \beta}_{ij}, \boldsymbol \rho_i$ are modal coefficients, and $*$ is the complex conjugate operator. 
Assuming that the other modes are sufficiently damped by PSSs as a result of which their effect dies down quickly, the goal is to add damping to only the inter-area oscillation modes. The dominance of the inter-area modes is defined based on the magnitude of the modal coefficients ${\boldsymbol \alpha}_{ij}$. For example, consider a power system including four generators (namely $|\mathbb N_{\rm G}|=4$), with three inter-area modes (namely $\kappa = 3$). Let the residues ${{\boldsymbol \alpha}}_{11}$, ${{\boldsymbol \alpha}}_{21}$, ${{\boldsymbol \alpha}}_{31}$, ${{\boldsymbol \alpha}}_{32}$, ${{\boldsymbol \alpha}}_{42}$ be classified as dominant residues because they satisfy $|{{\boldsymbol \alpha}}_{{i}i}|{\geq}\mu$, where $\mu$ is a pre-specified threshold. In other words, it is assumed that the inter-area modes ${\boldsymbol \lambda}_{1}$, ${\boldsymbol \lambda}_{2}$ are substantially excited by the incoming disturbance while the third inter-area mode has much poorer participation in the states. From the indices of the dominant modes, one can construct the two sets
\begin{equation}\label{eq:K}
 \mathcal G_{1} = \{1,2,3\}, \quad \mathcal G_{2} = \{3,4\},
\end{equation}
indicating that the generators in the first group participate dominantly in ${\boldsymbol \lambda}_{1}$, and those in the second group participate dominantly in ${\boldsymbol \lambda}_{2}$. This grouping information is then used to decide the topology of communication. In general, the rule is that the generators inside the $l$-th group should communicate with each other for suppressing the amplitude of oscillations excited by the mode ${\boldsymbol \lambda}_{l}$. 
The mode ${\boldsymbol \lambda}_{3}$  for the above example is poorly excited, and therefore, is ignored in the control design.
The structure of $K_{\rm G}$ for this 4-machine example is then constructed as \req{pss0} with the structure constraint \req{defSS} defined by the group set \req{eq:K}. For this example, the controllers can be written as
\begin{eqnarray}
 \mathcal K_1 :~u_1 &\hspace{-2mm} = &\hspace{-2mm} K_{{\rm G},11}\tilde{x}_1 + K_{{\rm G},12}\tilde{x}_2 + K_{{\rm G},13}\tilde{x}_3, \nonumber \\ 
 \mathcal K_2 :~ u_2 &\hspace{-2mm} = &\hspace{-2mm} K_{{\rm G},21}\tilde{x}_1 + K_{{\rm G},22}\tilde{x}_2 + K_{{\rm G},23}\tilde{x}_3, \nonumber \\ 
 \mathcal K_3 :~ u_3 &\hspace{-2mm} = &\hspace{-2mm} K_{{\rm G},31}\tilde{x}_1 + K_{{\rm G},32}\tilde{x}_2 + K_{{\rm G},33}\tilde{x}_3 + K_{{\rm G},34}\tilde{x}_4, \nonumber \\ 
 \mathcal K_4 :~ u_4 &\hspace{-2mm} = &\hspace{-2mm} K_{{\rm G},43}\tilde{x}_3 + K_{{\rm G},44}\tilde{x}_4. \nonumber 
\end{eqnarray}
Due to the sparse structure of the communication network, the controllers are implemented in a distributed way as opposed to all-to-all communication that would be equivalent to centralized implementation. Finally, the non-zero entries of $K_{\rm G}$ are computed by using sub-optimal  structured LQR algorithms such as $L_1$-sparse optimal control via ADMM \cite{lin2013design}.

\begin{figure}[t]
  \begin{center}
    \includegraphics[clip,width=160mm]{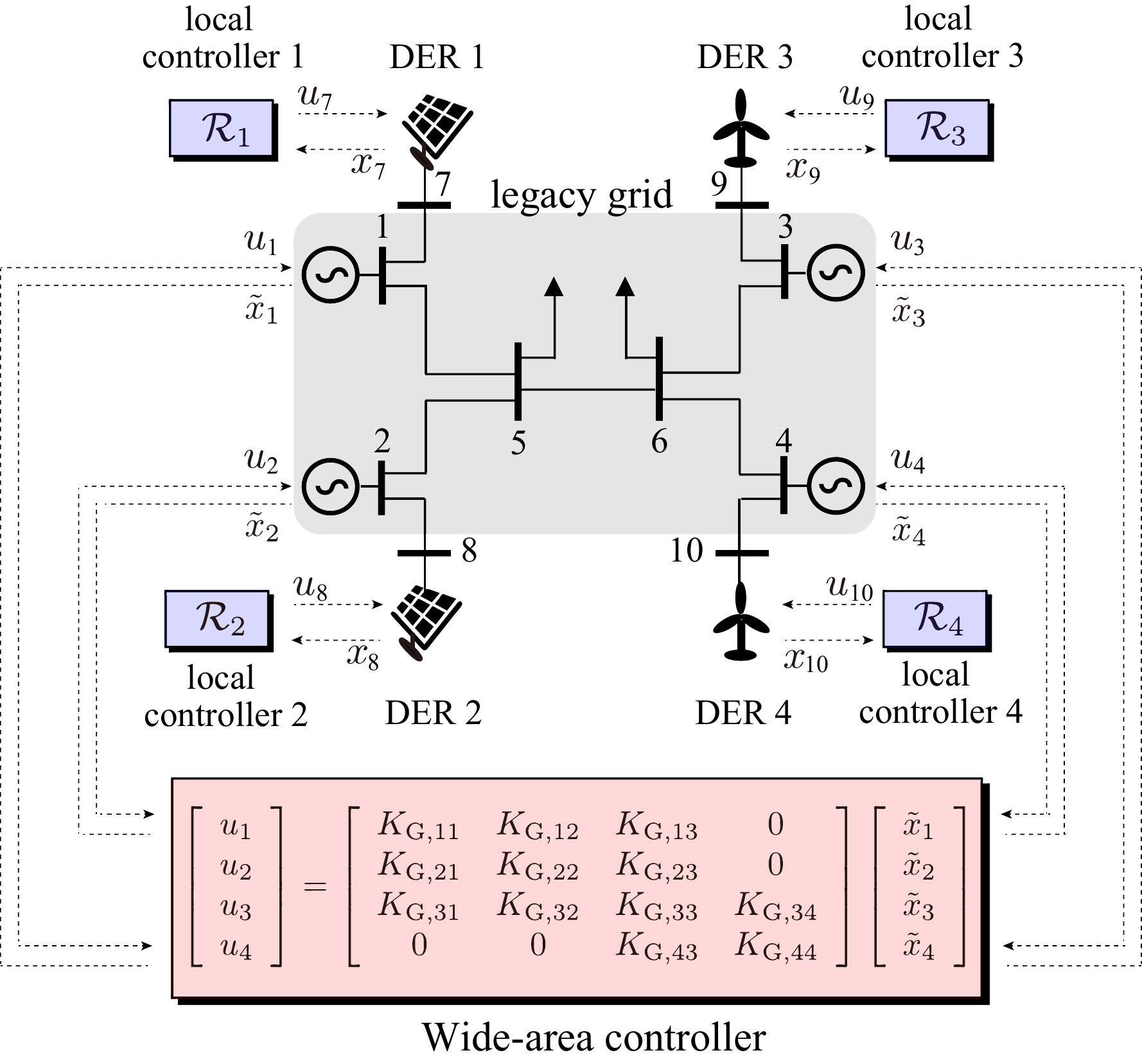}
   \caption{Grid control architecture showing the co-existence of local and wide-area controllers for a 4-machine power system with four DERs. The local controller $\mathcal R_k$ is designed in \req{retrocon}. The wide-area controller $K_{\rm G}$ in \req{pss0} is designed for the two groups $\mathcal G_1 = \{1,2,3\}$ and $\mathcal G_{2} = \{3,4\}$. The sparse structure of WAC avoids the need for all-to-all communication between the generators.}
    \label{fig_archtecture}
  \end{center}
\end{figure}

  \subsection{Combined Control Architecture for Tomorrow's Grids}
  The relative advantage between retrofit control and WAC lies in their effectiveness towards different types of faults and the fault locations. Retrofit control of DERs, for example, is more effective than WAC when a fault changing the DER initial state from its equilibrium occurs either at or closer to the DER bus. This is because the former has a high controllability on the DER states, while the latter, being actuated through the AVR and PSS of synchronous generators, has much lower controllability on the DER states. Similarly, WAC is much more effective than retrofit control when a fault happens at non-DER buses. Thus, to accommodate all types of fault scenarios the control architecture for future grids must be a combination of two layers  -  first, a completely decentralized retrofit control mechanism for each individual DER, and second, a distributed, peer-to-peer communication based wide-area control between the synchronous generators. The modularity of retrofit controllers will enable smoother integration of renewables into the grid in a plug-and-play fashion without jeopardizing stability, or without having to retune the existing PSS gains. WAC, on the other hand, will be necessary to balance out the increasing dynamic interdependence of grid components that may be geographically distant but are electrically close due to more transmission lines being built. The combination of these two control layers is shown in the form of an architectural diagram in Figure \ref{fig_archtecture}.

\section{Numerical Simulations}

\begin{figure}[t]
  \begin{center}
    \includegraphics[clip,width=165mm]{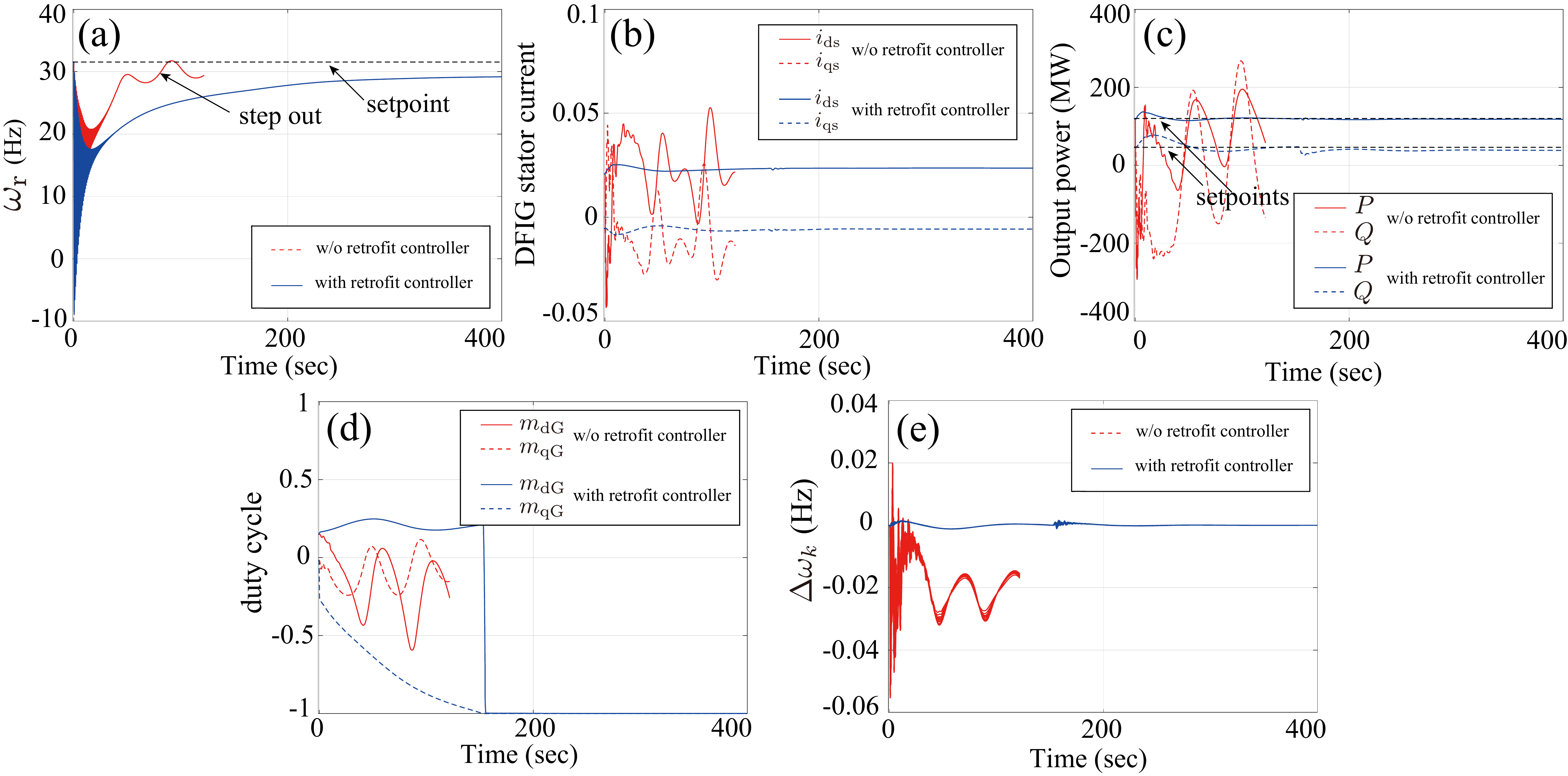}
    \caption{(a) Trajectories of  the DFIG rotor speed, (b) the DFIG stator currents, (c) the active and reactive power injected from the wind farm, (d) the GSC duty cycles, and (e) the frequency deviation of  all synchronous generators. }
    \label{sim01}
  \end{center}
\end{figure}

\begin{figure}[t]
  \begin{center}
    \includegraphics[clip,width=165mm]{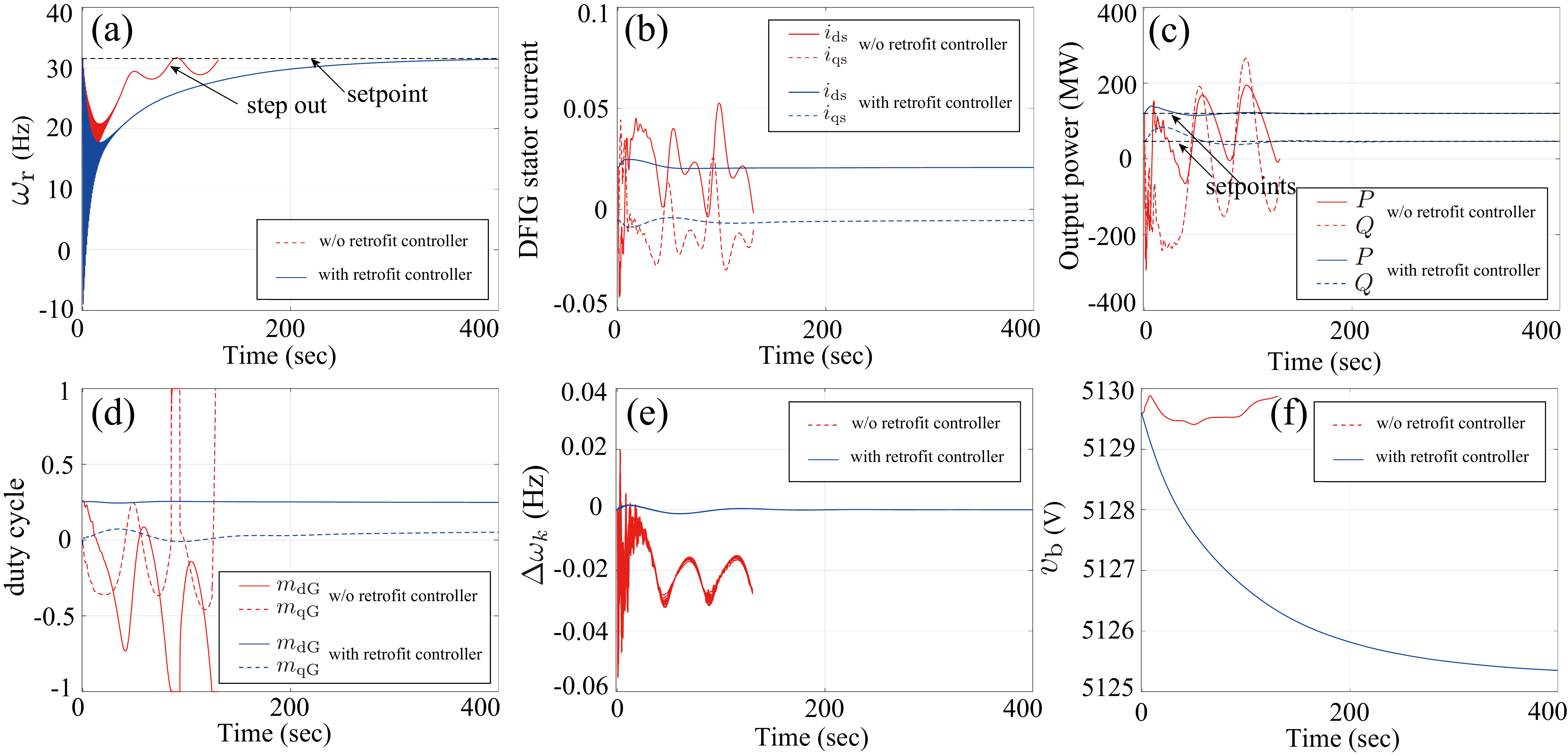}
    \caption{(a) Trajectories of  the DFIG rotor speed, (b) the DFIG stator currents, (c) the active and reactive power injected from the wind farm, (d) the GSC duty cycles, (e) the frequency deviation of all synchronous generators, and (f) battery voltage. }
    \label{sim02}
  \end{center}
\end{figure}

\begin{figure}[t]
  \begin{center}
   \includegraphics[clip,width=160mm]{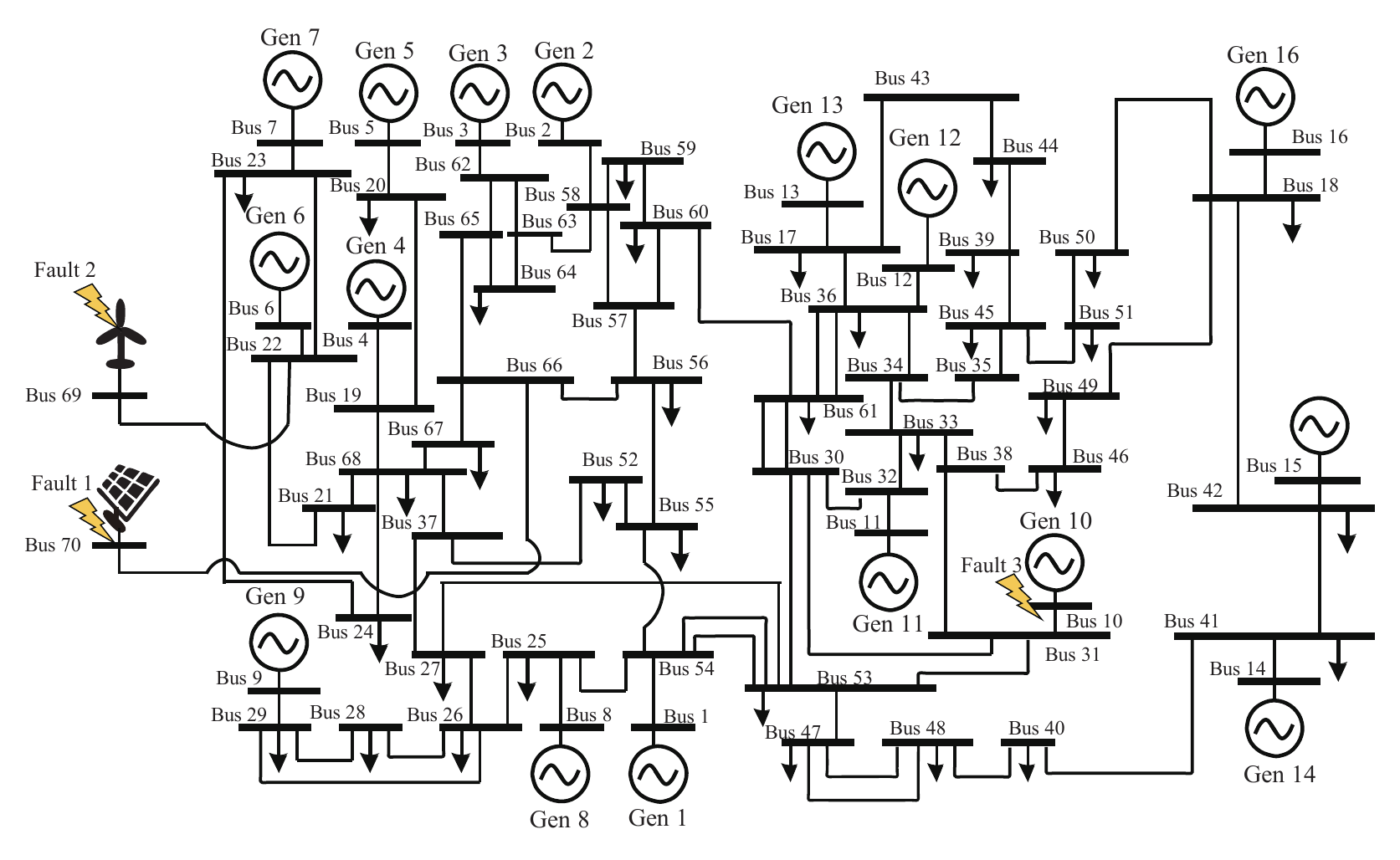}
    \caption{IEEE 68-bus, 16-machine power system model with two DERs. A single wind farm (with Bus 69) is added to Bus 22 while a single solar farm (with Bus 70) is added to Bus 66. }
    \label{tomorow_bus}
  \end{center}
\end{figure}

\begin{figure}[t]
  \begin{center}
    \includegraphics[clip,width=120mm]{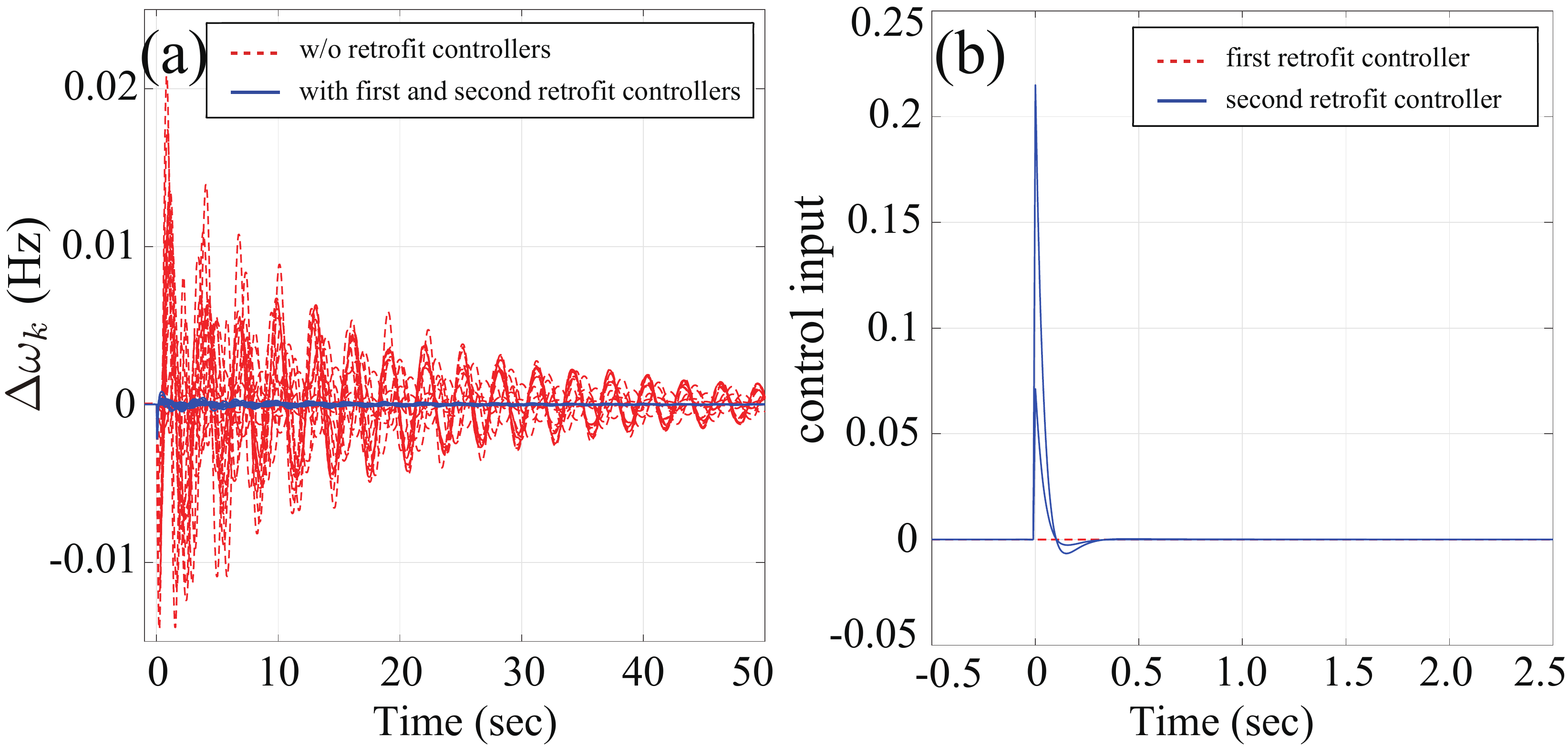}
    \caption{(a) Trajectories of the frequency deviation of all synchronous generators. The first and second retrofit controllers are implemented to the wind and solar farms at Bus 69 and 70, respectively. (b) Trajectory of the control input $u_k$ generated by the first and second retrofit controllers when Fault 1 happens.}
    \label{sim04}
  \end{center}
\end{figure}

\begin{figure}[t]
  \begin{center}
    \includegraphics[clip,width=165mm]{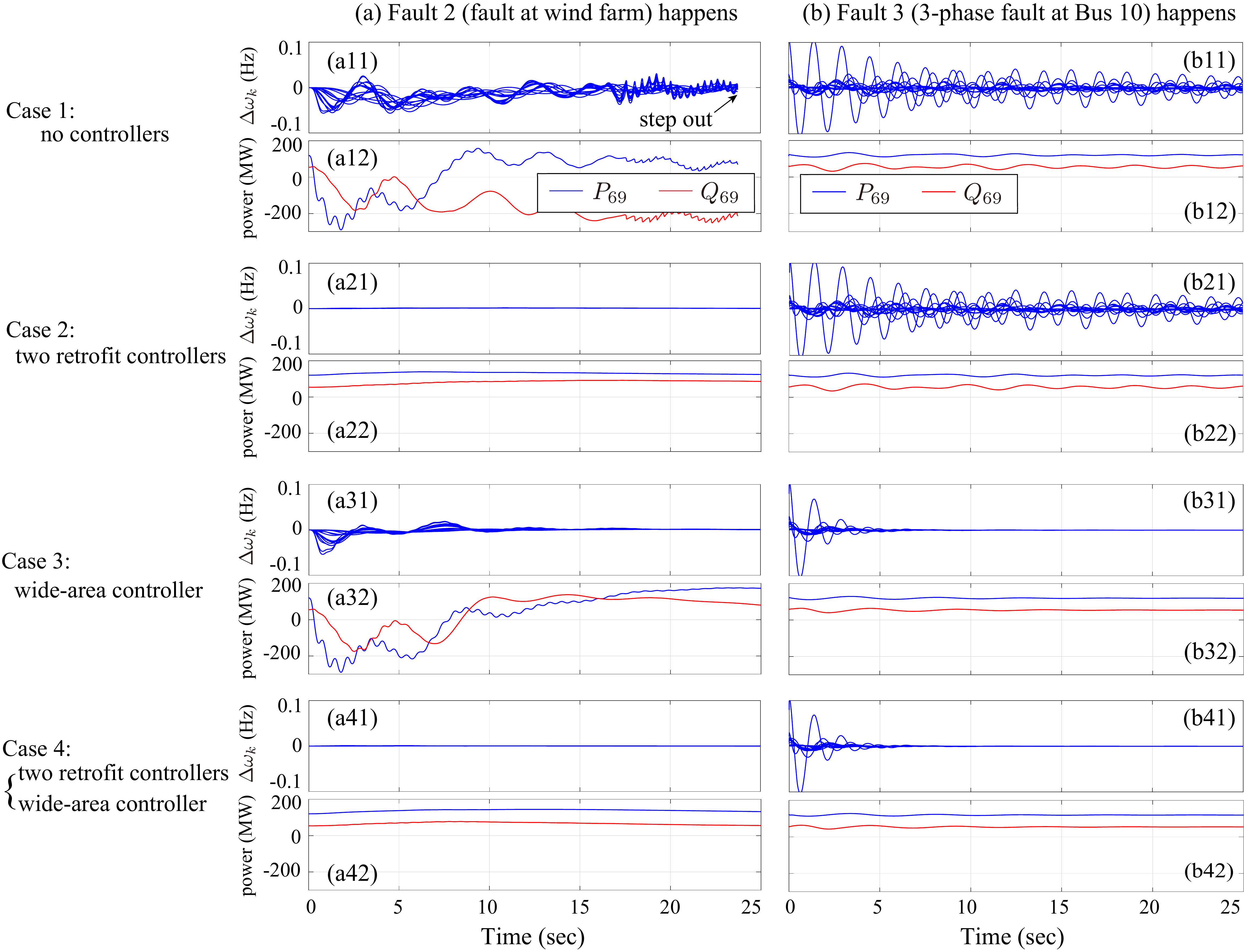}
    \caption{Trajectories of  the frequency deviation of all synchronous generators and power injecting from the wind farm when (top two panels) no additional controllers, (second two panels) two retrofit controllers, (third two panels) wide-area controller, and (bottom two panels) both of the two retrofit controllers and the wide-area controllers are used, respectively, for (left column) the fault at the wind farm and (right column) three-phase fault at Bus 10.}
    \label{sim05}
  \end{center}
\end{figure}

   

This section demonstrates the effectiveness of the combined retrofit and wide-area control through simulation of the IEEE 68-bus power system model with DERs. The Matlab codes for all of these simulations have been made public at the repository \cite{tomonori2018matlab}.
First, let a single wind farm be connected to Bus 69, as shown in Figure \ref{fig:example}, and the effectiveness of the retrofit control is investigated. Let $\gamma_{\rm W} = 60$.
The behavior of this wind-integrated power system after a fault, modeled as an impulsive change in the angular velocity of the low-speed shaft of the wind turbine, is simulated. Let $\omega_{\rm l}(0) = 0.3\omega_{\rm l}^{\star}$. In Figures \ref{sim01}(a)-(e), the red lines show the trajectories of the DFIG rotor speed $\omega_{\rm r}$ in \req{turbine_model}, the DFIG stator currents $i_{\rm ds}$ and $i_{\rm qs}$ in \req{wind_current}, the active and reactive power $P$ and $Q$ in \req{wind_power} injected by the wind farm, the GSC duty cycles $m_{\rm dG}$ and $m_{\rm qG}$ in \req{con_duty}, and the frequency deviation $\Delta \omega$ in \req {model_sync01} of all synchronous generators.
The fault causes a decrement in the rotor speed $\omega_{\rm r}$, and  induces oscillations in the DFIG stator currents because of the resonance phenomenon described earlier. Both active power $P$ and reactive power $Q$ injected to the grid from the wind farm start oscillating, which in turn induces oscillations in the generator frequencies. Thus, although the grid is small-signal stable the impact caused by the fault makes the system leave its domain of attraction, resulting in transient instability. 

The retrofit controller \req{retrocon} is next used to counteract this instability. Since the instability is caused by the oscillations of the DFIG currents  the state feedback gain $K_k$ is designed to attenuate these oscillations using a LQR controller.  In Figures \ref{sim01}(a)-(e), the blue lines show the trajectories of $\omega_{\rm r}$, $i_{\rm ds}$ and $i_{\rm qs}$, $P$ and $Q$, $m_{\rm dG}$ and $m_{\rm qG}$, and $\Delta \omega$ of the closed-loop system, respectively. By comparing the red lines and blue lines in Figures \ref{sim01}(b)-(c), it can be seen that the oscillations in the stator current and the output power are significantly mitigated by retrofit control. As a result, as shown in Figure \ref{sim01}(e), the oscillations in the frequency deviations are also reduced. However, the oscillations reappear after $t>150$ sec as the duty cycles $m_{\rm dG}$ and $m_{\rm qG}$ of the GSC in \req{con_duty} become saturated after this time. This is shown in  Figure \ref{sim01}(d). Since the duty cycles are inversely proportional to $v_{\rm dc}$, this saturation tends to occur when the DC link voltage is small. Furthermore, as shown in Figures \ref{sim01}(a) and (c), there exist negative offsets in $\omega_{\rm r}$ and $P$. This implies that some of the mechanical power and output wind power are less than their desired values due to the decrement of the low-speed shaft speed.

Both of these shortcomings can be resolved by adding a battery to the DC link. The battery can compensate for the DC-link voltage variation, and can discharge energy to make up for the insufficient power. Figure \ref{sim02}(f) shows that the battery voltage reduces as the battery discharges stored energy. Comparing the blue lines in Figures \ref{sim01}(a) and (c) with those in Figures \ref{sim02}(a) and (c), it can be seen that the rotor speed and the injected wind power now both converge to their respective setpoints. Furthermore, the duty cycles are no longer saturated as the DC-link voltage remains almost constant. The battery voltage converges to its setpoint value asymptotically. 

Over time new DERs will be added to the existing grid. To emulate this, next, a solar farm with its bus (denoted Bus 70) is added at Bus 66, as shown in Figure~\ref{tomorow_bus}. The system behavior is investigated after installing the second retrofit controller at this solar farm while retaining the first one at the wind farm. The design of the second retrofit controller follows the same procedure as the first. Figure \ref{sim04} shows the trajectories of all generator frequency deviations when a fault  happens at the solar farm (Fault 1). Figure \ref{sim04}(b) shows the control input calculated by the first and second retrofit controller. It can be seen that the first retrofit controller is inactive in this situation, meaning that it does not have any influence on the closed-loop response. 
The second retrofit controller, on the other hand, improves damping as soon as it is activated, as depicted by the blue solid lines in Figure \ref{sim04}(a). This shows that retrofit controllers can be added to or removed from the grid in a plug-and-play fashion without creating any sensitivity to other retrofit controllers. The design enjoys a natural decoupling property from one DER to another. 

When the fault happens at the DER bus then the retrofit controller alone is sufficient for mitigating oscillations. Any additional wide-area control action may not be necessary in that situation. To show this fact, a LQR-based wide-area controller is designed for all the 16 synchronous generators, and actuated through their PSSs following the fault at the wind farm (Fault 2).
Figures~\ref{sim05}a summarizes the different ways in which the power system reacts to this fault. Since the fault occurred at the DER bus, the retrofit controller at the DER in Case 2 successfully cancels its adverse effect, and mitigates the oscillations in both the generators and the DER, as shown in (a21) and (a22). If only WAC is used in this situation without any retrofit at the DER, which is Case 3 in the figure, then an interesting observation is made: the generators are damped well, but the DER response is still unacceptably oscillatory. This clearly shows that WAC has limited controllability on the DER states, and thus for this situation using only WAC is not going to suppress all oscillations. Retrofit control is absolutely imperative for this case.  The respective responses are shown in (a31) and (a32).
However, when the fault happens outside the wind farm, then the retrofit controller is completely inactive and the wide-area controller becomes necessary. To show this, a three-phase fault is induced at Bus 10 (Fault 3) with the fault clearing time  $0.07$ (sec). 
By comparing the subfigures (b11)-(b33), it can be seen that the retrofit controllers at the wind and solar farms are no longer effective whereas the wide-area controller successfully damps the power flow oscillations. 

Finally, Figures~\ref{sim05}(a41)-(a42) and (b41)-(b42) show the case when the two retrofit controllers and the wide-area controller are all used together in the system. The combination can now handle faults occurring at both DER buses and non-DER buses.
The simulations bear a clear message that future power grids must have both control architectures implemented on top of each other to enjoy their combined benefits.


\section{Conclusion and Future Works}
With proliferation of distributed renewable generation many interesting opportunities for control and
optimization are arising in power system research. On one hand, wide-area control is becoming essential to make the grid more resilient against blackouts, while on the other hand local plug-and-play type controllers are becoming essential for renewable energy sources. This article proposed a control architecture that combines these two types of controllers, highlighting various design and implementation challenges and solutions for both. The vision here is that this architecture will serve as a platform for control theorists and power engineers to work together, and create a sustainable and secure future for electric energy supply in every corner of the world.
A list of open questions is presented for future work.

1.  Not only the level of penetration of DERs, but also their  relative locations in the grid may have a significant influence
on power system stability. For example, with reference to the simulation example presented in the article, when a wind farm with output power 200MW is connected to Bus 22 of the IEEE 68-bus model, the power system model becomes unstable. But, instead, if a wind farm with output power 58MW is connected to Bus 42, the system becomes unstable. More work is needed to understand the system-level characteristics of power system models that decide how the spatial distribution of DERs may or may not preserve stability. 

2. The controllers outlined in this article are meant to address transient stability and damping performance. In an actual grid, however, many other parallel control mechanisms will also exist - for example, load frequency control (LFC) which maintains the grid frequency at the synchronous value despite fluctuations in loads. Typically, LFC is designed independent of WAC and DER control because of its slower time-scale. However, with gradual disintegration of the grid into smaller micro-grids this time-scale separation may become less dominant leading to a stronger coupling between the controllers. The positive and negative effects of this coupling needs more research.


3. Many open challenges exist for wide-area control as well. For example, instead of updating $A_{\rm G}$ and $B_{\rm G}$ in \req{lqr2} with new PMU data every few hours and redesigning $K_{\rm G}$ in \req{pss0}, a future option  can be to learn $K_{\rm G}$ directly after a contingency using online reinforcement learning, Q-learning, adaptive dynamic programming, and similar model-free learning methods. This online learning approach, in fact, will become important if the contingency changes the nominal $A_{\rm G}$ and $B_{\rm G}$ significantly. 
Furthermore, both $u_{\rm G}$ and $u_{\rm D}$ in \req{lqr2} must be used for WAC design in the future grid to accommodate the coupling effects between $\tilde{x}_{\rm G}$ and $\tilde{x}_{\rm D}$.
Some recent paper such as \cite{yousefian2016direct} have reported preliminary results along these lines where wind power controllers have been used in conjunction to conventional PSS controllers for taking wide-area damping control action. Much more work, however, is needed on this topic, especially with regards to the time-scale separation between $u_{\rm G}$ and $u_{\rm D}$, and also on the sensitivity between the retrofit and WAC parts of $u_{\rm D}$.


All of these questions deserve dedicated attention from researchers with backgrounds in control theory, power systems, signal processing, economics, and machine learning.

\section{Acknowledgements}
This research was supported by CREST, JST Grant Number JPMJCR15K1, Japan. The work of the second author was partly supported by the US National Science Foundation under grant ECCS 1711004. The authors would like to thank the fourth author's secretary Akiho Setoguchi for her help with drawing several illustrations used throughout the paper.

\bibliographystyle{IEEEtran}
\bibliography{reference}

\appendix
\clearpage

\subsection{A Brief Survey on Control of Synchronous Generators}
\label{sidebar-surveyCofSG}
Control of synchronous generators is generally classified into two groups: prime mover control and excitation system control \cite{kundur1994power}. The former aims at balancing the total system generation against system load and losses so that the desired frequency and power interchange are maintained. These controllers are further subdivided into {\it primary, secondary, and tertiary control}; for details see \cite{machowski2008power}. However, the prime-mover control is usually much slower than the time-scale of oscillation damping, and, therefore, is not discussed in this article. 
The excitation system control, on the other hand, is achieved by Automatic Voltage Regulator (AVR) for the regulation of voltage and reactive power, as well as by Power System Stabilizer (PSS) for the enhancement of power system stability and damping of power flow oscillations \cite{kundur1994power}. A long line of work exists for AVR/PSS design using tools from control theory - for example, see \cite{pal2006robust,chow2004power,sauer2017power,demello1969concepts,dudgeon2007effective,IEEEstandard}.
In today's grid, many of these controllers are being executed using real-time measurements from Phasor Measurement Units (PMUs) \cite{fardanesh1998multifunctional}, integrated with efficient state estimation techniques such as {\it forecasting-aided state estimation} \cite{huang2012state,ghahremani2011dynamic,farantatos2009pmu,do2009forecasting}.

\subsection{A Brief Survey on Local Control of Power Systems}
\label{sidebar-surveyLC}
 Local control of power systems relies on two main techniques - decentralized robust control \cite{chen1992decentralized,chen1993decentralized} and dissipativity-based control \cite{moylan1978stability,qu2014modularized}.
Applications of these two methods to power systems can be found in papers such as  \cite{zhao2014optimal,rinaldi2017third,bevrani2009robust,wang1998robust,siljak2002robust,cai1996robust,guo2000nonlinear,zecevic2004robust,etemadi2012decentralized} for the former, and \cite{fiaz2013port,wang2003dissipative} for the latter. However, these methods have the following drawbacks. 
One drawback is that the class of applicable systems is restrictive. In the decentralized robust control, interferences among generators are regarded as  norm-bounded uncertainty, and a stabilizing controller for any of the uncertainties are designed. Thus, basically, this approach is applicable to weakly-coupled systems. Power system models, in general, however do not fall in this category. The dissipativity-based control is limited to dissipative network systems only. Although conventional synchronous generator models satisfy this property, the approach cannot be easily extended to DER models. In fact, the papers \cite{fiaz2013port,wang2003dissipative} deal with multi-machine power systems without DERs. Another drawback is that these methods need a priori knowledge of the entire system model. 
For example, in the decentralized robust control, a small-gain type property bounded by the uncertainties, which cannot be computed without using an entire network system model, is required. On the other hand, in dissipativity-based control, the interconnection among subsystems should be appropriate such that the networked system is stable \cite{lozano2013dissipative}. However, the design of this interconnection structure is difficult to perform by local subsystem information. 

 In contrast, the retrofit control proposed in \cite{ishizaki2016distributed} is applicable to a much more general class of stable systems that do not require any of the above specifications. The design and implementation can be done independently from the information of any other grid components, that is, the retrofit controller has a plug-and-play capability. Moreover, in contrast to heuristic tuning of local controllers as presented in literature \cite{wu2008decentralized,liang2013stability}, this method can theoretically guarantee the overall closed-loop system stability; details are described in the section Section \ref{subsec-localC}.

\subsection{A Brief Survey on Wide-Area Control}
\label{sidebar-surveyWAC}
 Recent papers have proposed the design of WAC for power oscillation damping based on optimal control \cite{zolotas2007study,chaudhuri2012damping}, LMIs and conic programming \cite{jabr2010sequential}, model predictive control \cite{jain2015model},  model reduction and control inversion \cite{zima2005design}, and adaptive control \cite{chow2012adaptive}. A tutorial on these different methods has recently been reported in \cite{chakrabortty2013introduction}. The paper \cite{chaudhuri2010wide} has proposed wide-area control that can be resilient to failures. One drawback of these methods is that they usually result in a centralized implementation that can be computationally challenging. Centralized control is also not very resilient to cyber-attacks \cite{chaudhuri2010wide,deng2017false}. Thus, in recent years, power system operators are inclining more towards distributed wide-area control, where the communication graph among controllers is sparse \cite{heniche2002control}, \cite{dorfler2014sparsity}. The work in \cite{heniche2002control} uses geometric measures for selection of control loops, whereas \cite{dorfler2014sparsity} uses a sparsity-promoting LQR-based optimal control strategy. LQR is often chosen as the central design tool as it provides flexibility in damping selected ranges of frequencies. A real-time version of the sparse LQR design has been proposed in \cite{jain2017online} using spectral decomposition of online PMU measurements. An advantage of this method over \cite{dorfler2014sparsity}, \cite{heniche2002control} is that this approach can provide sparser wide-area controllers than those methods, with a comparable closed-loop performance; please see \cite{jain2017online} for more detailed discussion of this point based on numerical simulations.

\section{Construction of Admittance Matrices}
\label{sidebar-admittance}

\begin{figure}[t]
  \begin{center}
    \includegraphics[clip,width=160mm]{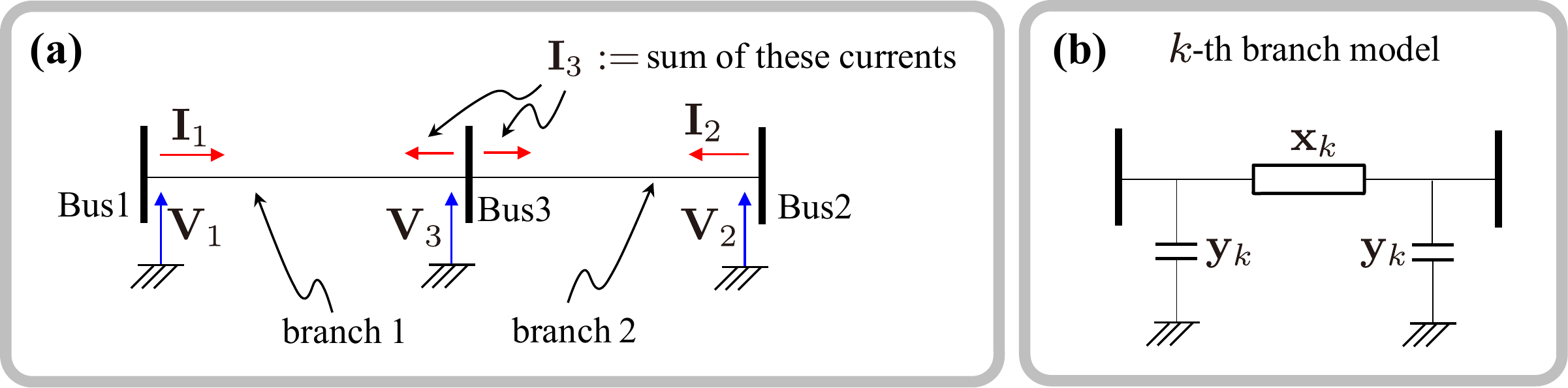}
    \caption{(a) Simple power system consisting of three buses. (b) The $k$-th branch model so-called $\Pi$-type circuit.}
    \label{fig_small}
  \end{center}
\end{figure}

A brief tutorial of constructing the admittance matrix ${\bf Y}$ from tie-line data typically shown in literature, for example \cite{pal2006robust} is described.
For this purpose, a power network consisting of three buses, as shown in Figure \ref{fig_small} (a), is considered, where ${\bf V}_{k}$ is the $k$-th bus voltage and ${\bf I}_{k}$ is the current injected from the bus.
Typically, each branch is modeled as a so-called $\Pi$-type circuit as shown in Figure \ref{fig_small} (b), where ${\bf x}_k$ is the branch impedance and ${\bf y}_k$ is the shunt admittance representing a capacitance between the transmission line and ground. The values of ${\bf x}_k$ and ${\bf y}_k$ for all branches are given data.

The bus voltage and current must satisfy Kirchhoff's law, that is,
\begin{equation}\label{kirch_sidebar}
 {\bf I}_{1:3} = {\bf Y}{\bf V}_{1:3},
  \quad {\bf I}_{1:3} := [{\bf I}_1, {\bf I}_2, {\bf I}_3]^{\sf T},
  \quad {\bf V}_{1:3} := [{\bf V}_1, {\bf V}_2, {\bf V}_3]^{\sf T},
\end{equation}
where ${\bf Y} \in \mathbb C^{3\times 3}$ is the admittance matrix. 
Note that \req{kirch_sidebar} is equivalent to \req{inter} when $N=3$ by taking conjugate of \req{kirch_sidebar} and multiplying each $k$-th row of \req{kirch_sidebar} by ${\bf V}_k$. Thus, ${\bf Y}$ in  \req{kirch_sidebar} is the matrix that needs to be constructed. The calculation of its $(1,1)$-element, denoted by ${\bf Y}_{11}$, is as follows. Note that \req{kirch_sidebar} should hold in a special case when ${\bf V}_{2} = {\bf V}_3 = 0$. Then, the first line in \req{kirch_sidebar} becomes ${\bf I}_1 = {\bf Y}_{11}{\bf V}_1$. On the other hand, when   ${\bf V}_{2} = {\bf V}_3 = 0$, it follows from Figure \ref{fig_small} (a) that  ${\bf I}_1 = ({\bf y}_1 + {\bf x}_1^{-1}){\bf V}_1$. Thus, ${\bf Y}_{11} = {\bf y}_1 + {\bf x}_1^{-1}$. Repeating this for all elements, the admittance matrix is constructed as
\begin{equation}
 {\bf Y} =
  \left[
  \begin{array}{ccc}
   {\bf y}_1 + {\bf x}_1^{-1}& 0& -{\bf x}_1^{-1}\\
   0& {\bf y}_2 + {\bf x}_2^{-1}& -{\bf x}_2^{-1}\\
   -{\bf x}_1^{-1}& -{\bf x}_2^{-1} & {\bf y}_1 + {\bf y}_2+  {\bf x}_1^{-1} + {\bf x}_2^{-1}\\
  \end{array} 
  \right]. 
\end{equation}

\subsection{Brief Tutorial on Power Flow Calculation}\label{sidebar_pflow}
Given the number of buses $N$ and an admittance matrix ${\bf Y}$, the power flow calculation, namely a procedure of finding bus voltage ${\bf V}_k^{\star}$, active power $P_k^{\star}$, and reactive power $Q_k^{\star}$ for $k \in \{1,\ldots, N\}$ satisfying
\begin{equation}\label{inter_star}
 0 = ({\bf YV_{1:N}^{\star}})^* \circ {\bf V}_{1:N}^{\star} - (P_{1:N}^{\star} + jQ_{1:N}^{\star}), \quad
  \left\{
  \begin{array}{l}
   {\bf V}_{1:N}^{\star} := \left[{\bf V}^{\star}_1 ~ \cdots ~ {\bf V}^{\star}_N\right]^{\sf T}, \\
   P_{1:N}^{\star}+jQ_{1:N}^{\star} := \left[P_{1}^{\star}+jQ_{1}^{\star} ~ \cdots ~ P_{N}^{\star}+jQ_{N}^{\star}\right]^{\sf T},
  \end{array}
  \right.
\end{equation}
where $\circ$ is the element-wise multiplication and $*$ is the element-wise complex conjugate operator, is explained. Note that there exist $4N$ decision variables, namely $P_k^{\star}, Q_k^{\star}$ and both the magnitude and angle of ${\bf V}_k^{\star}$ for $k \in \{1,\ldots, N\}$, while \req{inter_star} consists of $2N$ algebraic constraints, namely its real part and imaginary part. Hence,
there exist an infinite number of solutions satisfying \req{inter_star}. In order to choose a practical solution, $2N$ additional
constraints reflecting component characteristics are typically
considered in addition to \req{inter_star}. In this setting, a standard power flow calculation is formulated as follows. 

Given a natural number $N$ and ${\bf Y} \in \mathbb C^{N \times N}$ and $h_k(\cdot, \cdot, \cdot): \mathbb C \times \mathbb R \times \mathbb R \rightarrow
\mathbb C$ for $k\in\{1,\ldots,N\}$, find ${\bf V}^{\star}_k$, $P^{\star}_k$ and $Q^{\star}_k$ satisfying \req{inter_star} and 
\begin{equation}\label{add_con}
 0 = h_k({\bf V}_k^{\star}, P_k^{\star}, Q_k^{\star}), \quad {\rm
  for}~ k \in \{1,\ldots, N\}. 
\end{equation}

The detail of the additional constraint $h_k(\cdot, \cdot, \cdot)$ for generators, loads, non-unit buses, wind farms, solar farms, and energy storage systems are described below. Let
  $\mathbb N_{\rm G}$, 
  $\mathbb N_{\rm L}$,
  $\mathbb N_{\rm W}$
  $\mathbb N_{\rm S}$,
  $\mathbb N_{\rm E}$,
  $\mathbb N_{\rm N}$ be the index sets of the buses connecting to generators, loads, wind
 farms, solar farms, energy storages, and that of non-unit buses. Those sets are supposed to be disjoint, and 
$\mathbb N_{\rm G} \cup \mathbb N_{\rm L} \cup  \mathbb N_{\rm W} \cup  \mathbb
N_{\rm S} \cup \mathbb N_{\rm E} \cup  \mathbb N_{\rm N} = \{1,\ldots,N\}$. 

The non-unit buses must satisfy 
\begin{equation}\label{con1}
  0 = P_k^{\star}+jQ_k^{\star},\quad k\in \mathbb N_{\rm N}. 
\end{equation}
For the loads and energy storage systems, their steady-state active/reactive power $\bar{P}_k + j\bar{Q}_k$, is known in advance. In view of this, $h_k(\cdot, \cdot, \cdot)$ for those units is
\begin{equation}\label{con11}
  0 = P_k^{\star}+jQ_k^{\star} - (\bar{P}_k + j\bar{Q}_k),\quad k\in \mathbb N_{\rm L} \cup \mathbb N_{\rm E}. 
\end{equation}
The buses associated with $\mathbb N_{\rm N}$, $\mathbb N_{\rm L}$, and $\mathbb N_{\rm E}$ are called {\it PQ-buses} because the steady-state values of their active (P) and reactive (Q) power are given for the power flow calculation. 
For wind farms, solar farms and generators, their steady-state active power and bus voltage magnitude are usually known.
However, because the power loss through the transmission lines are not known a priori, active and reactive power of at least one component must be unspecified. Let this component be the generator connecting to the $k_{\rm s}$-th
bus for a given $k_{\rm s} \in \mathbb N_{\rm G}$, and assume 
\begin{equation}\label{con2}
 0 = |{\bf V}_{k_{\rm s}}^{\star}| - \bar{V}_{k_{\rm s}}, \quad
 0 = \angle {\bf V}_{k_{\rm s}}^{\star} - \bar{\theta}_{k_{\rm s}},
\end{equation}
for given $\bar{V}_{k_{\rm s}} \in \mathbb R$ and $\bar{\theta}_{k_{\rm
s}} \in \mathbb R$. Without loss of generality, $\bar{\theta}_{k_{\rm
s}} = 0$. The $k_{\rm s}$-th bus is called the {\it slack bus}, or sometimes {\it swing bus}. 
The other generators, wind farms, and solar farms are assumed to satisfy 
\begin{equation}\label{con3}
 0 = P_{k}^{\star} - \bar{P}_{k}, \quad
 0 = |{\bf V}_{k}^{\star}| - \bar{V}_{k}, \quad
 k \in \mathbb N_{\rm G} \cup \mathbb N_{\rm W} \cup \mathbb N_{\rm S}\setminus \{k_{\rm s}\}, 
\end{equation}
for given  $\bar{P}_{k} \in \mathbb R$ and $\bar{V}_{k} \in \mathbb R$. 
The buses associated with $\mathbb N_{\rm N}$, $\mathbb N_{\rm L}$, and $\mathbb N_{\rm E}$ are called {\it PV-buses} because their steady-state values of the active (P) power and voltage magnitude (V) are given for the power flow calculation. 

Finally, the power flow calculation is summarized as follows. Find ${\bf V}^{\star}_{1:N}$, $P^{\star}_{1:N}$ and $Q^{\star}_{1:N}$ satisfying \req{inter_star} and \req{con1}, \req{con11}, \req{con2}, and \req{con3}. This is what exactly done in the {\bf Power Flow Calculation} step described in Section \ref{sec:model}. 


\subsection{Relationship between the Standard Models of a Synchronous Machine}\label{sidebar-relation}
\begin{figure}[t]
  \begin{center}
    \includegraphics[clip,width=125mm]{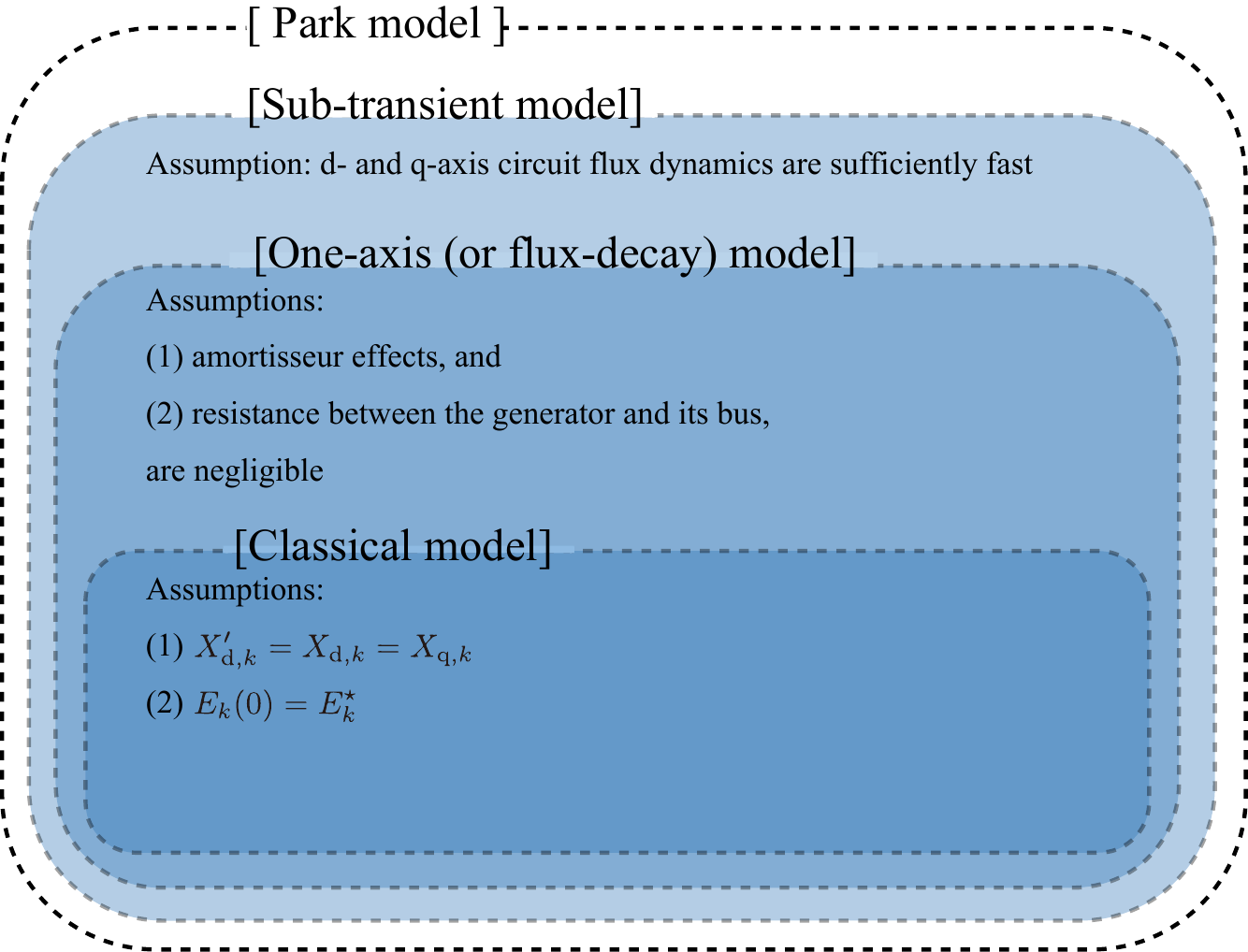}
    \caption{Relationship among four standard synchronous machine models}
    \label{relations}
  \end{center}
\end{figure}

Four standard models of a synchronous machine are commonly used in power system modeling, depending of the desired resolution of the model \cite{kundur1994power}. These models are called the {\it Park model}, the {\it sub-transient model}, the {\it one-axis model}, and the {\it classical model}. The mathematical relationship among these models is summarized here.
The Park model is one of the most well-known synchronous machine model, which is the combination of the motion dynamics and the
electro-magnetic dynamics representing the flux variation of d- and
q-axis circuits, excitation winding, and some amortisseur windings. The
equations of those two dynamics can be found in Sections 3.9 and 3.4.9 in
\cite{kundur1994power}, respectively. 
Assuming that the d- and q-axis circuit flux dynamics are sufficiently fast, the Park model can be simplified to the {\it sub-transient model} with four coils on the rotor \cite{pal2006robust}. The model consists of the motion dynamics (a second-order system) and the fourth-order system representing flux variation of excitation windings, one d-axis amortisseur winding, and two q-axis amortisseur windings.
By further assuming that the amortisseur effects are negligible and the resistance between the generator and its connecting bus is negligible, the sub-transient model can then be simplified to \req{model_sync01}-\req{model_sync02}. This model is called the {\it one-axis model}
\cite{dib2009globally,bergen1981structure} in the sense that its electro-magnetic dynamics represent the flux variation of only the excitation winding. 
In \cite{souvik2016time}, a further simplified one-axis model under the assumption 
$X_{{\rm q},k} = X_{{\rm d},k}'$ in \req{model_sync01}-\req{model_sync02} is presented. 
When $X_{{\rm d},k} = X_{{\rm q},k}$, which can be satisfied for
round rotor machines due to the symmetrical air gap between d- and
q-axes \cite{machowski2008power}, the electro-magnetic dynamics of the simplified one-axis model is reduced to
$\tau_{{\rm do},k} \dot{E}_k = -E_k + V_{{\rm fd},k}$. 
Thus, supposing that the initial value of the internal voltage $E_k$ is
its steady-state value $E_k^{\star}$ and $V_{{\rm fd},k}(t) \equiv E_k^{\star}$, it
clearly follows that $E_k(t) \equiv E_k^{\star}$ for all $t$. In that case, the
simplified one-axis model can be simply written as 
\begin{equation}
  \hspace{0mm}\left\{\hspace{-0.5mm}
  \begin{array}{rcl}
   \dot{\delta}_k &\hspace{-2mm}=&\hspace{-2mm} \bar{\omega} \Delta \omega_k,\\
   \Delta \dot{\omega}_k &\hspace{-2mm}=&\hspace{-2mm} \frac{1}{M_k}\left(P_{{\rm m},k} - d_k
    \Delta \omega_k - \frac{|{\bf V}_k|E_k^{\star}}{X_{{{\rm d},k}}} \sin (\delta_k - \angle {\bf V}_k)\right),\vspace{1mm}\\
   P_k + jQ_k  &\hspace{-2mm}=&\hspace{-2mm} 
    \frac{E_k^{\star}|{\bf V}_k|}{X_{{{\rm d},k}}} \sin (\delta_k - \angle {\bf V}_k) + j\left(
    \frac{E_k^{\star}|{\bf V}_k|}{X_{{{\rm d},k}}} \cos (\delta_k - \angle {\bf V}_k) - |{\bf V}_k|^2\right).
  \end{array}
 \right.
\end{equation}
This model is called the {\it classical model} \cite{kundur1994power,machowski2008power}. 
The relationship among the four models is shown in Figure~\ref{relations}. 

\end{document}